\documentclass[a4paper]{ar-1col}
\usepackage[utf8]{inputenc}
\usepackage{natbib}
\usepackage{graphicx}
\usepackage{lscape}
\usepackage{amsmath}
\usepackage{amssymb}
\usepackage{color}
\definecolor{darkblue}{cmyk}{1.0,0.4,0,0.5}
\definecolor{darkred}{rgb}{0.6,0.0,0.0}
\definecolor{titlecolor}{cmyk}{0.4,0.4,0.4,0}
\definecolor{headcolor}{cmyk}{0,1.0,1.0,0.30}
\definecolor{shadecolor}{cmyk}{0.03,0.03,0.12,0.0}
\definecolor{fignumcolor}{cmyk}{1.0,0.4,0,0.5}
\definecolor{marginrulecolor@val}{cmyk}{0,0,0,0.30}
\definecolor{textboxcolor@val}{cmyk}{0.12,0.04,0.08,0.0}

\usepackage{hyperref}
\hypersetup{
    colorlinks=true,
}

\newcommand\bb[1] {   \mbox{\boldmath{$#1$}}  }

\newcommand\bcdot{\bb{\cdot}}
\newcommand\btimes{\bb{\times}}

\usepackage{upgreek}
\begin{document}

% Page header
\markboth{Hennebelle $\&$ Grudi\'c }{Physics of the IMF}

% Title
\title{The Physical Origin of the Stellar Initial Mass Function}

%{The Physical Origin of the Stellar Initial Mass Function}

%Authors, affiliations address.
\author{Hennebelle, P.$^1$, Grudi\'c M.$^{2,3}$ 
\affil{$^1$ Universit\'{e} Paris-Saclay, Universit\'{e} Paris Cité, CEA, CNRS, AIM, 91191, Gif-sur-Yvette, France; email: patrick.hennebelle@cea.fr}
\affil{$^2$ Carnegie Observatories, 813 Santa Barbara St, Pasadena, CA 91101, USA, email: mgrudic@carnegiescience.edu}
\affil{$^3$ NASA Hubble Fellow}}

%Abstract

%Abstract
\begin{abstract}
Stars are amongst the most fundamental structures of our Universe. They comprise most of the baryonic and luminous mass of galaxies, synthethise heavy elements, and inject mass, momentum, and energy into the interstellar medium. They are also home to the planets. Since stellar properties are primarily decided by their mass, the so-called stellar initial mass function (IMF) is critical to the structuring 
of our Universe. We review the various physical processes, and theories which have been put forward as well as the numerical simulations which have been carried out to explain the origin of the stellar initial mass function. Key messages from this review are:
\begin{itemize}
\setlength{\leftskip}{-5.5mm}
\item Gravity and turbulence most likely determine the power-law, \\
high-mass part of the IMF.
\item Depending of the Mach number and the density distribution, several \\
regimes are possible, including $\Gamma _{IMF} \simeq 0$, -0.8, -1 or -1.3 \\
 where 
$d N / d \log M \propto M^{\Gamma_{IMF}}$. These regimes are likely universal, \\
however  the transition between these regimes is  not. 
\item Protostellar jets can play a regulating influence on the IMF by \\
 injecting momentum into collapsing clumps and unbinding gas.
\item The peak of the IMF may be a consequence of dust opacity and \\
 molecular hydrogen physics at the origin of the first hydrostatic \\ 
 core. This depends weakly on large scale environmental \\
 conditions such as radiation, magnetic field, turbulence or metallicity. \\
 This likely constitutes one of the reason of the relative universality of \\
the IMF. 
\end{itemize}
\end{abstract}

%\begin{abstract}
%Stars are amongst the most fundamental structures of our Universe. In particular, they synthetize the heavy elements, radiate and inject kinetic energy in galaxies. Last but not least they host planets. Since stellar properties are primarily decided by their mass, the so-called stellar initial mass function (IMF) is critical to the structuring 
%of our Universe. We review the various physical processes, and theories which have been put forward as well as the numerical simulations which have been carried out to explain the origin of the stellar initial mass function. Key messages from this reviews are:
%\begin{itemize}
%\item Gravity and turbulence are most likely determining the scale-free, high-mass part of the IMF.
%\item Depending of the Mach number and the density distribution, several regimes are possible  and this includes $\Gamma _{IMF} \simeq 0$, -0.8, -1 or -1.3 where 
%$d N / d \log M \propto M^{\Gamma_{IMF}}$. These regimes are likely universal, however
%the transition between these regimes is  not. 
%\item Protostellar jets probably play a  regulating influence on the IMF by injecting 
%kinetic energy in the collapsing clumps.
%\item The peak of the IMF is likely a consequence of dust opacity and molecular hydrogen physics at the origin of the first hydrostatic core. This latter weakly depends on large scale environnemental conditions such as radiation, magnetic field, turbulence or metallicity. This likely constitutes one of the reason of the relative universality of 
%the IMF. 
%\end{itemize}
%\end{abstract}

%Keywords, etc.

\maketitle

{
  \hypersetup{linkcolor=black}
  \tableofcontents
}

%Table of Contents
%\tableofcontents

% Heading 1
\section{INTRODUCTION}

In the history of our Universe, stars are playing a fundamental role in many respects. 
It is now well established that stars are responsible for synthethising 
the heavy elements such as carbon and oxygen from the primordial 
hydrogen and helium, which is a necessary step to get molecular complexity and eventually life in the Universe. Low mass stars are also hosting planets providing 
the necessary source of energy, at least on Earth, to render possible the existence of 
liquid water during the several Gyrs necessary to develop life. Massive stars on the 
other hand, exert a considerable influence on the surrounding gas but also at 
larger galactic scales through both radiation and mechanical energy injections in the 
interstellar medium (ISM). It is for instance well admitted that massive stars 
are responsible for setting the amount of galactic UV radiation and to set the star formation efficiency of giant molecular clouds. Finally the light emitted by stars 
or by ISM constituants such as dust, which have been heated by stars, remains 
the most important source of information at our disposal to study our Universe.

On the other hand, it is well established that the mass of stars is by far the most important parameters that determine their characteristic and evolution. In this respect  the initial stellar mass function (IMF) -- the distribution of birth masses of stars --  is certainly 
a fundamental quantity of our Universe. The goal of this review is to present the 
physical processes, the theories and the numerical simulations, which have been 
studied or carried out to understand how the IMF is established. 
Former reviews on this topic include \citet{bonnell2007}, \citet{Offner_2014_imf_universality} and \citet{lee2020}.
Several issues and facts 
are worth stressing and will serve here as a guideline. First of all, the entire known stellar mass range covers 
about four orders of magnitude, which likely implies that the range of spatial scales
involved in the whole star formation process is considerable. This  implies that giving the complexity of the baryonic physics, many physical processes are 
playing an important role in establishing the IMF. Second of all, two remarkable features of the IMF (see figure~\ref{fig:alphaplot}) are the peak it presents around 
0.3-0.5 M$_\odot$ suggesting the existence of a characteristic mass scale able to 
imprint the IMF, and the relatively uniform powerlaw exponent that extends
over almost 2 decades from about 1 to 100 M$_\odot$. This seemingly suggests the existence of a scale-free regime. Finally, as clearly seen from figure~\ref{fig:alphaplot}, the measured variations of the IMF remains limited, which 
has sometimes led to the conclusion that the IMF is {\it universal}. Whereas 
variations of the IMF seem to be nowdays well established, these variations may 
still appear limited given the large range of physical conditions in which 
stars form.

The plan of the review is as follows. In the second part, we briefly discuss the observational constraints that exist on the IMF as well as the analytical descriptions that have been proposed. We also discuss the efforts which have been made to
study the dense cores, thought to be the stellar progenitors, and to
infer the core mass function (CMF).
In the third part we present the physical processes that are relevant for star formation and more precisely which are believed to play a role to establish the IMF. 
The fourth part is dedicated to the analytical models that have been developed through years to explain the origin of the IMF. We explain the physical ideas and the 
physical processes that they rely on. We stress their successes and failures.  In the fifth part, we describe the numerical simulations which 
have been carried out to study the IMF, first presenting their finding and then discussing the link they have with the analytical models presented 
in the fourth section -- i.e., do they confirm or invalidate these models? 
A summary as well as future perspectives conclude the review.
%The sixth part discusses  future perspectives and conclude the review.

\section{Observational constraints on the IMF }
\label{sec:observations}

Historically, most of what is known about the IMF has come from observations \citep{Bastian_2010_imf_review,luhman2012,smith2020}. But when discussing IMF measurements it is important to note that the IMF itself is not directly observable, due to the different evolution of high- and low-mass stars. Systems old enough for low-mass stars to be on the main sequence ($\gtrsim 10\,\rm Myr$) are so old that massive stars would have already died. Systems younger than massive star lifetimes may not yet be finished forming stars, and the low-mass, pre-main-sequence stars require additional model assumptions to map their distribution on the Hertzsprung-Russell diagram onto the IMF. Hence inferences about the IMF are necessarily model-dependent, and the modeling uncertainties can be significant
\citep[e.g.][]{Bastian_2010_imf_review}. Apart from this fundamental uncertainty due to stellar evolution, there are many other practical biases and error sources that must be overcome to measure the IMF \citep[e.g.][]{kroupa_2013_imf_review,Hopkins_a}.

\begin{figure}
    \centering
    %\includesvg[width=\textwidth]{Figures/IMF_AlphaPlot.svg}
    \includegraphics[width=\textwidth]{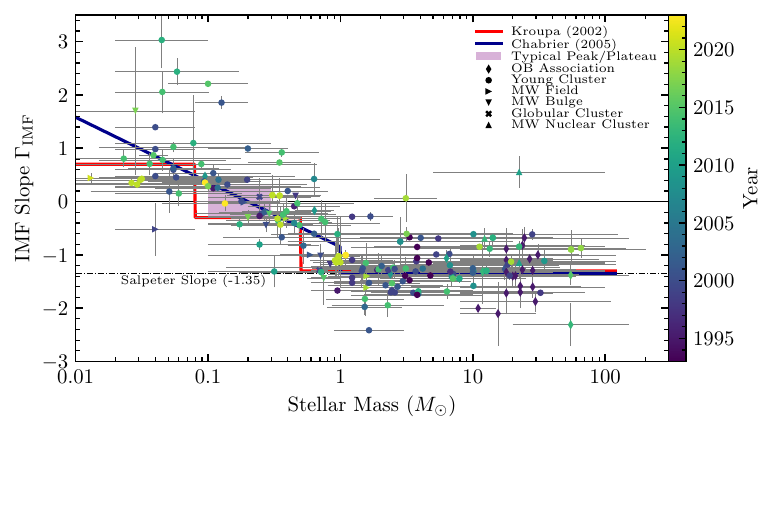}\vspace{-1.6cm}
    \caption{IMF slope as a function of the zero-age main sequence stellar mass range over which it is measured by numerous different studies. Horizontal bars give the stellar mass range over which a power-law was fitted; vertical bars plot the $\pm 1\sigma$ credible region. Note that these data are highly heterogeneous, and the error-bars mostly do not account for systematic errors. \href{https://data.obs.carnegiescience.edu/starforge/IMF_AlphaPlot.pdf}{Version with clickable hyperlink points}. The data compilation and code are available as supplementary material or at \href{https://github.com/mikegrudic/alphaplot}{these} \href{https://data.obs.carnegiescience.edu/starforge/alphaplot.tar.gz}{links}.}
    \label{fig:alphaplot}
\end{figure}

An encouraging development in recent decades has been the soundness of the modeling, analysis, and statistical practices used to estimate the IMF. Once common practice, it is increasingly rare for IMF studies to not quote at least some estimate of the statistical error with their measurement. Bayesian techniques that avoid binning and make full use of all information are in wider use \citep[e.g.][]{weisz_2013_imf_methods,Dib_2014_IMF_var_analysis}, enabled by advances in computation. It is also increasingly common to use more than one set of model assumptions to assess the impact of systematics, allowing statistical and systematic/modeling errors to be assessed separately \citep[e.g.][]{dario:2012.orion.imf,hosek2019}. And the statistics of binary and multiple systems are now better understood \citep{moe_distefano2016,offner_2022_multiplicity_review}, allowing the effects of unresolved multiplicity to be estimated. These developments have made IMF measurements from recent years generally much easier to interpret critically, although signficant modeling uncertainties remain.

A convenient way to parametrize the IMF measured in a certain stellar mass range is to linearize it as a power-law and report the inferred power-law index (the IMF slope $\Gamma_{IMF}$) \citep{salpeter_slope,scalo_1998_imf_review}
 \begin{eqnarray}
     {d N \over d \log M} \propto M {\cal N } \propto M^{ \Gamma_{IMF}}.
 \end{eqnarray}
With the convention adopted in this review, where the IMF ``slope" $\Gamma_{IMF}$ is the slope in a log-log histogram of stellar masses,  the \citet{salpeter_slope} slope is $\Gamma _{IMF}=-1.35$.
Figure \ref{fig:alphaplot} presents a compilation of IMF slopes reported in the literature for numerous different systems in the Local Group (see also \citealt{kroupa_imf,hillenbrand_2004_imf_proceedings}), for various subintervals across the mass range of stars and sub-stellar objects. For comparison we also plot the two presently most-popular parametrized models \citep{kroupa_imf,chabrier2003,chabrier_imf}. Note that Kroupa modeled the IMF as a piecewise function on this diagram, whereas Chabrier modeled the $<1M_\odot$ range as a log-normal distribution in stellar mass with a slope that various continuously. Studies comparing these models have generally found them to be similarly compatible with observations, as well as other parametrizations \citep{Parravano_2011_TPL_IMF, Maschberger_2013_IMF_parametrization}.

However it is not possible to draw a single curve through all data points in Figure~\ref{fig:alphaplot} that avoids tension with all measurements; there are many examples of mutually-incompatible slopes measured in similar stellar mass ranges. Due to the many uncertainties, systematic errors are not insignificant, so they certainly account for some of the variation in figure~\ref{fig:alphaplot} and relieve some of the tension, but not all of it \citep[e.g.][]{Dib_2014_IMF_var_analysis}. The strong hypothesis of a true IMF universality is unlikely.

Questions of universality aside, the IMF clearly has a remarkable degree of {\it regularity}: IMF variations may exist, but they are typically not large, at least in nearby galaxies where the IMF can be constrained. At least three fairly robust features are of interest. First, the ``peak"  of the IMF, which corresponds to the transition between $\Gamma_{IMF} < 0$ and $\Gamma_{IMF} > 0$, is usually found in a fairly narrow mass range of $\sim 0.1-0.3 M_\odot$. Second, the range over which the slope varies between -1 and 1 -- corresponding to the interval $\left[-\sigma^2,\sigma^2\right]$ for a log-normal form -- is about an order of magnitude. Hence the IMF samples a {\it broad} range of stellar masses, with a log-dispersion $\sigma \sim 0.5-0.6\,\rm dex$. And at high masses, the slope tends to be close to the value of -1.35 measured by \citet{salpeter_slope}, with relatively little variation in the mass range $1-150M_\odot$ except perhaps in the Milky Way nuclear cluster \citep{2010ApJ...708..834B,2013ApJ...764..155L} and some massive starburst clusters \citep{schneider2018, hosek2019}. %though in a large number of observations, shallower values are reported, with 
%$\Gamma_{IMF} \simeq -0.8$ being frequently inferred. These features -- the peak, variance, and high-mass slope -- are robust phenomena that a theory of the IMF must account for. % and, quoting \citet{kroupa_imf}:
%\begin{quote}
    %This apparent universality of the IMF is a challenge for star formation theory, because elementary considerations suggest that the IMF ought to systematically vary with star-forming conditions. 
%\end{quote}

%The various constraints we have on the IMF are briefly presented as well as the functional description that have been put forward. 
%The question of the apparent {\it universality} is discussed.
%The possible link with the core mass function (CMF) is presented and the meaning and limit of core definition are stressed. 
%The various constraints available on the CMF are presented. 
There are many different techniques for constraining the IMF, which we outline in the rest of this section, highlighting some significant recent developments.

\subsection{Resolved star counts}
Directly counts of stars of different masses remain a key technique for measuring the IMF within the Local Group, where the necessary resolution is available. Some interesting developments on this front include:

% The MW IMF is the reference distribution against which the IMF of both stellar clusters and external galaxies can be compared. Thus, deriving its accurate shape remains of the utmost importance. Multiple reviews of the IMF and its variations have been published over the last decade \cite{Bastian2010, Offner2014, Offner_2016_IMF_progress}. Significant progress in the last decade from mainly 2 sources: 
\begin{itemize}
    \item The Gaia mission has greatly extended the census of stars in the Solar neighborhood with measured spectra, and has provided high-precision parallaxes and proper motions. This permits more-reliable membership assignment for young star clusters in the Solar neighborhood, which can revise the inferred IMF significantly \citep{Luhman_2018_taurus_IMF}. The Milky Way field IMF can now be measured with much larger sample size \citep{mor2019,sollima2019}, down to the brown dwarf and substellar regime \citep{Kirkpatrick_2021_field_substellar_IMF}. This discipline is now effectively merged with galactic archaeology, because it is at level of precision where the specifics of how the Galaxy assembled and the IMF are tightly entangled. Trends in the IMF with age and metallicity, proposed as far back as \citep{schmidt:1963.imf.variations}, seem to persist under the scrutiny of this greatly-expanded dataset \citep{2023Natur.613..460L}.
%    \item and nearby star-forming regions \citep{Luhman_2012_lowmass_IMF}. JWST expected to provide more details and extend observations to lower masses.
    \item Various measurements in the most massive and dense Local Group clusters have indicated a relatively top-heavy IMF, $\Gamma _{IMF} \sim -0.8$ \citep{2013AJ....145...46L,2013ApJ...764...73P,schneider2018,hosek2019}. This result, in combination with the well-established top-heavy IMF of the Galactic nuclear cluster \citep{2010ApJ...708..834B,2013ApJ...764..155L}, may indicate some connection between the IMF slope and the extreme conditions in which such massive clusters formed. A key caveat does remain: such dense clusters may undergo dynamical evolution over short ($\lesssim 1\,\rm Myr$) timescales. This has two competing effects: mass segregation could make the mass function sampled in a limited field look top-heavy \citep{2013ApJ...764...73P}, but the preferential ejection of massive stars should also steepen the observed mass function over time \citep{banerjee.kroupa:2012.r136.imf}.
    \item The PHAT HST survey of the disk of M31 has produced a large, homogeneous catalogue of young, resolved star clusters, allowing \citet{weisz_2015_survey} to make the most precise measurement of the $\gtrsim 2\,M_\odot$ IMF slope to date: 
    $\Gamma_{IMF} = -1.45_{-0.06}^{+0.03}$. The posterior on the modeled intrinsic dispersion $\sigma_{\rm \Gamma}$ is consistent with 0, i.e. there is no strong evidence of major cluster-to-cluster variation, or random systematic errors. This is an important development because observations available in the Milky Way are subject to greater uncertainties in reddening, membership, etc, and it is difficult to distinguish between true intrinsic IMF variation and systematic effects \citep{massey_2003_imf_slope}.
    \item It is now possible to use large grids of N-body simulations to constrain the overall IMF in globular clusters, with general preference for an IMF that is bottom-light, but with a normal high-mass slope $\Gamma _{IMF}\sim -1.3$ \citep{2017MNRAS.472..744B,2020MNRAS.494.4226E,2023MNRAS.521.3991B}. The formation process, initial conditions, initial multiplicity properties, binary stellar evolution, and evolution in a galactic context all have major uncertainties, so the model dependence of such measurements should be assessed carefully.
 %Overall consensus seems to be a near-universal IMF in the Local Group (at least to stellar mass scales), \citet{Offner2014}.

 %Allowed variations around standard fitting functions \citep{kroupa_imf,chabrier_imf} in slope are roughly $\pm 0.2$ and in characteristic mass a factor 2. Details in \citet{Dib_2014_IMF_var_analysis, Dib_2017_IMF_variations_massive}.
 
 %Mention 30Dor, Arches \mike{Hosek 2019}, CMZ

 %\mike{Summarize key hallmarks of the IMF that any theory/model must account for, and also what is {\it not} a useful constraint (e.g. DO try to account for lack of turnover/peak variation, do NOT strive to get a cluster-level IMF of exactly -2.35)}
    
\end{itemize} 

\subsection{Unresolved populations}

Studying the IMF unresolved populations is inherently harder and must rely on careful modeling where several properties of the observed stellar population are degenerate with each other. However, they allow us to probe more extreme regions of the SF parameter space than that in the Local Group. 
\begin{itemize}
    \item Efforts to constrain the low-mass IMF in massive early-type galaxies \citep{smith2020} via high-resolution spectral modeling \citep[e.g.][]{conroy_vandokkum_ellipticals} have continued, and the trend toward greater mass-to-light ratios (i.e. a steeper $\approx 0.3-0.8M_\odot$ slope) in the central regions of more-massive galaxies persists \citep{2017ApJ...845..157N,gu:2022.ellipticals-imf}. \citet{gu:2022.ellipticals-imf} documents correlations between the mass-to-light ratio and galactic properties such as velocity dispersion, [Mg/Fe], and [Fe/H].
    %\item The ratio of $\rm^{13}CO$ and $\rm C^{18}O$ is sensitive to the high-mass IMF (among other factors), and has been used to study it in lensed, dusty starburst galaxies at $z\sim 2.3-3.1$ with ALMA \citep{2018Natur.558..260Z}. The observed abundance ratios, similar to $z\sim 0$ starbursts like Arp 220 and NGC 253, match models with quite shallow ($\Gamma _{IMF} \sim 0$) high-mass slopes, assuming a fixed $100 M_\odot$ maximum stellar mass.
    \item The Optical Gravitational Lensing Experiment (OGLE) has provided constraints on the low-mass ($\sim 0.01-0.5 M_\odot$) IMF in the Galactic bulge/bar region. \citet{wegg2017} found the IMF to be consistent with the standard IMFs measured in the local field. \citet{chabrier2023} found that these data specifically prefer the \citet{chabrier_imf} parameters for the log-normal parameterization over those of \citet{chabrier2003}, that this parametrization is preferred over the \citet{kroupa_imf} broken power-law, and that the central region may be bottom-heavy.
\end{itemize}

%Such models inferrede significantly "bottom heavy" IMFs in early elliptical galxies \citep{conroy_vandokkum_ellipticals,Conroy_2017_parametric_IMF, gu:2022.ellipticals-imf}.
%ULIRGs?

\subsection{The core mass function}
\label{core}
Prestellar dense cores \citep{motte98,ward2007,konyves2015}, often observed in the continuum, 
are believed to be the progenitors of 
stars. That is to say they may constitute the very final coherent gaseous reservoir
out of which stars build their masses. This opens the possibility that 
their mass distribution, the CMF, may be at the origin of 
the IMF. For this to be true, the correspondance between this material and the 
final stellar mass, must be sufficiently good and to a large extent, assessing this remains 
a challenge \citep[][]{pelkonen2021}. In particular, defining core boundaries is a complex question moreover since stars are generally found in multiple systems, cores likely lead to the formation 
of several objects. One of the main arguments in favor of the importance of cores in determining 
the shape of the IMF has  been  their relatively similar shapes
\citep[e.g. see Figure~16 of][]{konyves2015} as inferred for instance
in the Herschel Gould Belt survey \citep{andre2010}.
Both present 
a powerlaw or lognormal shape at high mass and a peak around 0.3 M$_\odot$ for the IMF 
and 0.5-1 M$_\odot$ for the CMF. 
The powerlaw of the high mass part inferred for the 
CMF appears to be compatible or close to the value $\Gamma_{IMF}$ usually measured 
for the IMF. 

Several studies have questioned the universality of this result. In recent
observations,  the CMF have been inferred in the 
active star-formation region W43-MM1 and W43-MM2$\&$MM3
\citep{Motte_CMF_slope_2018,pouteau2022} and it has been found that the 
exponent of the powerlaw is significantly flater than $\Gamma_{IMF} \simeq -1.3$ 
with typical values around -0.9 to -0.95 \citep[see however][for possible biases]{padoan2023}. Interestingly, as mentioned above, 
shallower values of $\Gamma_{IMF}$ have also been inferred in massive and dense 
local stellar clusters \citep[e.g.][]{hosek2019} suggesting a possible 
common origin.

Second of all, the reliability of the peak of the CMF has been investigated 
by \citet{louvet2021} where the cores have been extracted from 
observational and simulation data smoothed at various spatial resolutions.
It has been found that the identified cores strongly vary with resolution and 
in particular that the peak of the CMF shifts to higher masses as the spatial resolution 
is reduced. This result questions the existence of a clear resolution 
independent core definition as well as the existence of a CMF peak.  
It suggests that the peak of the CMF estimated from
the Herschel Gould Belt survey may be  overestimated. Interestingly, it 
has been recently suggested that the mass function of bound cores 
(seen in C18O) in Orion may present a peak around 0.1 M$_\odot$
\citep{takemura2023}. 

To conclude, whereas the correspondance between the CMF and the IMF needs to be further investigated, the cores 
certainly contribute to the star formation process and as such models of star formation studying the IMF 
should be able to reproduce the CMF as it certainly constrains the gas physical properties.

 %Lot of theoretical effort to expalin and "map" the CMF to the IMF (e.g., \citet{guszejnov_2015_CMF_IMF}). However, there is no widely accepted definition on what a "core" is, the structures picked out by observations do not necessarily represent objects with a clear physical definition 

\section{Physical processes}
%\mike{Make cartoon?}
In this section we review the various physical processes that are relevant in the context of the star formation process and the establishment of the IMF. We proceed 
broadly in {\it chronological} order, that is to say discussing the processes in 
the sequential order with which they become relevant. All the discussed 
topics are complex phenomena which request lengthy developments that are beyond 
the scope of the present review where only the most important aspects for the origin of the IMF are considered. Complementary information can be found for instance in the recent review by 
\citet{girichidis2020}. The equations that are generally employed to perform theoretical studies of star formation are the usual fluid equations that apply to a self-gravitating, radiative and magnetized gas. They can be found for 
instance in \citet[][]{commercon2011, krumholz2012, tomida2013}.

\subsection{Thermal structure of the cold ISM}
\label{cooling}

\begin{figure}
    \centering
    \includegraphics[width=\textwidth]{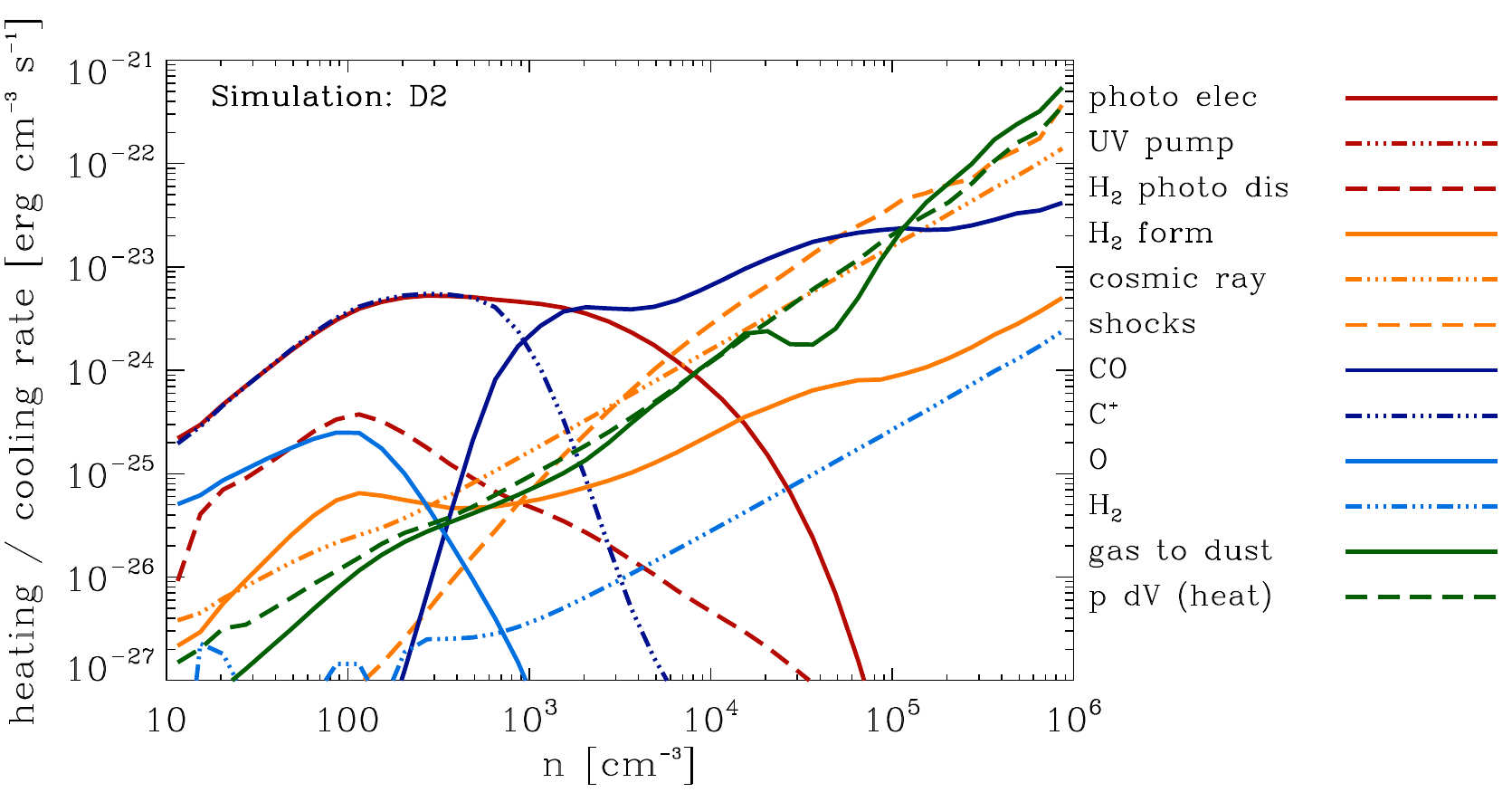}
    \caption{Heating and cooling rates of different physical processes as a function of density in a Smoothed Particle Hydrodynamics simulation of a $10^4 M_\odot$ giant molecular cloud with radius $6 \rm pc$, with compisition typical of a Solar neighborhood molecular cloud. Reproduced with permission from \citet{glover.clark:2012.molecules.for.sf}; see their figure 8 for full details.}
    \label{fig:heating.cooling}
\end{figure}

\begin{table}[]
    \centering
    \begin{tabular}{c|c|c|c|c}
     Process  & Relevant $n_{\rm H}$ at $Z_\odot$ & Key species & Rate per H & Refs. \\
     \hline \hline
     Photoelectric heating & 10$^{-2}$-10$^4$ cm$^{-3}$ & Grains, FUV & $\propto Z_{\rm d} G_{\rm 0}$ & 1,2,3 \\
     Cosmic ray heating & 10$^{-2}$-10$^4$ cm$^{-3}$ & Cosmic rays, H, H$_2$ & $\propto \xi_{\rm CR}$ $f_{\rm atten}\left(N_{\rm H}\right)$ & 4,5,6 \\
     Gravitational PdV work & $10^5 + \rm cm^{-3}$ & Gas & $\propto n_{\rm H}^{1/2} T$  & 7,8,9\\ 
     Turbulent dissipation & $10^2-10^6 \,\rm cm^{-3}$ & Gas & $\propto \sigma^3/l$ & 10\\ 
     \hline 
     C$^+$ fine structure & $1-10^3\,\rm cm^{-3}$ & C$^+$, H, H$_2$ &  $\propto Z_{\rm C} \exp\left(-91/T\right) n_{\rm H}$ & 3,11 \\
     C, O fine structure & $1-10^2\,\rm cm^{-3}$ & C, O, H, H$_2$ &  See refs. & 3,14 \\
     CO rot. lines (thin) & $10^2-10^3\,\rm cm^{-3}$ & CO, H$_2$ & $\propto Z_{\rm C} n_{\rm H} T^{3/2}$ & 12,13\\
     CO rot. lines (thick) & $10^3-10^4\,\rm cm^{-3}$ & CO, H$_2$ & $\propto \|\nabla \mathbf{v}\| n_{\rm H}^{-1} T^4 $  & 12,13 \\
     H$_{\rm 2}$ transitions & None (at $ Z_\odot$) & H$_{\rm 2}$, HD & See refs. & 14,15 \\
     Dust-gas collisions  & $10^4+\rm cm^{-3}$& Gas, grains & $\propto Z_{\rm d} n_{\rm H} T^{3/2} \left(T-T_{\rm d}\right) $ & 16,17,18 \\

    \end{tabular}
    \caption{Summary of processes determining the thermal structure of the cold ISM relevant to fragmentation and the IMF. For each process we give the estimated range of densities at which the process is important in Solar neighborhood ambient conditions, the scaling law for the rate per H nucleus, and useful references for detailed formulae. Note that the density range in which different processes dominate will generally vary with metallicity and radiation field. $H_{\rm 2}$ transitions are unimportant at $Z_{\odot}$, but are the main coolant at $\lesssim 10^{-2} Z_\odot$ until dust coupling is efficient. 
    %An interactive app for experiments with ISM cooling physics is available \href{https://github.com/mikegrudic/ismulator}{here}. 
    $T$, $T_{\rm d}$: gas and dust temperatures, $G_{\rm 0}$: FUV radiation field, $\xi_{\rm CR}$: cosmic ray ionization rate, $f_{\rm atten}$: cosmic ray attenuation factor, $\alpha_{\rm turb}$: GMC turbulent virial parameter, $\sigma$: ISM velocity dispersion, $l$: Scale at which $\sigma$ is defined, $Z_{i}$: relative mass fraction of species $i$. 1: \citet{bakes.tielens:1994.photoelectric}, 2: \citet{weingartner.draine:2001.photoelectric}, 3: \citet{draine:ism.book}, 4: \citet{indriolo:2012.cr.rate}, 5: \citet{neufeld.wolfire:2017.cr.attenuation}, 6: \citet{padovani:2023.cr.gmcs}. 7: \citet{low76} 8: \citet{Masunaga98} 9: \citet{grudic:2023.opacity.limit} 10: \citep{maclow:1999.turb.dissipation} 11: \citet{wiesenfeld:cplus.cooling}, 12: \citet{2017ApJ...843...38G}, 13: \citet{whitworth:2018.co.cooling}, 14: \citet{glover.jappsen:2007.chemistry.network} 15: \citet{glover.abel:2008.h2.cooling}, 16: \citet{hollenbach.mckee.1979.molecular.cooling}, 17: \citet{goldsmith:2001.co.depletion}, 18: \citet{krumholz:2011.molecular.gas}.}
    \label{tab:cooling}
\end{table}

Stars form in the cold, molecular phase of the ISM, where cooling times are generally shorter than other dynamical timescales, so the temperature  $T$ is determined mainly by the balance of heating and cooling processes. This temperature may be important for the IMF because it determines e.g. the Mach number of turbulence and all resulting density statistics, and the Jeans mass at a given density. There are multiple heating and cooling processes that scale steeply with $T$, limiting the overall degree of temperature variation. Therefore, the star-forming ISM is often approximated as isothermal, e.g. at $\sim 10 \rm K$ in conditions found in the Solar neighborhood. This can be a useful approximation, but for the purposes of IMF theory it is important to note that the temperature {\it will} generally vary with the gas density and chemical composition, the properties and abundance of dust, and the intensity and spectral energy distributions of photons and cosmic rays. The thermal balance of the ISM is a complex and extensive subject addressed by various dedicated texts, e.g. \citet{tielens:ism,osterbrock.ism.book,draine:ism.book}. 

The situation in Solar neighborhood conditions is summarized in Figure~\ref{fig:heating.cooling} from \citet{glover.clark:2012.molecules.for.sf}, which plots the volumetric heating and cooling rates of the various processes active in a GMC as a function of density. Table \ref{tab:cooling} lists some of the key processes determining the thermal structure of the cold, star-forming ISM, and the densities at which different processes are important (again, in Solar neighbourhood conditions). The temperature of the cold ISM up to moderate ($\sim 10^4 \rm cm^{-3}$) densities is determined by heating from the grain photoelectric effect, cosmic rays, and turbulent dissipation, and cooling from C$^+$ fine-structure and CO rotational transitions. This shapes the decrease in temperature from a few $100 \rm K$ at $n_{\rm H} \sim 10\,\rm cm^{-3}$ to $10-20 \rm K$ at $10^4 \rm cm^{-3}$, which will affect the turbulent density statistics that are important for gravo-turbulent IMF models. At higher densities, gas cooling due to gas-grain coupling becomes increasingly important, and $T$ approaches the dust temperature  $T_{\rm d}$. 

At lower metallicity the hierarchy of different heating/cooling processes can be different. CO cooling is never significant at $\lesssim 10^{-2} Z_\odot$ due to the lesser shielding and maximum abundance. The dust-coupled phase at low metallicity also occurs at higher density, roughly $n \propto Z_{\rm d}^{-1}$. Due to the overall lack of coolants and dust shielding, the equilibrium temperature will generally be significantly higher at low $Z$: state-of-the-art thermochemical MHD models typically find a typical factor of $\sim 2-3$ increase in $T$ at fixed $n_{\rm H}$  going from Solar conditions to $10^{-2} Z_\odot$ \citep{bialy2019,guszejnov2022, kim:2023.ism.model}. This implies that the Jeans mass at a given density or pressure will be greater at low $Z$, potentially affecting fragmentation. At very low metallicity ($\lesssim 10^{-3}-10^{-5} Z_\odot$) there will be effectively no dust-coupled phase at all, and H$_{\rm 2}$ is the only effective low-temperature coolant \citep{sharda.krumholz:2022.imf}. The first stars would have formed in this dust-free regime, and the inevitably-higher temperature and Jeans mass naturally results in a more top-heavy IMF \citep{klessen.glover:2023.pop3.sf}.

\subsubsection{Dust-coupled regime}
At sufficiently high density and dust abundance, $T \sim T_{\rm d}$, dust-gas collisions become so frequent that the temperature is determined by the balance of dust cooling ($\propto \kappa T^4$, $\kappa \propto T^\beta$, $\beta \sim 1-2$) with radiative or mechanical heating, and the overall temperature evolution can be determined by patching together the various asymptotic regimes \citep{grudic:2023.opacity.limit}. If absorption of IR radiation by dust is the main heat source, then the gas will remain essentially isothermal at a certain radiative equilibrium temperature:
\begin{equation}
T_{\rm abs} = \left(\frac{T_{\rm IR}^\beta u_\mathrm{IR}}{a}\right)^\frac{1}{4+\beta} \approx 7  \mathrm{K} \left(\frac{T_{\rm IR}}{20 \,\rm K}\right)^\frac{1}{3} \left(\frac{u_{\rm IR}}{1 \,\rm eV\,\rm cm^{-3}}\right)^\frac{1}{6},
\end{equation}
where $T_{\rm IR}$ is the effective blackbody temperature of the incident IR SED, $\beta \sim 1-2$ is the power-law scaling exponent of the dust opacity law $\kappa \propto T^\beta$, and $u_{\rm IR}$ is the radiation energy density. The galactic background, reprocessed emission powered by incident light absorbed in the outer parts of the cloud, and emission from nearby protostars can all contribute to the radiation field (Section~\ref{sec:radiation}). If $PdV$ work due to gravitational collapse dominates over radiative heating, then the balance of heating and cooling gives
\begin{equation}
\label{eq:Tff}
    T_{\rm PdV} \approx 8 \mathrm{K} \left(\frac{C_{\rm ff}}{Z_{\rm d}}\right)^\frac{1}{3+\beta}\left(\frac{n_{\rm H}}{10^{10} \rm cm^{-3}}\right)^\frac{1}{2\left(3+\beta\right)},
\end{equation}
where $Z_{\rm d}$ is the mass fraction of dust scaled relative to the Solar neighborhood, and $C_{\rm ff} \sim 1$ encodes the rate of collapse relative to freefall \citep{Masunaga98}. Note that for $\beta \sim 1-2$ this implies that $T \propto n_{\rm H}^\frac{1}{8}-n_{\rm H}^\frac{1}{10}$, i.e. nearly but not exactly isothermal.

\subsubsection{Opacity limit density and minimum Jeans mass} 
\label{opacitylim}
Once the rate of $PdV$ work exceeds the maximum rate of radiative diffusion of cooling radiation from the collapsing core, the gas can no longer collapse quasi-isothermally. Instead, cooling becomes inefficient and the gas will evolve almost adiabatically; this transition in thermodynamic evolution is known as the opacity limit \citep{low76,rees1976,silk:1977.opacity.limit}. Let us stress that such high densities are not met in the ISM except in collapsing regions. 
The density of the adiabatic transition $n_{\rm ad}$ depends mainly on dust opacity and the local radiation energy density \citep{Masunaga98,Vaytet17,grudic:2023.opacity.limit}:

\begin{equation}
    n_{\rm ad} \approx 5.7\times 10^{10}\, \mathrm{cm}^{-3} \times \min \left(Z_{\rm d}^{-\frac{10}{4\beta+7}},  Z_{\rm d}^{-2/3}\left(u_{\rm rad}/\mathrm{100\, eV\,cm^{-3}}\right)^\frac{2-\beta}{6}\right),
\end{equation}
where $Z_{\rm d}$ is the Solar-scaled dust opacity parameter\footnote{Specifically, in our parametrization $Z_{\rm d}=\sigma_{\rm d}\left(10 \rm K\right)/\sigma_{\rm 0}$ where $\sigma_{\rm d}\left(T_{\rm d}\right)$ is the Planck-mean dust emission cross section per H nucleus at dust temperature $T_{\rm d}$, and $\sigma_{\rm 0} = 4\times 10^{-26}\,\rm cm^{2}\,\rm H^{-1}$ is a fiducial value typical of widely-used dust models.} and $u_{\rm rad}$ is the local radiation energy density; this density is mainly sensitive to $Z_{\rm d}$, so the opacity limit occurs at higher density in low-dust conditions. This expression approximates the transition between the radiation-dominated and mechanical heating-dominated regimes.

The minimum Jeans mass reached at $\sim n_{\rm ad}$ is then \citep{grudic:2023.opacity.limit}:
\begin{equation}
    M_{\rm J,min} \approx 2\times 10^{-3} M_\odot \max\left(Z_{\rm d}^{-\frac{1}{4 \beta + 7}}, Z_{\rm d}^{1/3} \left(u_{\rm rad}/\mathrm{100\, eV\,cm^{-3}}\right)^\frac{2\beta+5}{24}\right).
    \label{eq:compression_mj}
\end{equation}
Hence, if the radiation field is weak or dust is scarce, the minimum Jeans mass is extremely robust to metallicity and dust properties, varying by at most a factor of $\sim 3$ over the range $\beta\sim 1-2$ and $Z_{\rm d}\sim 10^{-4}-1$. $M_{\rm J,min}$ only scales more steeply with environmental conditions in the radiation-dominated regime, in which $ M_{\rm J,min} \propto Z_{\rm d}^{1/3} u_{\rm rad}^\frac{2\beta +5}{24}$; this would apply mainly to dust-rich, radiation-dense environments.

The existence of a minimum Jeans mass has led to the idea that this 
represents the mass of the smallest objects that may form under the influence 
of gravity and thus this may explain for instance the mass of the brown dwarfs. 
This argument should be considered with care because when a 
fragment forms, this leads to the formation of
a thermally supported core, as explained below and collapse does not proceed immediately. 

\subsubsection{First hydrostatic core and second collapse}
\label{first_core}
As expressed above, $M_{\rm J,min}$ is the smallest self-gravitating mass of gas that is linearly unstable.
As the opacity limit is crossed, a quasi-hydrostatic structure is formed 
\citep{larson1969,masunaga_eqs_highgamma_ref,Vaytet17}: the first hydrostatic core (FHSC) also called the first Larson core. In order for the FHSC to collapse promptly to protostellar density a fragment must become somewhat more massive, so that its central temperature exceeds $T_{\rm dis}\sim 1500\rm K$, triggering a runaway collapse as energy goes into dissociating H$_{\rm 2}$. Indeed the dissociation 
of the H$_{\rm 2}$ molecule requires about 4.5 eV which corresponds to the 
thermal energy of a gas at several thousands of Kelvin degrees. This is considerable
and it constitutes a very substantial source of cooling that is able 
to maintain the temperature nearly constant and around $T_{\rm dis}$ (with an effective adiabatic index
$\Gamma_{ad} \simeq 1.1$) during the second collapse 
and up to the formation of the second hydrostatic core. To properly describe the gas evolution at these densities, it is necessary to take into account 
 the translational, rotational and vibrational degrees
of freedom of molecular hydrogen  as well as to
 include the dissociation of molecular hydrogen, and the
ionizations of hydrogen and helium \citep{black1975,saumon1995,tomida2013}. 
Note that there are still significant uncertainties due to the 
unknown  H$_{\rm 2}$ ortho:para ratio \citep{vaytet2014} and dust properties in collapsing cores \citep{guillet2020,2023MNRAS.518.3326L}.

Assuming the evolution past $n_{\rm ad}$ is isentropic, and that the adiabatic index of H$_{\rm 2}$ transitions from $\gamma_1=5/3$ to $\gamma_2=7/5$ at $T_{\rm ex} \sim 150 \rm K$  \citep{Vaytet17,hennebelle_2019}, the minimum mass that can undergo second collapse, is
\begin{equation}
\begin{split}
    M_{\rm FHSC} &\approx k_{\rm L} M_{\rm J}^{\rm min} \left(\frac{T_{\rm ex}}{T_{\rm ad}}\right)^\frac{3\gamma_1 - 4}{2\gamma_1 -2 }\left(\frac{T_{\rm diss}}{T_{\rm ex}}\right)^\frac{3\gamma_2 - 4}{2\gamma_2 -2} \\
    &\approx   0.03\,M_\odot \max\left( Z_{\rm d}^\frac{2}{4 \beta + 7}, Z_{\rm d}^{1/3} \left(u_{\rm rad}/\mathrm{100\, eV\,cm^{-3}}\right)^\frac{4\beta+1}{48}\right)
    \label{Mfhsc}
\end{split}
\end{equation}
which is again quite insensitive to the opacity parameter $Z_{\rm d}$ except in radiation-dense, dust-rich environments. %More realistically, $\gamma$ has a more complex behavior that, as recalled above depends on the ratio of H$_{\rm 2}$ spin isomers \citep{saumon1995}, and the evolution from the adiabatic transition to the second collapse will be given by the adiabat for that equation of state.

Bound cores that do not grow to $M_{\rm FHSC}$ can still collapse, but over a Kelvin-Helmholtz timescale, an order-of-magnitude longer than the $\sim 10^3-10^4 \rm yr$ accretion timescale generally required to reach $M_{\rm FHSC}$. During this time a low-mass fragment can continue to accrete, or be accreted or disrupted by a more-massive core or protostar, so the formation of protostars with $M\lesssim M_{\rm FHSC}$ from normal fragmentation is likely rare. A low-mass truncation in the IMF at $\sim M_{\rm FHSC}$ is therefore expected, and indeed is found in numerical simulations. Recently \citet{leeh2018b} and \citet{hennebelle_2019} 
advocated for the importance of the FHSC in setting the peak of the IMF as described in Section~\ref{FHSC}.

\subsection{Supersonic turbulence}
\label{turbulence}
Turbulence is ubiquitous in the ISM and comprehensive reviews have been dedicated 
to this topic \citep{maclow_star_formation_ism,elmegreen_2004_turbulence,McKee07,HF12}.
Here we focuss the discussion on the aspects which are most fundamental in the
context of star formation. 

\subsubsection{Density PDF}
\label{dens_PDF}
In a strongly compressible medium, such as isothermal gas, 
supersonic turbulence leads to a broad density distribution. This is because 
the fluid particles experience series of compressions induced by 
the ram pressure of the turbulent eddies. Several models have proposed 
that the turbulent density fluctuations constitute the seeds for 
the formation of future stars and may strongly influence the stellar mass spectrum
\citep{padoan1997,hc08,hopkins2012} as explained in Section~\ref{gravo_turb_sup}. 

Since the pioneering works of \citet{vazquez94} 
and \citet{nordlund1999}
 various simulations  of hydrodynamic supersonic
turbulence, have established that the  density PDF is well
represented  by a log-normal form 
\citep[e.g.][]{kritsuk2007,Federrath08}
given by
\begin{eqnarray}
\label{Pr0}
{\cal P}(\delta) &=& {1 \over \sqrt{2 \pi \sigma_0^2}}
\exp\left(- { (\delta - \bar{\delta})^2 \over 2 \sigma_0 ^2} \right) , \;
 \delta = \ln (\rho/ \rho_0 ), \\
 \bar{\delta}&=&-\sigma_0^2/2 \;
,  \; \sigma_0^2 \simeq \ln (1 + b^2 {\cal M}^2),
\nonumber
\end{eqnarray}
where ${\cal M}$ is the Mach number,  $\rho_0$ is the mean density, 
$\sigma_0$ is the variance of $\delta$
and $b \simeq 0.5-1$.
%\mike{depending on the specifics of the turbulent driving mechanism \citep{federrath_sim_2010}}
The origin of this log-normal PDF has been sought in the
central limit theorem and the series of statistically independent shocks that swept
the fluid particles. This picture has been recently further investigated 
by \citet{rabatin2022} which have calculated the density PDF that would result from 
a finite number of shocks leading to improve agreement with simulations. 

Another useful formulation of the density PDF has been given by 
\citet{hopkins_isothermal_turb} who has transposed the PDF 
obtained by \citet{castaing1996} to describe the velocity PDF of intermittent incompressible flows subject to a log-Poisson cascade. 
%The proposed functional form is given by 
%\begin{eqnarray}
%    \nonumber
%    {\cal P}(\delta) = \sum _{m=1} { \lambda ^m e ^{-\lambda} \over m !} { u ^{m-1} e^{-u} \over (m-1) !}, \\
%    u = { \lambda \over 1 + T} - {\delta \over T}, u \ge 0.
%\end{eqnarray}
%where $\lambda = \sigma_{0 } / (2 T^2) $. In this expression 
%where $\lambda = S_{\delta } / (2 T^2) $. In this expression 
%$S_{\delta }$
%is the variance of $\delta= \ln \rho$ and 
%$T$ is a free parameter. 
Confronting to large suite of numerical simulations, 
\citet{hopkins_isothermal_turb} found that it provides excellent and robust fit. 

%\mike{From here can talk about how the wide range of densities imprints a range of Jeans masses that could }

\subsubsection{Scale dependence of flow quantities}
\label{scale_v}
A very fundamental property of turbulence is that the flow quantities 
such as the mean velocity difference between two spatial locations at distance 
$\delta l$ from each other, is scale 
dependent. Typically, it is generally assumed that in an homogeneous and 
isotropic flow \citep{kolmogorov1941}, 
\begin{eqnarray}
    {\rho \delta v^2  \over \tau_{cross}} \simeq {\rho \delta v^3  \over\delta l} \simeq \epsilon,
    \label{ener_flux}
\end{eqnarray}
where  $\tau_{cross}=\delta l/ \delta v$ is the crossing time and $\epsilon$ is the energy flux 
through scale. It is injected by some process, generally at large scales, and 
is usually dissipated at small scales through viscosity or other types of dissipation.  
This implies that between injection and dissipation, within the so-called 
inertial domain, one has $\rho ^{1/3} \delta v \propto \delta l^{1/3}$. 
In incompressible fluids, this leads to the well known scaling relation 
$E(k) \propto k^{-n+2}$, where $E(k)$ is the energy spectrum, $k$ is the 
wavenumber and $n=2+5/3=11/3$. Numerical simulations of supersonic isothermal 
flows have obtained similar values \citep{kritsuk2007,federrath_slope}.
The velocity powerspectrum, which in incompressible flows, is identical to 
$E(k)$, has been found to be a little steeper and closer to $n \simeq 1.9$. 
Observationally, the velocity dispersion in molecular clouds
has been observed to follow $v_{rms} \simeq 1 {\rm km \, s^{-1}} (R / 1 {\rm pc})^\eta$
where $\eta = (n-3)/2 \simeq 0.4-0.5$ \citep{Larson81,HF12}.

As for the density PDF, in various gravo-turbulent theories, 
the velocity scale dependence is  playing a role in establishing the mass spectrum 
of the stellar progenitors either because of the turbulent  
support as inferred in Section~\ref{gravo_turb_sup} \citep{hc08,hopkins2012} or,
see Section~\ref{pn2022}, through the distribution of 
converging flows  \citep{padoan_nordlund_2002_imf}.

\subsection{Gravity, Jeans-instability and gravo-turbulence}
Gravity in combination with the various {\it supports}, i.e. thermal, magnetic and 
turbulent is obviously the most important driver of star formation.  

\subsubsection{Some fundamental aspects of gravity}
\label{fund_grav}
The most natural and well known timescale associated to gravity is the freefall time given by 
\begin{eqnarray}
    \tau_{ff} = \sqrt{ 3 \pi \over 32} {1 \over \sqrt{G \rho} },
\end{eqnarray}
which is used as a reference in many problems and essentially describes the time it takes for a cloud of density, $\rho$, and subject to its gravity only, to form a singularity.

In the absence of sufficient support, gravitational collapse occurs. Whereas the dynamics of pure self-gravitating gas is far beyond the scope of the present review, 
few results are worth mentionning. First, because the gravitational field 
derived from the gradient of a potential and because this potential is directly related 
to the density field through the Poisson equation, the gravitational force is stronger 
where the gradient of the density field is shorter. This implies that gravity 
tends to amplify anisotropies. Thus gravity tends to naturally form strongly anistropic structures such as sheets and filaments \citep[e.g.][]{lin1965,Smith_2014_filaments}.   
Second, as the collapse proceeds, density power-laws, $r^{-\alpha}$, develop. The most commonly reported exponents, at least in simulations and models, are $\alpha=2$ and $\alpha=3/2$. They can be both understood 
in relatively simple terms. Since $\rho v_r = \dot{M} / (4 \pi r^2)$, 
if the mass flux, $\dot{M}$ is sufficientlty stationary, a
relation between $\rho$ and $v_r$ follows. Yet a good approximation 
for $v_r$ is inferred from energy conservation leading to $v_r \simeq \sqrt{2 G M(r)/r}$. 

If the gas mass inside the sphere of radius, $r$, is 
dominated by a central object, $M(r) \simeq M_*$, we get $v_r \propto r^{-1/2}$ and thus $\alpha=3/2$. 
In the other case, $M(r) \propto r^{3-\alpha}$, $v_r \propto r^{1-\alpha/2}$ leading to the equation $-\alpha + (1-\alpha/2) = -2$ 
which admits $\alpha=2$ as a solution. These powerlaw density profiles have been inferred both from analytical self-similar solutions \citep{larson1969,penston_1969,Shu77} and from collapse calculations \citep{foster1993}.

A density profile $\rho \propto r^{-\alpha}$ leads to a powerlaw density PDF. 
The number of fluid elements located between  radius $r$ and $r+dr$ is 
given by $d  N \propto r ^2 dr$ but since $dr \propto \rho ^{-1 - 1/ \alpha} d \rho$, we get 
\begin{eqnarray}
{\cal P}(\rho) = {d N \over d \log \rho} = \rho ^{-3 / \alpha}. 
\label{pdf_grav}
\end{eqnarray}
For $\alpha=2$, this leads to ${\cal P}(\rho) \propto \rho^{-3/2}$, while for $\alpha=3/2$, we obtain 
${\cal P}(\rho) \propto \rho^{-2}$. The former behaviour is reported for high density gas in several simulations
\citep[e.g.][]{Kritsuk11}.

The density PDF, ${\cal P}(\rho) \propto \rho^{-3/2}$, has been proposed to play an important role regarding the IMF as exposed in Section~\ref{density_pdf}.

\subsubsection{Gravity and thermal support}
\label{sec_jeans_mass}
Thermal pressure is one of the support that can resist the gravitational force impeding 
gravitational collapse   for  
spatial scales that are smaller than the so-called Jeans length
\begin{eqnarray}
    \lambda_{\rm Jeans} = \sqrt{\pi} {C_s \over \sqrt{G \rho} },
\end{eqnarray}
where $C_s$ is the sound speed. Equivalently if the mass contained inside a sphere of diameter 
$\lambda_{\rm Jeans}$ is smaller than the Jeans mass given by 
\begin{eqnarray}
    M _{\rm Jeans} = {\pi^{5/2} \over 6} {C_s^3 \over \sqrt{G^3 \rho} } \approx 1.3 M_\odot \left(\frac{n_{\rm H}}{10^5\,\rm cm^{-3}}\right)^{-1/2} \left(\frac{T}{10\,\rm K }\right)^{3/2}.
\label{jeans_mass}
\end{eqnarray}
thermal pressure can prevent gravitational collape. 
This last expression reveals that a fundamental distinction has to be made depending 
on the effective adiabatic index, $\Gamma_{\rm ad}$, where $P \propto \rho^\Gamma_{\rm ad}$. Since 
$C_s \propto \rho^{\Gamma_{\rm ad}/2  - 1/2}$, we see that if $3 \Gamma_{\rm ad} -4 > 0 $, the Jeans mass 
increases with density. This implies that if $\Gamma_{\rm ad} > 4/3$, the gravitational collapse 
 ends after some density enhancement because thermal support unavoidably dominates gravity if the 
gas contracts sufficiently, eventually leading to an equilibrium, e.g. in a hydrostatic stellar interior.

On the other hand, if $\Gamma_{\rm ad} < 4/3$, the Jeans mass decreases with density showing that when 
collapse has started,  thermal support becomes continuously weaker and the number of Jeans mass
increases opening the possibility to the formation of several fragments.
Obviously, it does not imply that equilibrium is not possible and various hydrostatic solutions are known. The simplest  one, although due to instability never realised in realistic conditions,  is probably the so-called singular isothermal sphere (SIS)
\begin{eqnarray}
    \rho(r) = {C_s^2 \over 2 \pi G r^2}.
\end{eqnarray}
As discussed in Section~\ref{fund_grav}, an $r^{-2}$ density profile also develops during collapse. Let us stress that the physical 
origin of these two $r^{-2}$ profiles is entirely different.

Importantly  $\rho \propto r^{-2}$ implies that $M(r) \propto r$, meaning 
that most of the mass is found in the outer radii of a self-gravitating cloud. The general class of hydrostatic solutions have been described for instance in \citet{bonnor1956}. The solutions have a finite density in the center unlike the SIS but approach a corresponding
SIS solution at large radii. Therefore usually the solutions are simply cut at some radius, and it has been shown that only when the density contrast between the inner and outer density is below a certain threshold that the solution is stable. This lack of clear boundary definitions, constitutes a serious difficulty regarding the physical definition of dense cores (Section~\ref{core}) in observations and in simulations. 

The behaviour $M(r) \propto r $ is a marked difference with the case $\Gamma_{\rm ad} > 4/3$
for which equilibrium solutions present a density that vanishes at some finite radius (in this respect the critical $\Gamma_{\rm ad}$ is $\Gamma_{\rm ad}=6/5$). In practice, 
this means that whereas the definition of a self-gravitating structure 
with $\Gamma_{\rm ad} > 4/3$ is usually straightforward, this is not at all the case 
for smaller $\Gamma_{\rm ad}$ and in particular for $\Gamma_{\rm ad}=1$. 
The influence of $\Gamma_{\rm ad}$ on the IMF is discussed in Section~\ref{isotherm}, \ref{gamma_sup43}
and \ref{gamma_inf43}.

\subsubsection{Gravity and turbulence}
Whereas they are often considered separately and even sometimes 
{\it opposed}, 
turbulence and gravity are tightly linked to each other leading 
to a concept sometimes named gravo-turbulence. 
It is for instance 
well established that turbulence is amplified and even
generated during 
gravitational collapse \citep{robertson2012,2020ApJ...903..136G,hennebelle2021}. 
On the other hand, turbulence plays a dual role with respect 
to the development of gravitational instability. Since 
as discussed in \ref{turbulence}, stronger turbulence 
promotes the development of a higher fraction of dense gas, 
turbulence tends to trigger gravitational instability and induces 
fragmentation in smaller Jeans mass. However, in the same time 
turbulence may stabilize clouds that would otherwise be unstable. 
This effect is  usually described as turbulent pressure 
\citep{bonazzola1987}
and is a bit 
loosely described by replacing the sound speed by the velocity dispersion 
in Equation~(\ref{jeans_mass}). Importantly, because the velocity 
dispersion increases with distance, the turbulent support
unlike the sound speed, also tends to increase with scale.
Whereas turbulent pressure is an elusive 
concept since turbulence decays in a crossing time, it must be understood
in a dynamical sense. A cloud which is largely dominated by 
turbulence is typically dispersed in one turbulent crossing time 
and therefore does not collapse.
Due to the dual roles played by turbulent, a broad mass spectrum is expected 
to be naturally produced by gravo-turbulence. 

The density PDF in a self-gravitating and turbulent fluid can be viewed as a superposition of a low-density, turbulent part and a high-density, self-gravitating part
\citep[e.g.][]{Kritsuk11}. This, in particular, implies that as expected, at high density 
the density PDF presents a powerlaw compatible with $\simeq \rho^{-3/2}$. This is because 
the very dense gas indeed forms under the influence of gravity. Interestingly, 
it has been found that the density fluctuations that develop on top of 
the mean density profile ($\rho \propto r^{-2}$), appear to present a PDF that 
is close to a lognormal PDF with a width that would correspond 
to a PDF of an  isothermal gas with $\mathcal{M} \simeq 6$ turbulence 
\citep{hennebelle_2019}.
The combination of high density gas associated to density turbulent fluctuations 
favors the formation of low mass objects during the collapse and may be 
at the origin of at least some of the brown dwarfs \citep[e.g.][]{bonnell_clark2008}.

\subsection{ Ideal and non-ideal MHD}
\label{mag}
Magnetic field is ubiquitous in the ISM and magnetic energy is broadly 
comparable to kinetic and gravitational energies \citep{crutcher2012} 
making magnetic field an important physical process to take into account
although its exact importance remains controvertial. 

Magnetic field affects the gas dynamics in several ways and for a detailed 
discussions we refer to recent dedicated reviews \citep{Hennebelle_2019_MHD_cloud_evol,Krumholz_2019_IMF_magnetic_field,zhao2020}. 

The Lorentz force
can be decomposed in two terms, namely the magnetic pressure and tension
\begin{eqnarray}
\bb{F_L}=\frac{(\bb{\nabla} \btimes \bb{B}) \btimes
  \bb{B}}{4\pi}=-\bb{\nabla}\left( \frac{B^2}{8\pi} \right) + \frac{(\bb{B}
\bcdot \bb{\nabla}) \bb{B} }{4\pi},
\end{eqnarray}
%It is obviously
%perpendicular to the magnetic field.

Both terms play an important role regarding the support 
that magnetic field exerts against gravity. 
To quantify the
importance  of the magnetic support, let us consider a
spherical, uniform density cloud of mass $M$ and radius $R$,
threaded by a uniform magnetic field of intensity $B$.  The magnetic
flux that permeats the cloud is $\Phi= \pi R^2 B$. If 
the magnetic field is well coupled to the gas, 
$\Phi$ is a conserved quantity.  The ratio of
magnetic over gravitational energies 
is thus
\begin{eqnarray}
{ E_ {\rm mag} \over E_{\rm grav}} = {B^2 4 \pi R^3 / 3 \over 8 \pi} \times {2  R \over 5 G M^2}
\propto {B^2 R^4 \over M^2} \propto \left( {\Phi \over M } \right)^2.
\label{emag}
\end{eqnarray}
Interestingly, the ratio of magnetic over gravitational energies does not
depend on the cloud radius.  This is for 
instance different for the thermal energy of an isothermal
gas, which becomes smaller and smaller compared to the gravitational
energy as the cloud collapses.  From Equation~(\ref{emag}), 
it appears that
there is a critical value of the magnetic field strength for which the
gravitational collapse is impeded when the cloud is 
compressed.  
 If the mass-to-flux ratio is smaller  than this
critical value, it  is said to be subcritical.
It is called supercritical when the mass-to-flux is larger than the
critical value.
It is usual to define the parameter $\mu = (M/\Phi) / (M/\Phi)_{\rm crit}$. 
A useful way to take into the influence of the magnetic field on the 
stability of a structure is to use the virial theorem which in the 
presence of magnetic field writes as 
\begin{eqnarray}
  \langle V_{\rm rms}^2\rangle    + 3\,  \langle C_s^2 \rangle  + {1 \over 2 } \langle V_a^2  \rangle = - E_{\rm pot} / M
\label{viriel_ceff}
\end{eqnarray}
where $V_a = B / \sqrt{4 \pi \rho}$ is the Alfv\'en speed. 
Note that in this expression, the surface terms to which we will not refer below have been dropped. 
An important relation which has been observed by several authors in numerical 
simulations, is that during collapse, the relation $B \propto \sqrt{\rho}$ tends
to hold \citep[e.g.][]{lee2019,guszejnov2020}. This can likely  be explained as follows
\citep[e.g.][]{basu2000}. Mass and flux conservation lead to $B \propto \Sigma $
where $\Sigma \propto \rho h$ is the column density and $h$ is the altitude above 
the equatorial plane. Mechanical equilibrium along field lines leads to $C_s^2 \propto 
\phi$ whereas with Poisson equation $\phi / h^2 \simeq 4 \pi G \rho$. Combining these
relations, we obtain $B \simeq C_s / \sqrt{4 \pi G} \rho^{1/2}$. This implies that in Equation~\ref{viriel_ceff}, the magnetic support behaves similarly to the sound speed. 
In particular, it is not expected to be scale dependent as it is the case for 
$\langle V_{\rm rms}^2\rangle $.

Due to the low ionisation in the dense ISM, the magnetic field is imperfectly coupled to the gas and non-ideal MHD corrections need to be taken into account. 
%as described by Equation~\ref{AD_elec}
The typical time scale
for  ambipolar diffusion can  be  expressed as 
\begin{eqnarray}
\tau_{\rm ad} \simeq {4 \pi \gamma _{ad}\rho \rho_i R ^2 \over B^2 },
\label{time}
\end{eqnarray}
where $\gamma_{ad}$ is the ion-neutral friction coefficient 
and $R$ is the  spatial scale that is considered. 
%Performing an  ionization equilibrium in the dense ISM, \citet{Shu_1987_star_formation}
% estimate that the ion density is about $\rho_i = C \sqrt{\rho}$, where
%$C=3 \times 10^{-16}$ cm$^{-3/2}$ g$^{1/2}$.
Note that performing an accurate estimate of the ionisation and of the charge carriers
is necessary to get accurate resistivities \citep[e.g.][]{zhao2020}, particularly 
at high densities. In order to estimate $\tau_{\rm ad}$, 
one usually considers that the cloud is in virial
equilibrium, $B ^2 / 4 \pi   \simeq M \rho G / R$ (within a factor of a few).
\citep{Shu_1987_star_formation} found that for densities on the order
of $10^5$ cm$^{-3}$, 
$ \tau _{\rm ad} / \tau_{\rm ff} \simeq 8$. This high value may explain why 
observed cores do not appear to be far from being critical. 
Much lower values of $ \tau _{\rm ad} / \tau_{\rm ff}$
may arise at larger densities \citep[e.g.][]{zhao2020} and this may have 
a critical impact regarding fragmentation and disk formation.  

Another fundamental aspect of magnetic field, particularly relevant in the 
context of collapsing cloud is the so-called magnetic braking.
Magnetic tension allows the propagation of torsional Alfv\'en waves, which  transfer
angular momentum through the cloud
\citep{Shu_1987_star_formation,zhao2020}.
 To estimate the typical time scale for magnetic braking, 
let us consider an intercloud medium
of density $\rho_{\rm icm}$ and a magnetic field  parallel to
the rotation axis. The waves propagate  at the Alfv\'en speed,
$V_a= B / \sqrt{4 \pi \rho_{\rm icm}}$ along a cylinder parallel to the magnetic field.
Magnetic braking is significant when the
waves have transmitted to the intercloud medium a substantial fraction of the cloud angular momentum and 
this occurs when the Alfv\'en waves have propagated at a distance from the cloud, $l$,
such that $l \times \rho_{\rm icm} \simeq R \times \rho_0$.  An estimate for the magnetic braking time,
in case where the magnetic field and the rotation axis are aligned is thus:
\begin{eqnarray}
\tau _{\rm br} \simeq {R \over V_a} {\rho_0 \over \rho_{\rm icm}}.
\label{brake}
\end{eqnarray}
The braking time increases when  $\rho_ {\rm icm}$ decreases because if the 
 density of the intercloud
medium is low, its inertia is also low and the transfer of angular momentum
is rather inefficient. The impact that the magnetic field may have on planet-forming disks
is expected to be very significant \citep{zhao2020} as recently confirmed by 
numerical simulations where disk populations are inferred \citep[][]{lebreuilly2021}.

\subsection{Tidal forces and tidal radius}
\subsubsection{Tidal forces}
Generally speaking, gravitational forces vary in space and this implies that finite size 
objects that sit in external gravitational potential, $\phi$, experience  gradients of external gravitational force.
These gradients are known as tidal forces and tend to modify the shape and the evolution of 
the objects. 
When the size of the object remains small compared to the characteristic scale of the external gravitational
potential, it is meaningful to expand the gravitational force around the center of mass leading to a tidal tensor
 \citep[e.g.][]{colman2019}
\begin{eqnarray}
\nonumber 
    g_i = T_{ij} x_j, \\
    T_{ij} = { \partial ^2 \phi \over \partial x_i \partial x_j }.
\end{eqnarray}
Since this tensor is symmetrical it can be diagonalised, which restricts the problem to three eigenvalues,
$\lambda_1$, $\lambda_2$ and $\lambda_3$.
 Poisson equation leads to $\lambda_{1}+ \lambda_{2} + \lambda_3 = - 4 \pi G \rho$. 
In spherical geometry, $\lambda_1 = \partial ^2 \phi / \partial r^2$, while $\lambda_2=\lambda_3= r^{-1} \partial \phi / \partial r $.
Thus for a density profil $\rho = \rho_0 (r / r_0) ^{-\alpha}$, this leads to 
$\lambda_1 =  4 \pi G \rho (1 - \alpha) / (3 - \alpha)$ and $\lambda_2 = \lambda_3 =  4 \pi G \rho / (3 - \alpha)$. 
Thus we see that if $\alpha < 1$ (resp $\alpha > 1$), the radial component of the tidal tensor, $\lambda_1$, is negative
(positive), 
implying that the tidal forces tend to compress (shear) the piece of fluid. 
In particular, this implies that a piece of gas that sits within a significant external potential, can 
be either more or less prone to collapse. 

\subsubsection{Tidal radius}
Given two stars of masses $m$ and $M$ and located at a distance $D$, the question arises
on which of these two stars a fluid element will eventually be accreted.  The classical estimate \citep[e.g.][]{binney2008} considers
that the critical radius, also named tidal radius, is given by the location where 
the gravitational force vanishes. Assuming that $m \ll M$, one can show that this 
radius, also nammed the Jacobi radius and used as a proxy for the tidal radius is 
given by
\begin{eqnarray}
    R_{tidal} \simeq R _J = \left(   {m \over 3 M}   \right) ^{1/3} D.
\label{rtidal}
\end{eqnarray} 
\\

In the context of the IMF, tidal forces have been advocated 
to play an important, although quite different, role respectivelly by \citet{bonnell2001} and by 
\citet{leeh2018b} and \citet{colman2019}.
The former propose that the tidal radius 
may contribute to establish  the powerlaw behaviour of the IMF by regulating the accretion rate in cluters
(Section~\ref{compet_accret})
whereas the 
latter argue that tidal forces contribute to set the characteristic mass of stars, i.e. the mass at which the 
IMF peak occurs, by preventing fragmentation in the neighborhood of an existing first hydrostatic core 
or young stellar object (see Section~\ref{FHSC}).

\subsection{Accretion}
\label{accretion}

The specific rate of accretion can be important for the IMF in several ways. The accretion timescale $t_{\rm acc} = M/\dot{M}$ is one the key timescales for protostellar evolution (see Section~\ref{protostellar}), which %together with the $\dot{M}$ itself 
for instance
determines the rate of accretion-powered feedback such as outflows and radiation. $t_{\rm acc}$ also determines the time available to gather mass before disruptive processes can intervene and terminate accretion, e.g. dynamical interactions or disruption of the host cloud by feedback. 
%And the scaling of $\dot{M}$ with affects the shape of the IMF, particularly the self-similar \citet{salpeter_slope} power-law regime \citep[e.g.][]{zinnecker_1982_competitive_accretion,bonnell2001, BallesterosParedes15,kuznetsova2017}.

The mass accreted  by a star can be roughly divided into two parts. Initially, the gravity in a protostellar core is dominated by gas self-gravity, and gas will accrete onto the protostar at a rate that is essentially independent of the mass of the star. \citet{Shu77} derived a spherically-symmetric, self-similar solution for collapse of a $\propto r^{-2}$ density profile from rest:
\begin{equation}
    \dot{M}= 0.975 f(A)\frac{C_{\rm s}^3}{G},
\end{equation}
where $f\left(A\right) \geq 1$ depends on the instability parameter $A = 4 \uppi r^2 \rho\left(r\right)G/C_{\rm s}^2$, with $A=2$ being the threshold of collapse, and  $f\left(A\right)\propto A$ for $A>>2$. 
For instance the self-similar collapse solution inferred by \citet{larson1969} has $A=8.86$ while 
in collapse calculations \citep[e.g.][]{foster1993,gong_ostriker_2015}, the value of $A$ which varies with time, is 
several times higher than 2 at the early stage and eventually drops below this value.
Note the lack of explicit dependence of $\dot{M}$ on the current stellar mass $M_{\star}$ -- there is always central region of supersonic flow where stellar gravity dominates and $v\approx \sqrt{GM_\star / r}$, but the rate-limiting step for accretion is the initial infall determined by gas dynamics and microphysics. Also note that $\dot{M}$ is simply $\propto M\left(<r\right)/t_{\rm ff}\left(<r\right)$ evaluated from the enclosed mass and mean density within any radius $r$. This collapse on the gas free-fall timescale is a distinguishing feature of ``core accretion" scenarios in general, including turbulent cores forming massive stars \citep[e.g.][]{Mckee_tan_2003_turbulent_core}.

In the opposite regime, the gravity of the star itself is the causal factor that diverts gas into the accretion flow. In this case the accretion rate may be written
\begin{eqnarray}
\dot{M} = \pi \rho V_{rel} R_{acc}^2,
\label{R_accretion}
\end{eqnarray}
where $R_{\rm acc}$ is the critical impact parameter for gravitational capture of a gas element. The canonical example is Bondi-Hoyle-Lyttleton (BHL) accretion \citep[e.g.][]{edgar2004}, where a point mass captures gas while moving at a speed $v_\infty$ relative to a uniform gas medium with density $\rho_\infty$ and sound speed $c_{\rm \infty}$.  In an isothermal medium, the rate of BHL accretion may be approximated by \citep[e.g.][]{1996A&A...311..817R}:
\begin{equation}
    \dot{M}_{\rm BHL} \approx 4\uppi \rho_{\infty} G^2 M^2 c_{\infty}^{-3} \left[\frac{\lambda^2 + \mathcal{M}^2}{\left(1+\mathcal{M}^2\right)^4}\right]^{1/2} %\approx 8\times 10^{-7} M_\odot \,\mathrm{yr}^{-1}\left(\frac{M}{M_\odot}\right)^2 \left( \frac{n_{\infty}}{10^4 \mathrm{cm}^{-3}}\right)\left(\frac{\sqrt{v^2_\infty+c_\infty^2}}{0.2 \,\mathrm{km}\,\mathrm{s}^{-1}}\right)^{-3},
    \label{eq:bhl}
\end{equation}
where $\mathcal{M}=v_{\infty}/c_{\rm \infty}$ and $\lambda \sim 1.1$ is measured from hydrodynamics simulations.

Naturally the behaviour is more complex in real turbulent, inhomogeneous, magnetized, giant molecular clouds. The time dependence of the ambient flow makes the accretion stochastic, but scale-free simulations still find the mean accretion rate is proportional to the BHL rate \citep{krumholz:2006.bhl.accretion}. Magnetic fields can also support gas against infall, and this effect can be modeled by the usual substitution $\mathcal{M}^2 \rightarrow \mathcal{M}^2 + k \mathcal{M}_{\rm A}^2$ in Equation~\ref{eq:bhl} \citep{lee:2014.mhd.bhl.accretion}. %If a gas element falls toward multiple stars with overlapping spheres of influence, as in a star cluster, it will eventually accrete onto the star that catches it within its tidal radius, hence $R_{\rm acc} \propto R_{\rm tidal} \propto \left(M_\star/M_{\rm cl}\right)^{1/3}$, hence $\dot{M} \propto \dot{M}_{\rm cl} \left(M_\star/M_{\rm cl}\right)^{2/3}$ \citep{bonnell2001}.
%In this case, $\dot{M} \propto \dot{M}_{\rm cl} \left(M_\star/M_{\rm cl}\right)^{2/3}$ \citep{bonnell2001}.

In turbulent, self-gravitating flows, the effective values of $v_{\infty}$ and $\rho_{\infty}$ must also be considered scale-dependent, and $v_{\infty}$ can be much smaller if there is correlation between the stellar and gas kinematics. Therefore $\dot{M}_{\rm BHL}$ can be much larger than the fiducial value estimated in Equation~\ref{eq:bhl}, and simple hydrodynamics experiments  have found that this mode of accretion can contribute appreciably to the growth of massive stars in dense star clusters \citep{bonnell2001,bonnell2006,BallesterosParedes15,kuznetsova2017} possibly shaping the high-mass IMF 
(see Section~\ref{compet_accret}). It has not yet been shown explicitly that accretion proceeds in this way once other important processes such as feedback and self-consistent MHD turbulence are accounted for, and \citet{tan:2014.pp6.massive.sf} estimated that even a conservative amount of outflow feedback momentum could present a severe obstacle to accretion of gas beyond the natal dense, turbulent core. On the other hand, \citep{Smith09}
and \citet{gong_ostriker_2015} do not find bound gas cores massive enough to form massive stars, requiring a more extended accretion scenario. Similarly, the multi-physics simulation in \citet{grudic2022} found massive stellar accretion timescales too long to be accounted for by accretion from dense cores alone.

%\mike{mention smith 2009    }

% The initial ``seed" of a star is generally agreed to form from the collapse of a self-gravitating gas fragment, and the mass initially arriving at the surface of the protostar will originate in the parent fragment. However, a larger reservoir of mass may also be accessible for accretion because the star can continue to capture gas from the surrounding gas cloud that was not initially bound to the gas progenitor of the star. 
% - Larson 1978;1982, Zinnecker 1982

% \begin{equation}
%     \dot{M}\approx 4 \uppi \rho \frac{\left(G M_\star\right)^2}{\left(c_{\rm s}^2 + v^2\right)^{3/2}}
% \end{equation}

% - write down BHL formula, cite works investigating it in messy GMCs

% - point out that BHL formula predicts tiny $\dot{M}$ (Krumholz 2005) that would take too long to form a star, but you can't ignore the size-linewidth relation which enhances the rate by orders of magnitude (Bonnell 2006)
% - mention how BHL accretion from an initial mass spectrum can produce Salpeter-like slope - Zinnecker 1982, Bonnell 2007, Hsu 2010

\subsection{Stellar evolution and feedback}
\label{protostellar}
%\begin{figure}
    %\centering
%    \includegraphics[width=0.5\textwidth]{Figures/hosokawa_inutsuka_2009.png}%\includegraphics[width=0.3\textwidth]{Figures/kuiper_yorke_2013.png}
%    \caption{Radial evolution of protostellar models accreting at various fixed rates according to simulations of spherical accretion \citep{hosokawa2009}. The top panel plots the protostellar radius as a function of mass while the bottom panel plots the central temperature. Bold curves indicate models in which D burning is accounted for, and the corresponding fainter curves are models without D burning.}% {\it Right:} Evolution of a massive protostar accreting at $\sim 2\times 10^{-3} M_\odot\,\mathrm{yr}^{-1}$ from \citet{2013ApJ...772...61K}; the labeled phases I-IV correspond to pre-bloating, bloating, Kelvin-Helmholtz contraction, and main-sequence evolution, and the dashed line is a model accreting at exactly $\sim 2\times 10^{-3} M_\odot \,\mathrm{yr}^{-1}$. Discrepancies between the different models are mainly due to the different assumptions about the radiative efficiency of the accretion flow (i.e. the star's energy and entropy budget).}
%    \label{fig:hosokawa2009}
%\end{figure}

Once collapse to protostellar densities has occurred, the protostar's evolution may affect the IMF because the stellar properties determines the star's various rates of mass, momentum, and energy feedback. The stellar mass-radius relation determines the strength of accretion-powered processes such as protostellar jets and outflows and radiation. And once on the main sequence, a star's evolution and energetics of radiation and stellar winds are largely determined by its main-sequence mass. Here we will focus on the aspects of stellar evolution that are most relevant for determining feedback rates. 

%\subsubsection{Pre-main-sequence evolution}
Once the dissociation of H$_{\rm 2}$ is complete, quasi-isothermal collapse can no longer proceed, and the protostar will again assume a hydrostatic, approximately-polytropic structure with a certain outer radius $R_{\rm \star}$ and a central temperature
\begin{equation}
    T_{\rm c} \approx 0.5 \beta_{\rm c} \frac{\mu m_{\rm p}}{k_{\rm B}}\frac{G M}{R}% \approx 7 \times 10^{6}\mathrm{K}\left(\frac{M}{M_\odot}\right)\left(\frac{R}{R_\star}\right)^{-1}, % a_{\rm T}=0.5 for n=1.5, =0.8 for n=3
    \label{eq:tcore}
\end{equation}
where $\mu \sim 0.613$ is the mean molecular weight for ionized stellar interiors and $\beta_{\rm c}$ is ratio of gas to total pressure in the center (the latter expression assumes $\beta_{\rm c} \sim 1$, valid except for $\gtrsim 10 M_\odot$ stars). Once nuclear burning has begun, $T_{\rm c}$ will be vary only weakly due to the steep dependence of the reaction rate on temperature, first at $\sim 1 \, \rm MK$ for D burning and then $\gtrsim 15 \, \rm MK$ for the various H burning processes. A hydrostatic protostar's internal dynamical time is much shorter than any other relevant timescale, and evolution will take place over the timescales required for the star's mass and energy to change, e.g. the accretion time %$ t_{\rm acc} \simeq 1\,\mathrm{Myr}$
\begin{equation}
    t_{\rm acc} = \frac{M}{\dot{M}} = 10^5 \,\mathrm{yr}\left(\frac{M}{M_\odot}\right)\left(\frac{\dot{M}}{10^{-5} M_\odot\,\mathrm{{yr}^{-1} } }\right),
\end{equation}
and the Kelvin-Helmholtz time
\begin{equation}
    t_{\rm KH} = \frac{G M^2}{R_\star L_\star } = 3\times 10^5 \left(\frac{M}{M_\odot}\right)^2 \left(\frac{R_\star}{R_{\odot}}\right)^{-1} \left(\frac{L_\star}{10 L_\odot}\right)^{-1} \rm yr,
\end{equation}
where $L_{\rm \star}$ is the the rate of energy transfer from the interior of the protostar. At early times, when $t_{\rm acc} << t_{\rm KH}$,  emission from the star is dominated by the accretion luminosity:
\begin{equation}
\label{Lacc}
L_{acc} = f_{acc} {G M \dot{M} \over R_\star} = 30 L_{\rm \odot}\, f_{\rm acc} \left(\frac{\dot{M}}{10^{-6} M_\odot\,\mathrm{yr}^{-1}}\right)\left(\frac{M}{M_{\odot}}\right)\left(\frac{R_{\rm \star}}{R_{\odot}},\right)^{-1}
 \end{equation}
and the protostar grows in mass and radius with an energy and entropy budget determined by the energetics of the gas as it reaches $R_{\rm \star}$. Eventually, when $t_{\rm KH} \lesssim t_{\rm acc}$, or equivalently $L_\star \gtrsim L_{\rm acc}$, the star contracts with a luminosity $L_\star \sim 4\uppi R_{\rm \star}^2\sigma_{\rm SB}T_{\rm eff}^4$. The overal sequence is one of initial growth in radius followed by contraction to the star's eventual main-sequence radius \citep{palla.stahler.1991:ps.evol,nakano:2000.ps.evol,hosokawa2009}. This evolution in radius must be accounted for to model the effects of accretion-powered feedback.

\subsubsection{Protostellar outflows}

Only a certain fraction of the gas captured by a protostar  will end up accreting onto the protostar itself: the remainder should be ejected in a magnetocentrifually-powered outflow \citep{1982MNRAS.199..883B,shu:1998.xwind,pelletier.pudritz.diskwind}. This naturally explains the ubiquity of bipolar protostellar outflows in star-forming, as evidenced by optical Herbig-Haro objects at the shocked interface with the ambient medium, cavities imaged in cold molecular emission, and warm molecular emission from the jet. For a review of observations of protostellar outflows see \citep{bally:2016.protostellar.jets}. Very recently, JWST has imaged the warm molecular component from protostellar jets with order-of-magnitude finer resolution than Spitzer \citep{ray:2023.jwst.jets}.

Some details of the jet launching mechanism remain uncertain: MHD disk wind models have been studied extensively, but typically under the assumptions of axisymmetry and steady inflow, which is unlikely to fully describe the conditions in accreting protostellar envelopes. The magnetic field structure and the degree of misalignment between the disk and the large-scale magnetic field, can both affect the strength and collimation pattern of the outflows \citep{2019MNRAS.485.5532G}.

Current 3D models cannot simultaneously resolve the jet launching zone near the star and the larger cloud environment and star cluster, so IMF calculations have adopted sub-grid prescriptions describing the velocity, mass load, and collimation pattern of the jets. For a magnetocentrifugal wind, the natural parametrization for the jet speed is
\begin{equation}
    v_{\rm w} = f_{\rm K}\sqrt{\frac{G M}{R_{\rm \star}}},
\end{equation}
and one may assume a certain fraction $f_{\rm w}$ of the mass that accretes near the protostar is launched in the wind. These parameters $f_{\rm w}$ and $f_{\rm K}$ are somewhat constrained by observations: it is possible to measure the momentum present in protostellar outflows, and to compare with the corresponding luminosity of the young stellar object launching the jet \citep[e.g.][]{maud:2015.outflows}. Typical parameters adopted by star cluster formation simulations are $f_{\rm w} \sim 0.1-0.3$ and $f_{\rm K} \sim 0.3-1$ \citep{Cunningham_2011_outflow_sim,Federrath_2014_jets}, and a collimated angular pattern of mass and momentum deposition is typically assumed \citep[e.g.][]{matzner.mckee:1999.jets}. The addition of protostellar outflows to numerical models tends to have a dramatic effect on the predicted stellar mass spectrum, due to their effects on the transport of both gas and radiation (\S\ref{jets}).

\subsubsection{Stellar winds}
\label{sec:winds}
% driven by lines
% scales with metallicity
%
Winds from intermediate-  and high-mass stars can contribute appreciably to the stellar feedback from a young star cluster, potentially affecting the growth of individual stars \citep[e.g.][]{rosen:2021.winds, rosen:2022.winds} or the unbinding of gas from the cluster's natal clump \citep{2013MNRAS.436.3430D,rogers.pittard:2013.winds}, both of which may affect the IMF. Metal ion transitions contribute most of the radiative force driving O star winds, so these are expected to be metallicity-dependendent, generally scaling roughly as $\dot{M} \propto Z^{0.7}-Z^{0.9}$ \citep{vink:2001.massloss}. The onset of the Wolf-Rayet phase increases the mass-loss rate by about an order of magnitude. The terminal velocity of the wind is generally proportional to the stellar escape speed, with $v_{\rm wind}=A\sqrt{2GM/R_\star}$, $A\sim 1.3$ for $T_{\rm eff} < 21\,\rm kK$ and $A\sim 2.6$ for $T_{\rm eff} > 21 \rm kK$ \citep{lamers:1995.winds}. A recent set of evolutionary tracks and mass-loss histories for single massive stars of varying metallicity is presented in \citet{boost_tracks}.

%The exact rates of stellar mass loss are uncertain. The most widely-used mass-loss prescription for main-sequence O stars is \citet{vink:2001.massloss}, but this appears to overpredict the mass-loss rate of early-type O stars by a factor of ~2-3, and by an even larger factor for later O and B stars, the so-called ``weak-wind problem" \citep{smith:2014.massloss}. Even without these challenges in the modeling of isolated stellar winds, some other, more-fundamental issues complicate the modeling of winds from massive protostars and stars. During massive protostar formation, radiatively-driven and magnetocentrifugally-driven wind launching may all occur in concert with ongoing accretion and stellar evolution, a complex scenario that cannot be modeled in spherical symmetry. At most, current numerical star formation models implement all of these processes separately in an operator-split fashion \citep{grudic2021}, but in reality they all correspond to one coupled, multi-scale gas flow. And, crucially, most massive stars are in binaries close enough for their interactions to affect their evolution \citep{offner_2022_multiplicity_review}. Hence binarity can result in fundamentally-different evolutionary pathways and mass loss histories \citep{2012Sci...337..444S}, whose effects on feedback have yet to be explored in the context of the IMF and star cluster formation.

\subsubsection{Stellar radiation}
\label{sec:radiation}
Radiation from stars and protostars can influence the IMF in numerous ways, affecting both the thermal structure and the dynamics of the star-forming gas flow. In the absence of massive stars, the radiation field is dominated by accretion-powered radiation (Equation \ref{Lacc}), with a typical light-to-mass ratio of order $10-100 L_{\odot}\,M_\odot^{-1}$. Once massive ($\gtrsim 8 M_\odot$) stars form, they quickly reach the main sequence and start moving along it as they accrete, and their fusion-powered radiation comes to dominate radiation field in the protocluster \citep{grudic2022}. A fit to the zero-age main sequence luminosity of the \citet{boost_tracks} $9-500M_\odot$ stellar models gives 
\begin{equation}
    L_{\rm MS} = 4.5\times 10^{5} L_{\odot} \left(\left(\frac{M}{20 M_\odot}\right)^{-1.3} + \left(\frac{M}{20 M_\odot}\right)^{-0.33}\right)^{-3.3},
\end{equation}
with no strong dependence on metallicity. Radiation from a massive star emerges primarily as UV, but if the star is accreting then the photons will be absorbed by dust and neutral gas, and the emergent radiation will be primarily at $\sim 10\rm \mu m-1mm$ wavelengths. In the IR-dominated regime, the structure of the radiation field is determined by the gas/dust density profile and radius of the photosphere \citep{chakrabarti:2008.ir.radiation}.

Whatever the source of the radiation, (proto-) stellar emission will nearly always dominate over the initial ambient radiation field near the protostars, determining the dust temperature and hence the thermal structure and Jeans mass dense gas in protostellar disks and envelopes, which can be an important regulator of the IMF \citep{offner2009,krumholz2011,guszejnov2016,hennebelle2022}. Radiation pressure of optically-thick infrared radiation on dust in optically-thick accreting envelopes can also influence the upper end of the IMF by moderating massive stellar accretion \citep{larson1971}, although it cannot stop it entirely if the gas flows in a disky or filamentary geometry \citep{krumholz:2009.massive.sf,kuiper2010}. This IR radiation effect pressure depends on dust abundance and opacity, and hence metallicity.

Lastly, radiation plays an important role in cutting off the overall gas supply for star formation, unbinding gas from the cloud through some combination of cloud-scale radiation pressure on dust grains and gas pressure-driven HII region expansion, depending on the density and dust abundance/opacity of the cloud \citep{krumholz.matzner:2009.hii.regions, 2016ApJ...819..137K}. Once the cloud is dispersed, both fragmentation and accretion cease and the stellar mass spectrum is set.

% parameters: 4.45638874e+05  2.06921545e+01 -1.26669191e+00 -3.32754858e-01  -3.31039117e+00

%and the stellar light-to-mass ratio approaches $\sim 10^3 L_\odot M_\odot^{-1}$ for a standard, well-sampled IMF.

\subsubsection{Supernovae}

Most stars more massive than $\sim 8 M_\odot$ will end their lives in an energetic ($\sim 10^{51} \rm erg$) core-collapse supernova, ejecting most of their mass and leaving behind a remnant. This form of feedback is regulates galactic star formation and enriches the ISM, setting the mass and chemical composition of galaxies \citep{naab_ostriker_galform_review}. SNe regulate the level of turbulence in a galaxy, and hence cloud properties and the initial conditions for star formation \citep{2020ApJ...904...58L,2022ApJ...936..137O}. But the IMF in a given star cluster is not likely affected by SNe {\it from that cluster}, because the soonest they can happen is after the $\sim 3\, \rm Myr$ lifetime of a very massive star. This can be further delayed depending on the specifics of which stellar masses explode and binary mass transfer \citep{2023arXiv230409350H}, and the additional assembly time required for massive stellar accretion \citep{grudic2022}. Protostellar outflows, radiation, and stellar winds are all active during this time, so cloud-scale experiments starting from a starless state find that supernovae tend to come too late to strongly influence the dynamics of star formation on GMC or star cluster scales \citep{grudic_2018_mwg_gmc}, or are even relatively ineffective in molecular cloud environments \citep{Walch15,2016MNRAS.463.3129G}.  In observations, $\lesssim 3\,\rm Myr$ old, exposed star clusters are common, implying that early feedback alone is sufficient to end the embedded phase of a cluster in many cases \citep{2015MNRAS.449.1106H}.

\section{Overview of IMF theories - Success and limit }
\label{sec:theory}
Numerous analytical theories have been developed through years to explain the physical origin of the IMF. 
They broadly fall in two categories, the theories that have attempted to explain the peak of the IMF, that is to say explain the origin of 
a characteristic mass, and the theories that tried to explain the mass spectrum of higher mass stars. 
%It is interesting to notice that, in a sense, the two theoretical efforts present opposite goals
%whereas the former struggle to identify a process that is sufficiently 
%robust to produce a characteristic mass, the latter, on the contrary rely on scale-free processes
%to explain the broad stellar mass spectrum say from 1 to 100 solar masses. 

\subsection{How to explain a broad mass spectrum of stars?}
The exact shape of the IMF for $M > 1 $ M$_\odot$ is still discussed. 
It is possibly a power-law, a lognormal or a combination of the two. At first sight 
it does not seem to present a characteristic scale or mass and is more accurately described 
as a scale-free process something that for instance turbulence and also gravity tend to produce. 
The challenge is therefore to understand how exactly this may work and which process, if any, is dominant. 
The first models which have been proposed 
\citep{hoyle1953,larson1973,elmegreen1983,zinnecker1984} were based on the general idea that star 
forming clouds
would recursively fragment with equal and independent probability at each step.

Nowadays, the proposed explanations fall, broadly speaking,
in two categories whose main differences rely on the importance of an initial {\it deterministic} mass accretion reservoir. 
Theories based on nearly stochastic accretion
 \citep{basu2004,maschberger2014},
 for instance competitive accretion \citep{bonnell2001}, assume that most of the mass accreted 
 by a star, is not provided by a preexisting gas reservoir but largely the result 
 of nearly random processes. 
 On the other-hand, gravo-turbulent theories \citep{inutsuka2001,padoan_nordlund_2002_imf,hc08,hopkins2012}
 emphasize the role of a coherent mass of gas, a reservoir, 
 that is typically set by turbulence and {\it already exists} when the protostar first forms.

 While these scenarios have often been presented as being exclusive from each other, it is likely the case 
 that they both contribute to the final stellar mass spectrum.

\subsubsection{Competitive accretion}
\label{compet_accret}

Competitive accretion is probably the first  model, based on stochastic accretion 
to have been developed. It has been proposed 
by 
\citet{zinnecker_1982_competitive_accretion}
and then further elaborated by
\citet{bonnell2001} \citep[see also][]{hsu2010,maschberger2014,BallesterosParedes15,kuznetsova2017,kuznetsova2018}.
In this scenario, the gas accretion onto a star increases with its mass, because 
of the stronger gravitational influence, it has on the surrounding medium. 
Thus the more massive stars attract more gas than the less massive
ones, and therefore tend to become even more massive.

The accretion rate is given by Equation~\ref{R_accretion}, and
% \begin{eqnarray}
% \dot{M}_* = \pi \rho V_{rel} R_{acc}^2,
% \label{accret}
% \end{eqnarray}
% where $\rho$ is the gas density, $V_{rel}$ is the relative velocity between gas and stars
%  whereas $R_{acc}$ is the accretion radius.
to estimate the different factors and make specific predictions, 
\citet{bonnell2001} have considered two cases, namely where the gravitational
potential is dominated by the gas or by the stars as expected in an older cluster. 

\paragraph{Gas dominated potential}
\citet{bonnell2001} considered
that the gas density is proportional to $r ^{-2}$, where $r$ is the radial coordinate. They further
assume that the stellar density, $n_*$, also follows $n_* \propto r^{-2}$. The accretion radius, $R_{acc}$ is
assumed to be equal to the tidal radius given
by Equation~\ref{rtidal} with $m=M_*$ and $M=M_{enc}$
%\begin{eqnarray}
%R_{tidal} \simeq 0.5 \left( {M_*\over M_{enc}} \right)^{1/3}  r,
%\label{Rtidal}
%\end{eqnarray}
where $M_{enc}$ is the cluster mass enclosed within radius $r$ and $M_*$ is the mass of an individual star.
%Indeed  at a distance, $r$, 
% such  that $r<R_{tidal}$  the gravitational force  due to the star, onto a fluid particle 
% is larger than the force due to the cluster itself. Thus the fluid particle is more likely 
% to be accreted onto the star.
Since the mass of gas  $M(r) \propto r$, the infall speed is about                                           $V_{in} \simeq \sqrt{G M(r) /r}$. Assuming that the stars have virialized
 velocities leads to                                                                  
$V_{rel} \simeq V_{in}$.                                                                                      The number of stars, $dN_*$, located between  $r$ and $r+dr$ is                                                   $dN_* = n_*(r) \times 4 \pi r^2 dr \propto dr$. Thus                                                             Equation~\ref{R_accretion}  leads to  $\dot{M}_* \propto (M_*/r)^{2/3}$.                             Therefore, after integration, asuming that stars remain at the same distance, $r$, during the accretion process,  we obtain     
 $M_* \propto r^{-2}$, $r \propto M_*^{-1/2}$ and consequently:                                                                               
\begin{eqnarray}                                                                                                                                      
dN_* \propto M_*^{-3/2} dM_*,                                                                                                                 
\label{dNacc1}                                                                                                                                        
\end{eqnarray}                                                                                                   implying $\Gamma_{IMF} = -1/2$.

\paragraph{Star dominated potential}                                                                           

Assuming the potential is dominated by stars located in the cloud center, the density would then be given by $\rho \propto r^{-3/2}$ (see Section~\ref{fund_grav}), 

%which corresponds to the expected density      distribution after the rarefaction wave has propagated away                                                                 \citet{Shu77}.                                                                          
The accretion radius is now supposed to correspond to the Bondi-Hoyle radius                                       since the star and gas velocities are not correlated anymore.                                          This leads to                                                                                                   $\dot{M}_* = \pi \rho V_{rel} R_{BH}^2$ where $R_{BH} \propto M_* / V_{rel}^2$ (see Section~\ref{accretion}), and it follows
that 
\begin{eqnarray}
\dot{M}_* \propto M_*^{2}.
\label{dMacc2}
\end{eqnarray}
 \citet{bonnell2001} show that under reasonable assumptions this would lead to
 $dN \propto M_*^{-2} dM_*$ or equivalently $d N / d \log M \propto M^{-1}$ implying
 $\Gamma_{IMF}=-1$. \\
 
 These stellar distributions
seem to be confirmed by the simulations presented in \citet{bonnell2001} and \citet{hsu2010}
which consist in placing randomly sink particles in a molecular cloud and letting them accrete.
In these experiments a power-law with $\Gamma_{IMF} \simeq -1$ develops.
 The predicted exponents,
$\Gamma_{IMF} = -0.5 $ and -1
are close, although slightly shallower than, the Salpeter exponent, $\Gamma_{IMF} \simeq -1.3 $.

\subsubsection{Stochastic accretion}
Several studies \citep{basu2004,myers2009,dib2010,maschberger2013,basu2015} 
have investigated the scenario in which the gas accretion onto 
the star is proportional to some powerlaw of the 
mass of the stars and is then stopped by a stochastic process which can 
be either the finite reservoir of mass,
 an outflow or an HII region  that would sweep up the
remaining gas within the vicinity of the accreting protostar or an 
N-body gravitational interaction that would eject the stars from the gas rich region.

The first models were proposed by \citet{silk1995} and Adams \& Fatuzzo (1996).
They start by relating the mass of the stars to the physical parameters of the cloud such as
sound speed and rotation and then assume that an outflow whose properties
are related to the accretion luminosity stops the cloud collapse.
Using the Larson relations \citep{Larson81}, all these parameters are linked to 
the clump masses. Using the mass spectrum of the clumps 
\citep[e.g.][]{HF12}, they inferred the IMF.

\citet{basu2004} have developed a  quantitative statistical approach.
It is assumed that initially, 
due to the large number of processes that control their
formation (and invoking the central limit theorem)
the dense core distribution is simply 
lognormal. 
The cores are then assumed to grow by accretion with an accretion rate that is
 proportional to their mass,
$\dot{M} = \gamma M \rightarrow M(t)=M_0 \exp(\gamma t)$,
leading  to $\log M = \mu = \mu_0 + \gamma t$. Finally,  accretion
is assumed to last
 over a finite period of time given by
$f(t)=\delta \exp(-\delta t)$.
The star mass distribution is thus obtained by summing over the
accretion time distribution.
\begin{eqnarray}
f(M) = \int _0 ^\infty { \delta \exp(-\delta t) \over \sqrt{2 \pi} \sigma_0 M } \exp \left( -{ (\ln M - \mu_0 -\gamma t)^2\over 2 \sigma_0^2}  \right) dt \\
= {\alpha \over 2} \exp( \alpha \mu_0 + \alpha^2 \sigma_0^2/2) M^{-1-\alpha} {\rm erf} \left( {1 \over \sqrt{2} } \left(\alpha \sigma_0 - { \ln M - \mu_0 \over \sigma_0}
\right)\right),
\nonumber
\end{eqnarray}
where $\alpha=\delta/\gamma$ and $\sigma_0$ determine the width of the initial dense
core distribution. Since  $\delta$ and $\alpha$ derived from the same physical
processes, their ratio is assumed  to be close to  unity
and thus $f(M)$ is expected to present a powerlaw  behaviour close to the observed IMF.

\citet{batebonell2005} have presented 
another related model  based on an
idea proposed by \citet{price1995}.  Some of the objects
that form by fragmentation within  clusters  are being ejected
through N-body gravitational interaction with the other objects.
Once ejected, the objects stop  accreting further gas.
Assuming a lognormal accretion rate and an exponential probability of being ejected,
 a mass distribution is obtained and it is shown  that it can fit the IMF for some reasonable 
 choices of parameters. \citet{myers2009}
 developed a model that rely on similar ideas
considering  the accretion coming from the surrounding
background. Adjusting two parameters, the observed
IMF is well reproduced (his figure 5).

 %\citet{hc08} \citet{hc2009}
 %\citet{hopkins2012}

\subsubsection{Gravo-turbulent theory in a turbulent density field}
\label{gravo_turb_sup}
Whereas in the accretion models, turbulence is not determinant, 
it is one of the essential physical processes for the gravo-tubulent theories presented below, although the  role it plays differs between models.

The first theory which combined turbulence and gravity was
proposed by \citet{padoan1997}. The authors consider
a lognormal density distribution (see Section~\ref{dens_PDF})
 and select the regions of the flow which  
are Jeans unstable.
By doing so, they get too steep an IMF (typically 
$dN/ dM \propto M^{-3}$) but nevertheless find a lognormal behaviour at small 
masses, a direct consequence of the lognormal density distribution,
 and a powerlaw one at large masses.

\citet{hc08} \citep[see also][]{chabrier2014} proposed 
an approach which consists in counting the mass of the gas
for which gravity supersedes  all supports, namely thermal, turbulent and 
magnetic, according to the virial theorem. 
In this theory, turbulence possesses a 
 dual role. On one hand it enhances star formation by locally compressing the gas 
 within shocks and this is accounted for in the density PDF.
 On the 
other hand, it also exerts a support to gravity through turbulent dispersion. 
%limits star formation efficiency because of the turbulent 
%dispersion of the flow, which is accounted for  through the selection of the 
%self-gravitating pieces of fluid. Note that the effectiveness of this 
%effect has sometimes be questionned based on the fact that initial turbulence does not significantly 
%delay the collapse. It is however clear that if the kinetic energy of a cloud largelly dominates the gravitational 
%one, this cloud is not going to collapse but to rather disperse.

The  theory is an extension of the \citet{pressschechter} statistical formalism, 
developed in cosmology.
Two significant  differences are $i)$ the density field, which in cosmology is characterized by
small and Gaussian fluctuations and by  a  lognormal PDF with very large 
fluctuations in molecular clouds, and $ii)$ the selection criterion, which is a simple scale-free density 
threshold in cosmology and scale-dependent, based on the virial theorem in the star formation case.

More precisely, fluid particles which are dominated by their 
internal gravity as specified by Equation~\ref{viriel_ceff}
are assumed to undergo gravitational collapse and form stars.
This leads to the definition of a critical mass, $M_R^c$ and of a critical density, $\rho_R^c$, which generally depend
on the spatial scale $R$.

%The turbulent rms velocity is generally assumed to be proportional to a power-law of the region size  (see Section~\ref{scale_v})

%\begin{eqnarray}
%$\langle V_{\rm rms}^2\rangle =  V_0^2 \times \left( {R \over  1 {\rm pc}} \right) ^{2 \eta}$,
%\label{larson}
% \end{eqnarray}
%with $V_0\simeq 1\, {\rm km\, s}^{-1}$ and $\eta \simeq 0.4$-0.5 \citep{Larson81}.

 The method entails the following steps. First,  using a prescribed window function, the density field is smoothed at a scale $R$. Then, the total mass contained in sphere which, at scale $R$,  present a density larger than the 
 critical density,  $\rho _R^c$,  is obtained by integrating the density PDF from $\rho_R^c$ to infinity. 
This mass, on the other hand, is also equal to the total mass of the stars  having a 
of mass lower than $M_R^c$. This leads to 
\begin{eqnarray}
 \int ^{\infty} _ {\delta_R^c} \bar{\rho} \exp(\delta)   {\cal P}_R(\delta)  d\delta
 = \int _0 ^ {M_R^c} M' \, {\cal N} (M')\,   P(R,M')\, dM'. 
\label{hc_eq1}
\end{eqnarray}
In this expression, $M_R^c$ is the mass which at scale $R$ is 
gravitationally unstable, $\delta_R^c = log(\rho_R^c/\bar{\rho})$ 
and $\rho_R^c= M_R^c / (C_m R^3)$, $C_m$ being a dimensionless coefficient of 
order unity. ${\cal P}_R$ is the density PDF, while 
$P(R,M')$ is the conditional probability to find a gravitationally  unstable mass, $M'$
embedded into $M_R^c$ at scale $R$, which is usually assumed to be equal to 1.
${\cal N}$ is the mass spectrum of the stars that eventually form.

The derivative of 
Equation~(\ref{hc_eq1}) with respect to the spatial scale $R$, leads to the mass spectrum 
\begin{eqnarray}
\label{n_general}
 {\cal N} (M_R^c)  &=& %\\ \nonumber 
 { \bar{\rho} \over M_R^c} 
{dR \over dM_R^c} \,
\left( -{d \delta_R ^c \over dR} \exp(\delta_R^c) {\cal P}_R( \delta_R^c) + \int _  {\delta_R^c}^\infty \exp(\delta) {d {\cal P}_R \over dR} d\delta \right).
\end{eqnarray}
While the second term is important to explain the mass spectrum of unbound clumps defined 
by a uniform density threshold, in most cases, it plays a minor role for 
gravitationally bound mass reservoir 
%defined by the Virial condition
 and can generally be dropped.

Using the lognormal density PDF as specified by Equation~\ref{Pr0}, and using 
the expression of $\rho_R^c$ and $M_R^c$, one obtains an expression 
for the mass spectrum.
 When the turbulent support is negligible with respect to the thermal one, 
with Equation~(\ref{n_general}) it can be shown that
 at large masses  the mass spectrum is identical to the \citet{padoan1997} result, 
i.e. $dN/d\log\, M \propto M^{-2}$, $\Gamma_{IMF}=-2$ . However, when the turbulent support is 
significant,
 \begin{eqnarray}
 {dN \over d\log\, M} \propto M^{-(n+1)/(2n-4)}, 
 \label{gamma_hc}
 \end{eqnarray}
 where the index of the velocity powerspectrum  $n$
is related to $\eta$ by the relation $\eta=(n-3)/2$ and where we remind that 
$v_{rms} \propto R^{\eta}$ . As $n \simeq 3.8-3.9$ in 
supersonic turbulence simulations (see Section~\ref{scale_v}),
turbulent dispersion leads to an exponent that is very close to the Salpeter $\Gamma_{IMF}=-1.3$ value.

Detailed comparisons 
with numerical simulations of supersonic isothermal turbulence were carried out in \citet{Schmidt10}. 
 The mass spectrum of {\it cores} supported 
either by pure thermal support or by turbulent plus thermal support has been computed and 
 show very good  agreement 
with the present theory, suggesting  that indeed, turbulent support is needed to 
yield the Salpeter index. Note that \citet{Schmidt10} use for the density 
PDF the one they measure in their simulations which is nearly, but not exactly lognormal.
Most importantly one should stress that indeed the theory presented by \citet{hc08} as well as these simulations were not truely self-gravitating as they do not entail collapse density PDF (Equation~\ref{pdf_grav}).
This is a major restriction, particularly for low mass stars, as discussed in Section~\ref{density_pdf}. \\

A complementary formulation has been proposed by 
\citet{excursion_set_ism,hopkins2012,hopkins2013} 
using excursion set theory \citep{zentner}. This approach consists in 
performing random walks
in the Fourier space of the density field. Essentially
a random point in the field is picked and  the gas density 
around that point is computed using a window of some specified radius. 
As this radius increases from zero to some maximum
 value, the mean density in the window can be compared 
 to the, scale dependent, critical density, also named the barrier, obtained from 
 virial theorem as specified in Equation~\ref{viriel_ceff}
 and as explained above contained the various supports.
 Note that \citet{excursion_set_ism}  propose a global model that includes spatial scales larger than
that of molecular clouds to describe the whole galactic disc. An appealing
concept is that the clouds are defined as density fluctuations that first cross the collapse threshold.
More precisely when the spatial scale is decreased, the first crossing is the first time that self-gravity becomes dominant. 
These density fluctuations present a mass spectrum that is slightly shallower than
$\Gamma_{IMF}=-1$, and is in fair agreement with observed GMC mass distributions. As the radius of the window function is further 
decreased, the density fluctuations cross the barrier several times. 
Selecting the fluctuations for which the barrier is crossed for the last time, 
it is found that 
their mass spectrum is almost identical to the one inferred in \citet{hc08}. 
The first type of structures, that is to
say the fluctuations that cross the barrier for the first time, have been 
interpreted to represent giant
molecular clouds, which  are not themselves embedded in a larger self-gravitating
cloud. The density fluctuations for which the barrier is crossed for the last time, 
 represent prestellar core progenitors.

\subsubsection{Gravo-turbulent theory in a collapsing cloud}
\label{density_pdf}

\begin{figure}
    \centering
    \includegraphics[width=10cm]{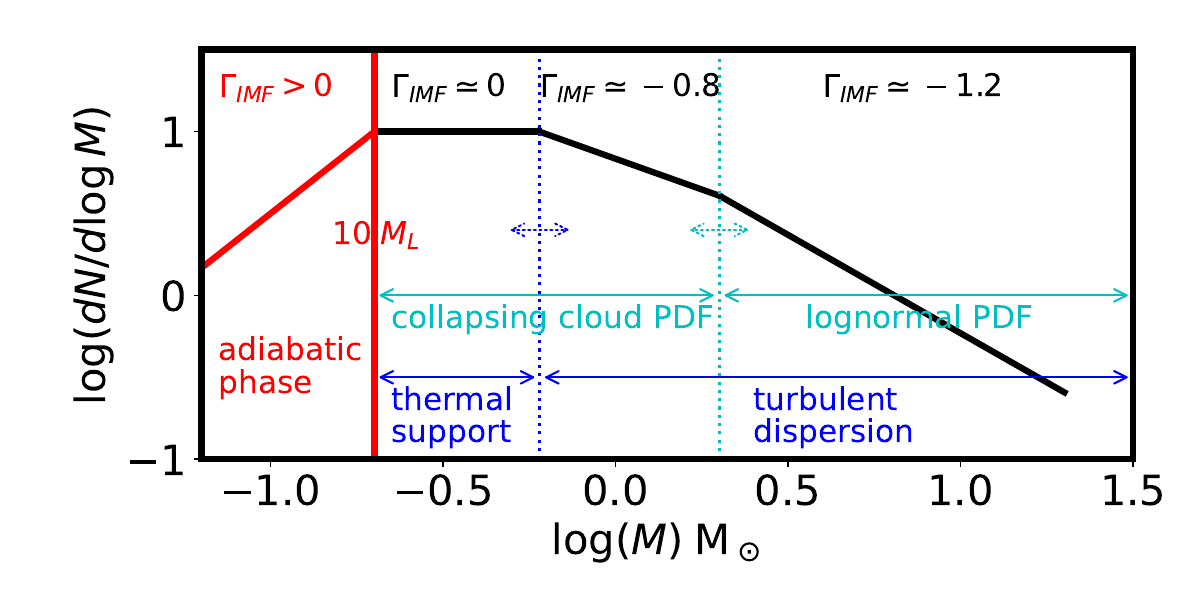}
    \caption{Schematic view of the gravo-turbulent theory predictions presented in Sects.~\ref{gravo_turb_sup},~\ref{density_pdf} and \ref{FHSC}. From right to left, we distinguish four regions. $\Gamma_{IMF} \simeq 
    -1.2$ is obtained when the density PDF is lognormal and when at the scale of the mass reservoirs, turbulent dispersion dominates over thermal support. %\citep{hc08,hopkins2012}. 
    $\Gamma_{IMF} \simeq -0.8$
    \label{fig:recap} is expected in regions where collapse has already occured, implying that the density PDF is given by Equation~\ref{pdf_grav}, and turbulence still dominates over thermal support. 
    On the other-hand, $\Gamma_{IMF} \simeq 0$  is predicted to occur when thermal support dominates. 
    %\citep{leeh2018a}. 
    Finally, when dust cooling becomes inefficient, objects of masses comparable to the mass of the FHSC, $M_L$ are expected to form with a peak at about 10 times this value leading 
    to $\Gamma_{IMF} > 0$. The transitions between the various regimes are not universal and depend 
    on the physical conditions. For instance, if Mach numbers or column densities are very large, the regime $\Gamma_{IMF} \simeq 0$ may not exist and a direct transition between $\Gamma_{IMF} \simeq -0.8$ and $\Gamma_{IMF} > 0$
    may occur.} 
\end{figure}

As seen from Equation~\ref{n_general}, the mass spectrum is a direct, linear, function of the density PDF. 
Therefore it is important to examine the prediction that are obtained for other density PDF, in particular 
the one, which develops during gravitational collapse as stated by Equation~\ref{pdf_grav}. The physical underlying idea is that as collapse proceeds new seeds to form self-gravitating objects develop.

We adopt a  simplified but enlightening approach \citep{leeh2018a}. 
Following Equation~\ref{viriel_ceff}, we simply write $M \propto R^{2 \eta + 1}$. 
$\eta=0$ corresponds to thermal or magnetic 
support whereas $\eta \simeq 0.4-0.5$ is typical of turbulent support. 
Since $\rho \propto M / R^3$, we get $\rho \propto M ^{ (2 \eta -2) / (2 \eta +1)}$.
Thus with Equations~\ref{n_general} and~\ref{pdf_grav}, and since both $M$ and $\rho$ 
are powerlaws of $R$, it is easy to see that ${\cal N}(M) \propto \sqrt{\rho} M^{-2}$ and 
therefore we obtain 
\begin{eqnarray}
\label{msp_grav}
{\cal N}(M) \propto M^ {-{ 1 + 5 \eta \over 1 + 2 \eta }}= M^ {-{ 5 n -13   \over 2 n  - 4 }}. 
\end{eqnarray}

If, at the scale of the accretion reservoir, thermal or magnetic support are dominant $\eta=0$ and 
thus ${\cal N}(M) \propto M^{-1}$ which implies $\Gamma _{IMF}=0$. On the other hand, 
if turbulence is dominant at the scale of the accretion reservoir, $\eta \simeq 0.5$ and 
${\cal N}(M) \propto M^{-7/4}$, leading to $\Gamma _{IMF}=-3/4$. The transition between the two regimes
is expected to occur roughly at the magneto-sonic scale, that is to say when $3 C_s^2 + V_a^2/2 \simeq V_{rms}^2$
(see Equation~\ref{viriel_ceff}).

These predictions are major deviations from the Salpeter value, 
$\Gamma _{IMF}=-1.33$ but they might explain 
lower values of $\|\Gamma _{IMF} \|$ obtained for lower masses, or even the shallow 
slopes observed for dense starburst clusters (c.f. Fig. ~\ref{fig:alphaplot}). 
As discussed in Section~\ref{simulation}, both behaviours, i.e. 
$\Gamma _{IMF}=0$ and $-3/4$ have been observed 
in numerical simulations.
These considerations, therefore open the road for a non universal 
IMF made of universal regimes but whose domains of validity depend on both the support and 
the density PDF. 

The value $\Gamma_{IMF}=0$, obtained from thermal support {\rm at the scale of the accretion reservoir}, 
is dominant, may constitute an {\rm effective} peak for the IMF, 
evenso it would not lead to a real decline of the number of objects with masses. It is more accurately 
defined as a {\it plateau}.
As discussed in Sections.~\ref{sec_jeans_mass} and ~\ref{FHSC}, positive $\Gamma_{IMF}$ will be induced by the adiabatic phase.

Figure~\ref{fig:recap} recaps the most important ideas presented in Sects.~\ref{gravo_turb_sup} and~\ref{density_pdf}.

\subsubsection{Gravo-turbulence: the influence of filamentary geometry}
It has since long been established that filaments are ubiquitous in the ISM
\citep{hacar2022}. For instance many cores are observed to be lying 
inside larger scale filaments \citep{pineda2022}. 

The fragmentation of self-gravitating 
filaments  as well as the possible role that this may have 
on the establishment of the IMF have thus been investigated. 
\citet{inutsuka2001} provided the first analytical calculation to predict
the IMF from the fragmentation of a filament. His calculation is very similar to the classical approach of \citet{pressschechter} except that it is applied to the filamentary geometry. 
%The fraction of collapsed region $f(M,\delta > \delta_c)$, defined as regions which have a density
%$\delta _c$ above the mean density  and having a mass scale $M$ is given by
%\begin{eqnarray}
%\nonumber
%   f(M,\delta > \delta_c) = \int ^\infty _{\delta_c} {d \delta \over \sqrt{2 \pi \sigma_M^2} }
%   \exp \left( {-\delta^2 \over 2 \sigma_M ^2} \right) \\
%   = \int _M ^\infty {\cal N} (M') {M' \over M_{line}} P(M | M') dM',
%   \label{int_mass_inut}
%\end{eqnarray}
%where $M_{line}$ is the lineic mass of the filament, $P(M | M')$ is the conditional
%probability of finding a collapsed region of mass scale $M$ inside a collapse region
%of mass scale $M'$ and $\sigma_M$ is the dispersion of the density fluctuations assumed to have a Gaussian distribution. 
%Through Parseval's theorem the density variance, $\sigma_M$, is
%\begin{eqnarray}
%    \sigma_ M ^2 \propto \int _{k_{min}} ^ {k_M} \delta _k(k')^2 dk' \propto
%    k_{min}^{n+1} -  k_{M}^{n+1}, 
%\end{eqnarray}
%where $k_{min}$ is the integral scale of the turbulence, likely comparable to the 
%length of the filament.
%Taking the derivative with respect to mass of Equation~\ref{int_mass_inut}, we obtain 
The predicted mass spectrum is given by  
\begin{eqnarray}
\label{fil_spectrum}
{\cal N}(M) = {dN \over dM } = -2 {M_{line} \over M } {\delta_c \over 
\sqrt{\pi} } \exp \left( - { \delta _c^2 \over 2 \sigma_M ^2} \right) 
{1 \over \sigma_M^3 } {d \sigma_M^2 \over dM} \propto {M^{-3-n} \over \sigma_M^3 },
\end{eqnarray}
where $M_{line}$ is the lineic mass of the filament and $\sigma_M$ is the density variance.
%To go further, the dependence of $\sigma_M$ on the mass, $M$, must be specified. 
%\citet{inutsuka2001} wrote that $\sigma_M^2 \propto \int _{-k_M} ^{k_M} \delta _k(k')^2 dk'$, where $k_M = 2 \pi M_{line} / M$. However since it is expected that the density fluctuations in molecular clouds are due to turbulence, $\delta_k^2$, the powerspectrum of the density field, is expected to be such that $\delta_k^2 \propto k^n$, with 
%$n \simeq -1.5$ (-5/3 would correspond to the Kolmogorov case). Thus $\sigma_M$ as 
%written above does not converge. Instead we believe that the expected expression for 
 %With this expression, Equation~\ref{fil_spectrum} leads to
%\begin{eqnarray}
%\label{fil_spectrum2}
%{\cal N}(M)  \propto  \delta_c  
% \exp \left( - { \delta _c^2 \over 2 \sigma_M ^2} \right) 
%{M^{-3-n} \over \sigma_M^3 }.
%\end{eqnarray}
With $n=-1.5$, which is close to the Kolmogorov case, we have ${\cal N}(M) \propto M^{-1.5}$ and therefore $\Gamma_{IMF}=-0.5$. 
Note that in the figure 1 of \citet{inutsuka2001} at early time 
an exponent close to $\Gamma_{IMF}=-0.5$ is obtained but at later time
an exponent closer to the 
expected Salpeter exponent appears. This is due to an evolution of the density variance,
 $\sigma_M^2$, which at later time is dominated by small scales because 
 the development of gravitational instabilities has been taken into account into the calculation. 
 Whether this may really lead to the formation of {\it new} structures is a matter of debate. 
 %However we believe
 %that this does not lead to the formation 
 %of {\it new} self-gravitating structures and  therefore $\sigma_M$ should remain nearly constant.
  Finally, let us also stress that since to obtain Equation~\ref{fil_spectrum}, an integral is performed between 
$M$ and $\infty$, this calculation represents groups of cores rather than individual 
cores, which request an integration between 0 and $M$ (see for instance Equation~\ref{hc_eq1}). \\

More recently, \citet{lee_H_C_2017} have developed an analytical model in which 
the formalism presented in Section~\ref{gravo_turb_sup} is adapted to 
the filamentary geometry. The key feature of the model is that it 
considers a continuous transition for scales that are much smaller 
than the filament radius, $R_f$ to scales larger than $R_f$. The criterion 
that density fluctuations must satisfied to collapse is based on the virial theorem 
as stated by Equation~\ref{viriel_ceff}. The effect of an elongated geometry is accounted for by using an ellipsoidal model for the gravitational energy. Thermal, turbulent and magnetic 
support are all included. Both the CMF and the group of core distribution have been calculated \citep[see figure~1 of][]{lee_H_C_2017}.
%\begin{figure}
%    \centering
%    \includegraphics[width=\textwidth]{Figures/Lee_2017_Fig1.png}\vspace{-0.5cm}
%    \caption{CMF from filaments
%    of various mass per unit length and magnetization. Solid curves display the CMF in non-magnetized filaments of 40 (yellow), 160 (light green), and 640 (dark green) M pc$^{-1}$. Dashed and dot-dashed light green curves show the CMF
%in 160 M pc$^{-1}$ filament having an Alfvenic Mach number of 2 and 1 respectively. 
%The dot-dashed dark
%green curve shows the CMF in a 640 M pc$^{-1}$ 
%filament with an Alfv\'en Mach of 1. The straight line shows
%the slope $\Gamma_{IMF} =-1.33$. Canonical values of 10 pc length and
%0.1 pc diameter are used for all filaments. Figure extracted from \citet{lee_H_C_2017}.}
%    \label{fig:lee2017}
%\end{figure}
%Figure~\ref{fig:lee2017} displays the predicted CMF as well as the mass function of the group of cores (that would correspond to the first crossing as discussed in \citet{hopkins2012}).

As expected, the CMF depends on the line mass
 and on the magnetic field strength. In particular, it is found that in the absence of
magnetic field, filaments with high line mass
fragment into many small cores. In the
presence of a moderate magnetic field and for sufficiently high line mass,
the CMF presents a peak around 0.5-1 M$_\odot$. This however requires the Alfvenic Mach number to be on the
order of 2-3. For lower values, no low-mass core would develop
and the peak of the CMF would be at values of 10 M$_\odot$ or more.
It should be stressed however that since the density PDF is assumed to be lognormal 
and collapse is not accounted for (see Section~\ref{density_pdf}), the peak of the distribution is likely not  physical. Importantly and interestingly, it is seen that the CMF in filaments with high line mass, turns out to present an exponent 
$\Gamma_{IMF}$ that is fairly close to the Salpeter value. As discussed in Section~\ref{gravo_turb_sup}, this is also what is inferred 
when filament geometry is not considered. It should therefore be stressed that whereas 
many stars appear to form in filaments, there is at this stage no clear evidence that 
the filamentary geometry is playing a particularly strong role regarding the IMF.

\subsubsection{Effect of MHD shock jump conditions}
\label{pn2022}

 The \citet{padoan_nordlund_2002_imf}  CMF/IMF model  relies on MHD shocks, which are primarily responsible for setting up the value of $\Gamma_{IMF}$ through the formation of  
compressed layers induced by ram pressure in a weakly 
magnetized medium.  The magnetic field is assumed to be parallel to 
the compressed layer and  therefore perpendicular to the gas velocity. 
The postshock density, $\rho_1$, the thickness of the layer, $\lambda$,
and the postshock magnetic field, $B_1$, can all be related to the 
Alfv\'enic Mach numbers, ${\cal M}_a=v/v_a$, 
%($v$ is the velocity and $v_a$ the Alfv\'en speed),
  and preshocked quantities, $\rho_0$ and $B_0$ according to
\begin{eqnarray}
{\rho_1 \over \rho_0} \simeq {\cal M}_a \; , \;  {\lambda \over  L } \simeq {\cal M}_a^{-1} \; , \; { B_1 \over B_0} \simeq {\cal M}_a \; ,
\label{eq_pn1}
\end{eqnarray}
where $L$ is the spatial extension of the incoming flow that gives rise to the shocked layer. Let us remind
that for 
hydrodynamical isothermal shocks,  $ \rho_1 / \rho_0 \propto {\cal M}^2$, $ L / \lambda \propto {\cal M}^2$. 
The dependence on ${\cal M}_a$ instead of ${\cal M}^2$ is due to the
magnetic pressure which is proportional to $B^2$. 

The typical mass of this perturbation is given by $M \simeq \rho_1 \lambda^3 $ leading to
\begin{eqnarray}
\nonumber
M \simeq   \rho_0 L^3  {\cal M}_a^{-2} \; {\rm  MHD}, \\
M \simeq \rho_0 L^3  {\cal M}^{-4} \; {\rm HYDRO} 
\label{eq_pn2}
\end{eqnarray}
As the flow is turbulent, the velocity distribution depends on the scale 
and $v \simeq L^{(n-3)/2}$.
With Equation~(\ref{eq_pn2}), one infers
$M \propto \left( { L / L_0} \right)^a $
where $L_0$ is the integral scale and $a=6-n$ in MHD or $a=9-2n$ in hydro. To get a mass spectrum, it is further assumed 
that the number of cores, $N(L)$ at scale $L$, 
is proportional to $L^{-3}$. These two  relations
lead to $\Gamma _{IMF} = -3/a$.
For $n=3.8$, 
one gets $\Gamma _{IMF} \simeq -1.36$ in MHD and $\Gamma _{IMF} \simeq -2.14$ in hydro.

In a second step,  a lognormal distribution 
of Jeans masses within the clumps is envisonned
 and 
the mass spectrum  is therefore multiplied by a 
distribution of Jeans masses, $\int _0^M p(M_J) dM_J$. The shape of the final mass spectrum  is very similar to 
the observed IMF as seen for example in the Figure 1 of \citet{padoan_nordlund_2002_imf}.

Note that this theory presents two significant difficulties.  
First of all, Equation~\ref{eq_pn1} implies that in the densest regions
where dense cores form, the magnetic field strength 
is proportional to the density and this is not in good agreement with what is 
observed both in numerical simulations and in
observations. As recalled in Section~\ref{mag}, in numerical simulations
 it is usually found  that at high  densities the trend $B \propto \sqrt{\rho}$
 is inferred. Second of all, the theory predicts that in the  hydro case $\Gamma _{IMF} \simeq -2$ and
 this is not compatible with hydrodynamical simulations as discussed in Section~\ref{simulation}.

\subsection{How to explain the characteristic mass of stars?}
\label{phys_peak}
The existence of a relatively well defined peak of the IMF 
is at first sight surprising because the ISM presents 
a wide range of processes and physical conditions. Below we 
review the various ideas which have been proposed and briefly present 
the arguments in favor or against their validity. 

\subsubsection{Possible role of the Jeans length in setting a peak}
\label{jeans_mass_peak}

The Jeans length \citep{Jeans_1902} has often been  envisaged to be playing a significant 
role regarding cloud fragmentation and even possibly to be at the origin of the peak of the IMF. 
One important difficulty however is that it presents a relatively strong dependence on density and temperature since as shown by Equation~\ref{jeans_mass},
$M_{jeans} \propto T^{3/2} \rho^{-1/2}$. Since both temperature and density 
are observed to vary substantially through star forming regions, this is 
in apparent contradiction with the apparent universality of the IMF but 
various propositions have been made to combine Jeans mass 
considerations with other physical arguments.

\citet{larson2005} has proposed that the transition between the molecular cooling and the dust cooling leads
favors fragmentation at the density at which this transition occurs. 
Indeed, the effective adiabatic index induced by molecular cooling is about 
$\Gamma_{\rm ad} \simeq 0.7$ whereas dust cooling leads to  $\Gamma_{\rm ad} \simeq 1.1$. 
Since the transition between the two cooling regimes occurs at about $10^5$ cm$^{-3}$, 
the Jeans mass would be compatible with a peak around 0.3-0.5 M$_\odot$ (see also \citet{elmegreen2008}). \citet{hc2009} have developed an analytical model for the CMF in which 
the two cooling regimes are taken into account but could not identify the corresponding peak. This also cannot be a {\it general} explanation for the IMF, because some very-dense clusters like Orion have an essentially normal-looking IMF peak, but must have formed in conditions orders of magntiude denser than the transition density \citep{sf_big_problems}. Nevertheless it cannot be discarded {\it a prior} as an explanation for the IMF in less-dense, more-typical Solar neighborhood clouds. This scenario has been explored in simulations and the result will be discussed 
in Section~\ref{gamma_inf43}.

%\citet{elmegreen2008} analyse the dependence of the Jeans mass in 3 different environments and concluded that it 
%weakly depends on density and temperature. 

The gravo-turbulent theories \citep{padoan1997,hc08} identify a peak in the reservoir distribution when they assume a lognormal 
density PDF, that is to say assume that a density and therefore a Jeans length is dominating the distribution. 
\citet{hc08} found that the peak of the reservoir mass function is given by $M_J^0 / {\mathcal M}^2$, where 
$M_J^0$ is the Jeans mass of the mean density and ${\mathcal M}$ the Mach number. 
In principle, this must lead to some significant variations of the peak since 
large variations of density and velocity dispersion are observed between the clumps. 
However, it has been proposed that due to Larson relations \citep{Larson81}, some  compensations may occur leading to less variability. Indeed, the CO clumps are observed 
to present a scaling $M \propto R^2$ or equivalently $\rho \propto R^{-1}$ and $\delta v \propto R^{0.4}$. This would lead to $M_J^0 / {\mathcal M}^2 \propto R^{0.3} \propto M^{0.15}$. The latter relation is a shallow dependence, which could therefore 
possibly explain a weak variation of the peak of the IMF from regions to regions. 
There are however two main caveats. First there are large deviations around 
Larsons relations which would result in large variations of the peak and second 
at high density the PDF is not lognormal any more 
(see Section~\ref{density_pdf}) and therefore these considerations would become invalid.

\subsubsection{Protostellar heating}
\label{protoheat}
When a star forms, the accretion luminosity heats the gas around and increases the temperature, therefore increasing 
the Jeans mass around the protostar. It has been proposed by 
\citet{bate2009}, \citet{krumholz2011} and \citet{guszejnov2016}
that this leads to a very shallow dependence of the Jeans mass into the density and therefore may 
constitute a possible way to explain the characteristic mass of stars. 
Below we follow the arguments presented in \citet{krumholz2015}. 

Let us consider a protostar of luminosity $L$, at a distance $r$, we have the relation
\begin{eqnarray}
\label{luminosity}
    L \simeq 4 \pi \sigma_{sb} r^2 T^4,
\end{eqnarray}
where we remind that $\sigma_{sb}$ is the Stefan-Boltzmann constant.
The  idea  is now to compare the Jeans mass, $M_J \propto C_s^3 \rho^{-1/2}$ 
with the mass, $M \propto \rho r^3$, enclosed in the radius $r$. 
If these two masses are comparable,  fragmentation is likely suppressed and therefore 
the mass should be accreted in a single object. Writing $M_J \simeq M$ and 
using Equation~\ref{luminosity}, we get that $M \propto L^{3/10} \rho^{-1/5}$.
Next it is assumed that the luminosity, $L$ is the accretion luminosity and thus 
$L \propto \dot{M}$, with the constant of proportionality set by the condition for deuterium burning (Equation \ref{eq:tcore}), and $\dot{M} \simeq M / \tau_{ff} \propto M \sqrt{\rho}$. 
Thus this leads to $M \propto \rho^{-1/14}$. The full expression obtained by 
\citep{krumholz2015} is (see his equation 13.38)
\begin{eqnarray}
\label{rad_peak}
    M = 0.3 M_\odot \left( {n \over 100 \, {\rm cm}^{-3}} \right)^{-1/14}. 
\end{eqnarray}
As can be seen, this mass has  a very shallow dependence in gas density and therefore 
little variation is expected across environments. Moreover its value is very 
close to the estimated peak of the IMF. The more-detailed semi-analytic radiative transfer calculation by \citet{krumholz2011} gives a mass scale approximated by \citep[see Figure~5 of][]{krumholz2011}
\begin{equation}
    M\approx 0.03 M_\odot \left(\frac{P}{10^{10} k_{\rm B}\,\rm K\, cm^{-3}}\right)^{-0.28} \left(Z_{\rm d}\right)^{-0.22},
\end{equation}
where $P$ is the confining pressure of the clump and $Z_{\rm d}$ is the dust opacity parameter. This is significantly more sensitive to ambient conditions than the corresponding analytic approximation, but still less so than the Jeans mass.

A possible difficulty of this theory is that the density or pressure that is involved in the 
mass estimate is not very clearly defined. When a protostar has formed, the density distribution around 
it, is likely an $r^{-2}$ powerlaw and in this configuration, the Jeans mass is not uniquely determined. 
%\mike{typically given in terms of the pressure confining a Bonnor-Ebert sphere}
Also when 
radiation starts, the reservoir is already globally collapsing and it requires a strong heating to stop.  
As discussed in Section~\ref{radiation}, numerical studies exploring this scenario have arrived at conflicting results.

%\mike{looks like you assumed very-different protostellar evolution tracks from the Offner 2009 model used by the Berkeley group; could this be an important uncertainty/source of discrepancy?}

%\mike{some other criticisms of the krumholz picture: }
%\begin{itemize}
%\item part of the \citet{krumholz2011} argument is that D burning sets a linear protostellar mass-radius relation (constant $T_{\rm c} \sim 10^6\,\mathrm{K}$). This is a VERY rough approximation and $R_\star$ never really evolves as steeply as $\propto M$ in the D burning phase (e.g. Fig. \ref{fig:hosokawa2009}). 
%\item Protostellar heating can make the IMF more top-heavy but not less top-heavy. Isothermal MHD studies \citet{haugbolle2018,guszejnov_2020_MHD_characteristic_mass} get a top-heavy IMF in normal star-forming conditions that doesn't seem to converge away. At best, protostellar heating is  a partial explanation.
%\end{itemize}

%\mike{Important subtlety: the Jeans mass has two competing terms: $\propto \rho^{-1/2}$ and $\propto T^{3/2}$, so a universal IMF dependent on the Jeans mass would require $\rho \propto T^3$. But higher-density ISM regions do host more-intense star formation and radiation/cosmic ray fields, hence higher temperatures, so it is difficult to dismiss outright the scenario where $\rho$ and $T$ covary so as to limit variations in $M_{\rm J}$.}
%\mike{+ various studies find that the IMF can be too top-heavy even in plausible cold isothermal conditions \citep{haugbolle2018,guszejnov_2020_MHD_characteristic_mass}, so protostellar heating can only make things worse there. }

\subsubsection{First hydrostatic core and tidal screening}
\label{FHSC}
\citet{leeh2018b} and \citet{hennebelle_2019} have proposed that the peak of the IMF may be related 
to the mass of the first hydrostatic core (Section~\ref{first_core}).
The proposed mechanism works in three steps. First, the formation of the first hydrostatic core
signs the transition from an isothermal to an adiabatic equation of state. The latter suppresses any further fragmentation and essentially stops the collapse. Second, the gas piles-up until the central temperature becomes sufficiently high for H$_2$ to dissociate. At this point the mass of the first hydrostatic core is about 0.03 M$_\odot$ and this constitutes the minimal mass that needs to be accumulated in order to produce a protostar. Third, the first hydrostatic core is surrounded by a collapsing, nearly isothermal envelop and can further accrete and grow. The surrounding gas can however also fragment and produce new objects instead of being accreted. However, both the first hydrostatic core and the collapsing envelope (which typically has a 
profil $\rho \propto r^{-2}$) exert tidal forcing on the surrounding medium and this stabilizes the gas around against fragmentation. \citet{hennebelle_2019} developed a model which predicts that 
this results in a central object that typically has a mass of about 10 times the mass of the first hydrostatic core, which is about 10$\times 0.03$ M$_\odot$= 0.3 M$_\odot$. The model considers concentric shells around 
the first hydrostatic core and compute the probability to find self-gravitating density fluctuations
with a mass  at least equal to the mass of the first hydrostatic core. The mass contained in the 
radius for which the probability to find $n$ density fluctuations, is then computed. 

A simple model has been proposed by \citet{colman2019}. The argument is as follows. Let us consider 
a central point mass such as a first hydrostatic core and let $M_{FHSC}$ be its mass. As collapse is proceeding 
this object is surrounded by a density envelope $\rho _{env} = A r^{-2}$, where $A$ is a constant. 
A simple condition for fragmentation to occur is that the self-gravity of the envelope is lower 
than the  tidal force induced by  the first hydrostatic core and this leads to 
$2 G M_{FHSC}  r_{tidal} ^{-3} = 4 \pi /3 G \rho_{env}$. We thus obtain 
\begin{eqnarray}
r_{tidal} = {6 M_{FHSC}  \over 4 \pi A}. 
\end{eqnarray}
The gaseous mass enclosed within $r_{tidal}$ is thus 
\begin{eqnarray}
M_{tidal} = \int _0 ^ {r _{tidal}} 4 \pi r^2 dr = 4 \pi A r _{tidal} = 6 \times M_{FHSC}. 
\label{Mtidal}
\end{eqnarray}
Assuming that the total mass,  $M_{tidal}$, is accreted onto the central first hydrostatic core, 
we thus obtained that the peak of the IMF would be about  $7 \times M_{FHSC} \simeq 0.2$ M$_\odot$.
Remarkably enough the tidal mass does not depend on the constant $A$ and this implies that 
it is likely relatively insensitive to the environmental conditions. 

 We stress that this theory to explain the peak of the IMF does not primarily rely 
 on Jeans mass consideration but rather on the mass of the FHSC which 
 as discussed earlier depends on the detailed microphysics of the ISM (see Equations~\ref{eq:compression_mj} and \ref{Mfhsc}) and more
 particularly on the dust opacity and of the H$_2$ physics. Moreover the dependence on the 
 metalliticy is expected to be very shallow (see Equations~\ref{eq:compression_mj}).
 
 Altogether,  
 the tidal screening theory appears to  weakly depend on large scale conditions and ISM composition which 
 makes it  a good candidate to explain the relative universality of the peak of the stellar initial mass function.

%\subsubsection{Protostellar jets Guszejnov et al. 2021?}
%Shall we discuss their possible influence on the peak in a dedicated session? 
%MIKE

\section{Current state of the art in numerical simulations - testing the theories}
\label{simulation}
A broad set of simulations have been performed with increasing accuracy, both regarding 
the physics and the numerical resolution. 
We discuss here the various types
of simulations from the simplest, in terms of physical processes,
to the most complete calculations. 
Whenever possible, we make the link with the analytical models discussed in the previous section.

%\begin{figure}
%    \centering
%    \includegraphics[width=\textwidth]{Figures/grudic_2022.png}
%    \caption{ Series of snapshots for a molecular cloud simualtion from \citet{grudic2022}.
%    The simulation includes radiation feedback from ionizing to far IR bands, protostellar jets, stellar winds, and supernovae. The cloud
%    get dispersed after about 8$\%$ of the gas has been converted stars. \href{https://www.youtube.com/watch?v=LeX5e51UkzI}{Movie}}
%    \label{fig:grudic2022}
%\end{figure}

\subsection{Numerical algorithms and caveats}
Before discussing the results of the various simulations 
%that have been performed to investigate the origin of the IMF 
we need to stress some of the most important 
limits and numerical uncertainties that must be addressed. 
For a description of the numerical techniques themselves, which are beyond the scope 
of the present paper we refer to other, more-comprehensive discussions
\citet{teyssier_commer2019} and \citet{grudic2021}. 
We stress that substantially diverse codes, ranging from Eulerian adaptive mesh refinement 
algorithm to  Lagrangian smoothed particle hydrodynamics, moving mesh, and mesh-free Godunov-type
codes are now employed in the star formation community. Comparisons between different codes 
have generally found fair agreement in hydrodynamics setups with simple EOS, provided sufficient resolution
\citep[e.g.][]{commercon2008,federrath_sim_2010}. However, further careful code comparisons will be useful as
complex, multi-physics simulation setups with subgrid feedback prescriptions become more common: the impact of
the further numerical choices required will deserve some investigation.

In the present context, two aspects of the simulations require a particular attention. First, and 
as it is most often the case, the issue of numerical resolution must be carefully addressed. 
This is most notably true for the question of the peak of the IMF as it is unavoidable that 
given the number of resolution elements \citep[e.g.][]{Bate_1995_accretion, truelove_1997_dens_condition, grudic2021},
there is a minimum mass which can be described with a given specific algorithm. 
It is therefore necessary to perform systematic convergence tests and to 
obtain resolution independent results to study the IMF with numerical simulations.
It must in particular be stressed that  in simulations that do not include  physical processes
(as discussed in Section~\ref{phys_peak})
able to generate a peak in the sink particle distributions (aiming at representing the IMF),   
 a numerical peak due to finite spatial resolution nevertheless appears and it is fundamental 
 to distinguish between {\it physical} and {\it numerical} peaks. 
 It should also be made clear that getting self-consistently a whole IMF in a single simulation remains very challenging.
 Indeed forming simultaneously very massive stars, say of 100 $M_\odot$, and in the same time resolving objects
 of $10^{-2}$ $M_\odot$ like the FHSC, requires resolving spatial or mass scales that differ by at least 4 to 5 orders
 of magnitude. Obviously dealing with the corresponding timesteps is a significant part of this challenge, in particular 
 because processes like non-ideal MHD, jets and winds lead to severe constraints. 
 
The second key issue is the usage of the Lagrangian sink particles 
that model the accretion, stellar evolution, and feedback processes occuring below the resolution limit \citep{bate1997,Krumholz_2004_sinks_in_eulerian,hubber2013,bleuler_teyssier_sinks,grudic2021}. These methods inevitably introduce ad-hoc constructs into the simulation that do not correspond to physical source terms and are not guaranteed {\it a priori} to produce numerically-converged results. This is not ideal, but will remain necessary until it becomes possible to resolve stellar radii in star cluster calculations. The choice 
of the numerical parameters can potentially have decisive effects on the results, so it is important to check that $i)$ the prescription chosen can at least robustly model exactly-known point-mass accretion solutions (see Section~\ref{accretion}), and $ii)$ the IMF produced is not affected by specific choices of 
numerical parameters.

\subsection{Barotropic equation of state and hydrodynamical simulations}
Historically the first simulations that have been carried out 
to infer a statistical distribution of stars (or more accurately of sink
particles) were hydrodynamical and were either 
purely isothermal \citep{klessen2001} or 
have been using a barotropic equation of state (eos) meaning 
that the pressure is a (often piece-wise) prescribed function of the density
$P = K_i \rho ^{\Gamma_{ad,i}}$ for $\rho_{i} < \rho < \rho_{i+1}$ \citep{Bate03}. This 
 constitutes an easy way to mimic cooling and radiative process.
Various prescriptions have been used, depending which physics 
was under investigation. Here we make the important distinction 
between {\it soft} and {\it hard} eos. The former corresponds 
to an effective adiabatic index, $\Gamma _{ad,i} < 4/3$. As recalled 
in Sec.~\ref{sec_jeans_mass}, $\Gamma_{ad}=4/3$ is indeed a
critical value. Collapse is completely halted for $\Gamma_{ad} > 4/3$.

\subsubsection{Isothermal-like calculations}
\label{isotherm}
Many isothermal, or effectively isothermal hydrodynamics IMF calculations have been performed. Here we report the results 
of several calculations that are either strictly isothermal or 
which are, for the aspect discussed in this section, equivalent to an  isothermal calculation.

\citet{girichidis2011} performed a series of calculations of 
100 M$_\odot$ collapsing clouds with various initial density profiles. 
When the cloud has an initially-flat density profile (top hat or Bonnor-Ebert 
sphere), the sink mass spectra that develop, tend to present a peak, typically 
located at a mass of a few $10^{-2}$ M$_\odot$, and for larger masses, a power-law behaviour.
The exponent of this latter has not be systematically measured. It is possibly compatible 
with $\Gamma _{IMF} = -1.35$ but for several cases, it appears to be in better agreement 
with $\Gamma _{IMF} = -1$ or possibly even larger values. 
These results are in good agreement with the ones presented in 
\citet{BallesterosParedes15} where  1000 M$_\odot$ clouds of various size and Mach numbers
have been investigated. They also found a %peak at about 0.1 M$_\odot$ and a
value of $\Gamma _{IMF} \simeq -1$.  

\citet{guszejnov_isothermal_collapse} and \citet{leeh2018b} also studied the collapse of
massive clumps and carried out runs that are  strictly isothermal. 
The results were qualitatively similar to
\citet{girichidis2011} and \citet{BallesterosParedes15}, but they performed
systematic numerical convergence tests and found that the peaks of the 
stellar distributions shifted toward smaller mass
when the numerical resolution is improved, a result first noticed by 
\citet{martel_numerical_sim_convergence} using particle splitting in SPH. 
Both studies have therefore concluded that in isothermal hydrodynamics, the peak of the 
sink mass spectrum is {\it entirely numerical}, and not physical. In principle, this result
could be due to an error in the sink particle implementation, rather than the actual mass spectrum
that would occur physically, but the fact that this is seen in SPH, AMR, {\it and} MFM
implementations suggests that it is robust. Physically, this result may be understood by the tendency of isothermal, self-gravitating filaments to form spontaneously in turbulent flows, and then to collapse to infinite density synchronously along their length without fragmenting \citep{Inutsuka_Miyama_1992}. Isothermal disks that form would also be prone to catastrophic fragmentation \citep{kratter10a}. The existence of a well-defined, converged mass spectrum from self-gravitating, isothermal turbulence is difficult to rule out {\it a priori}, but it has yet to be demonstrated in numerical models \citep[except][where numerical convergence for cores is claimed]{gong_ostriker_2015}. This is a serious problem for all analytic models that attempt to explain the IMF in terms of isothermal, self-gravitating hydrodynamics (c.f. Section \ref{sec:theory})

\setlength{\unitlength}{1cm}
\begin{figure*}[]
\begin{picture} (0,16)
\put(6.4,12){\includegraphics[width=5cm]{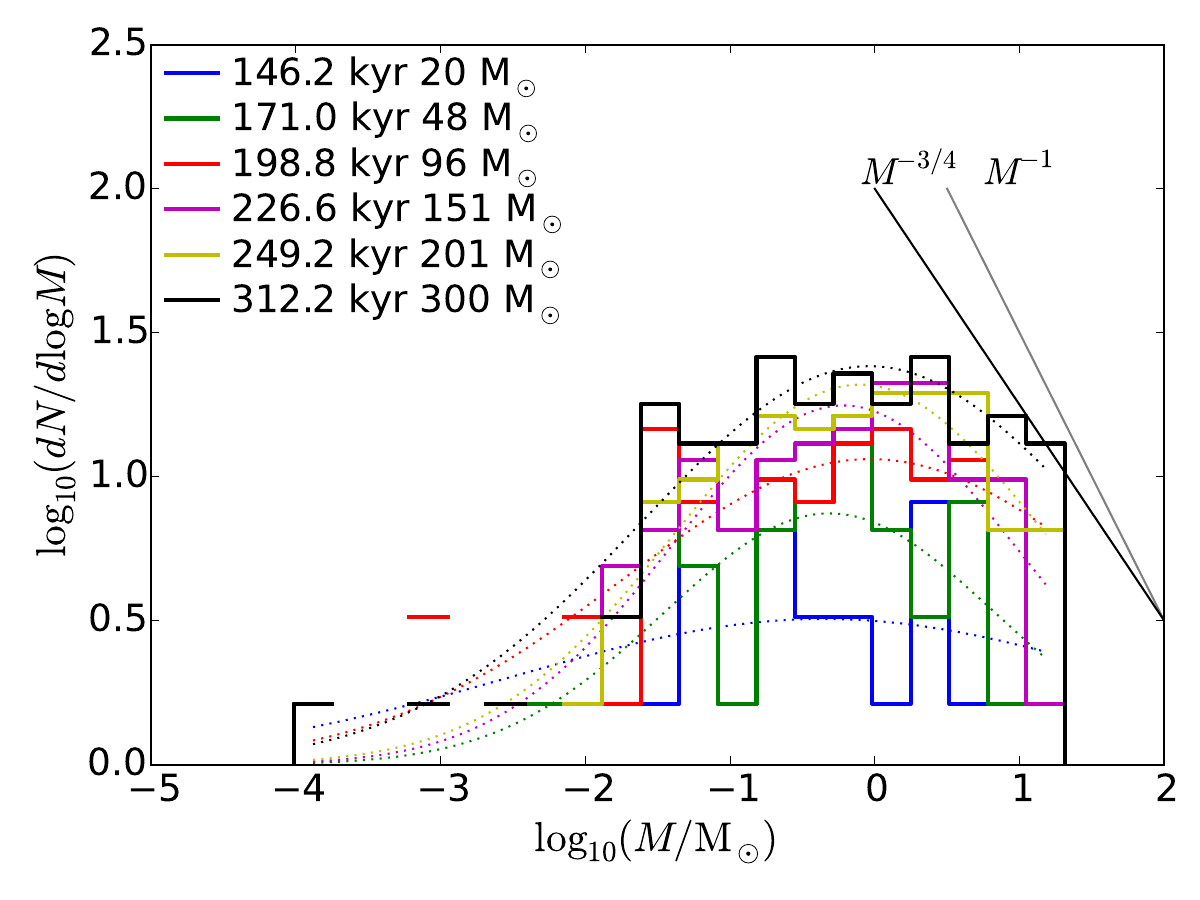}}
\put(1.2,12){\includegraphics[width=5cm]{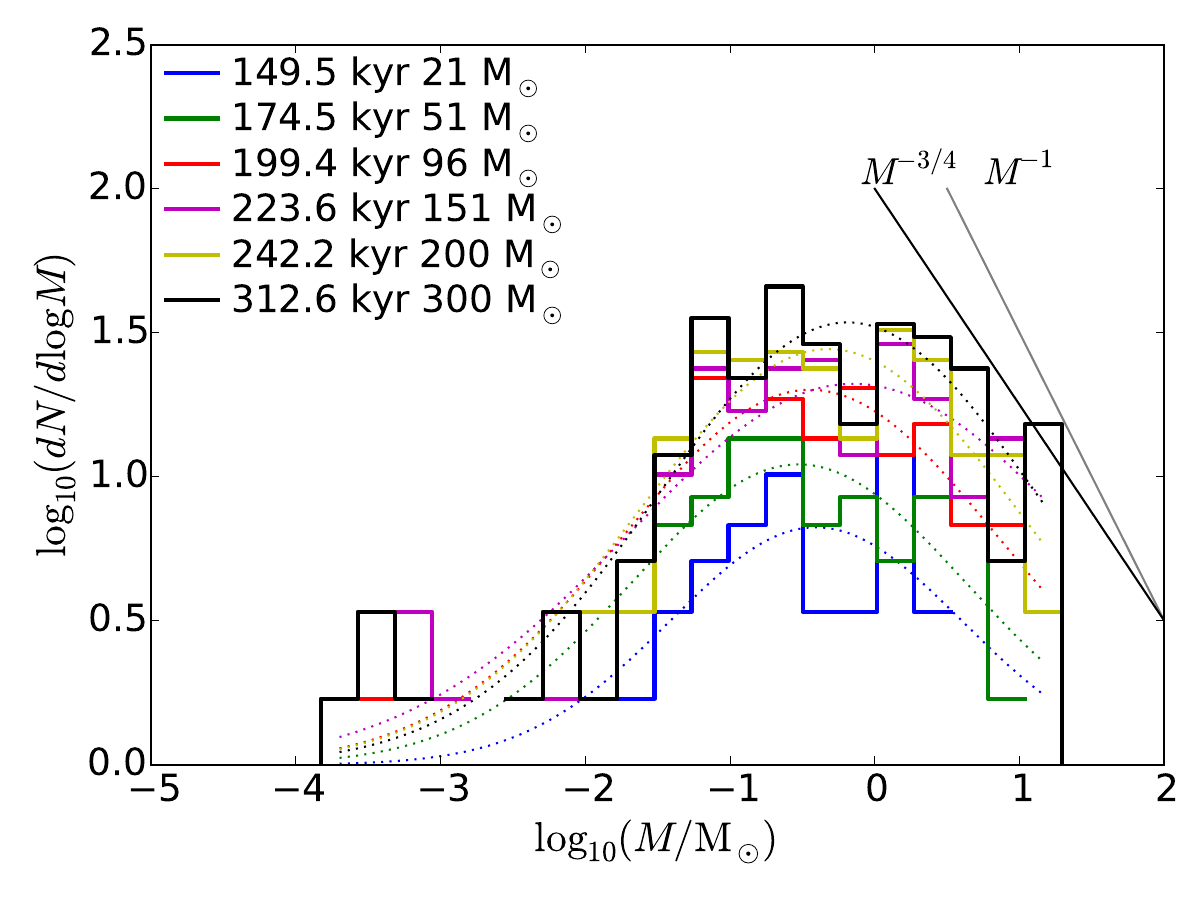}}
\put(-4,12){\includegraphics[width=5cm]{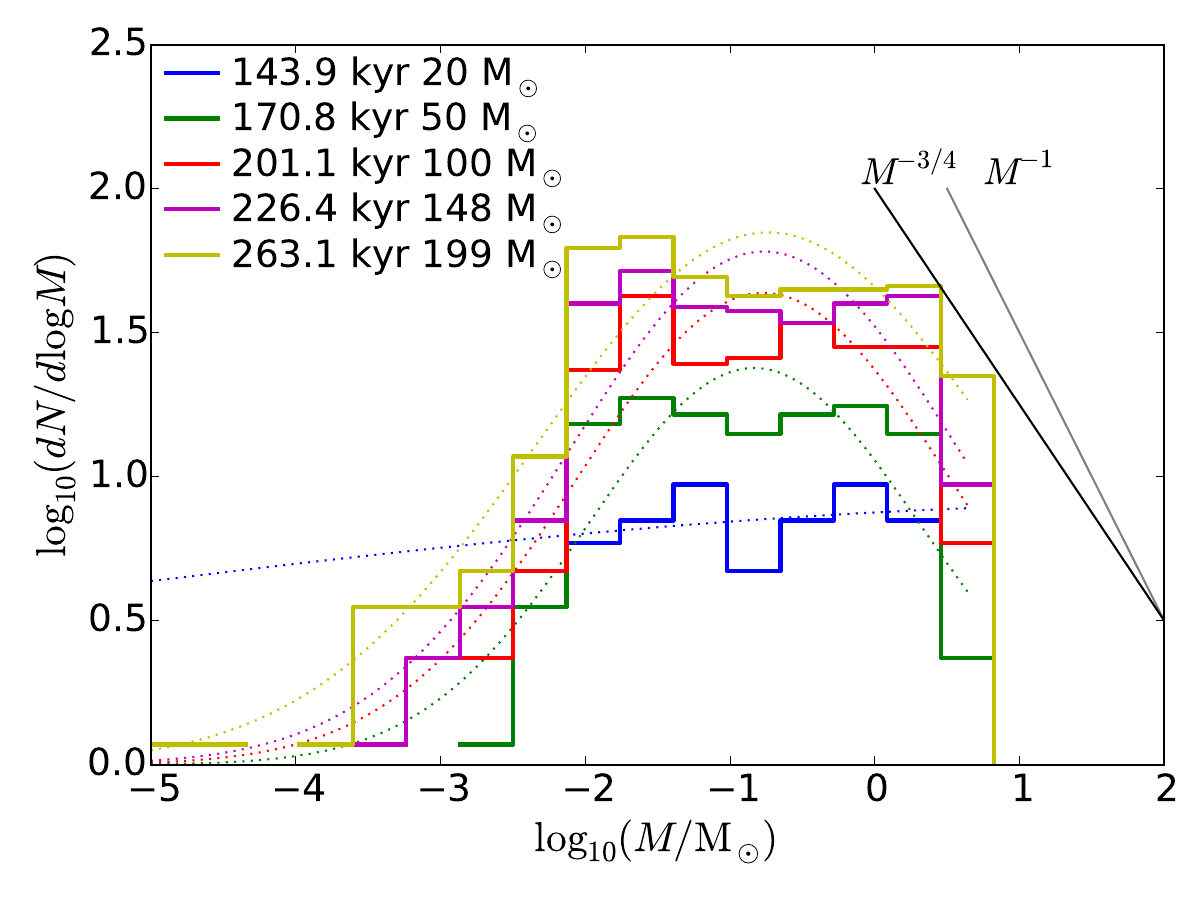}}
\put(3.6,15.7){A1++, 9 AU}
\put(8.6,15.7){A1+, 19 AU}
\put(13.6,15.7){A1, 38 AU}

\put(6.4,8){\includegraphics[width=5cm]{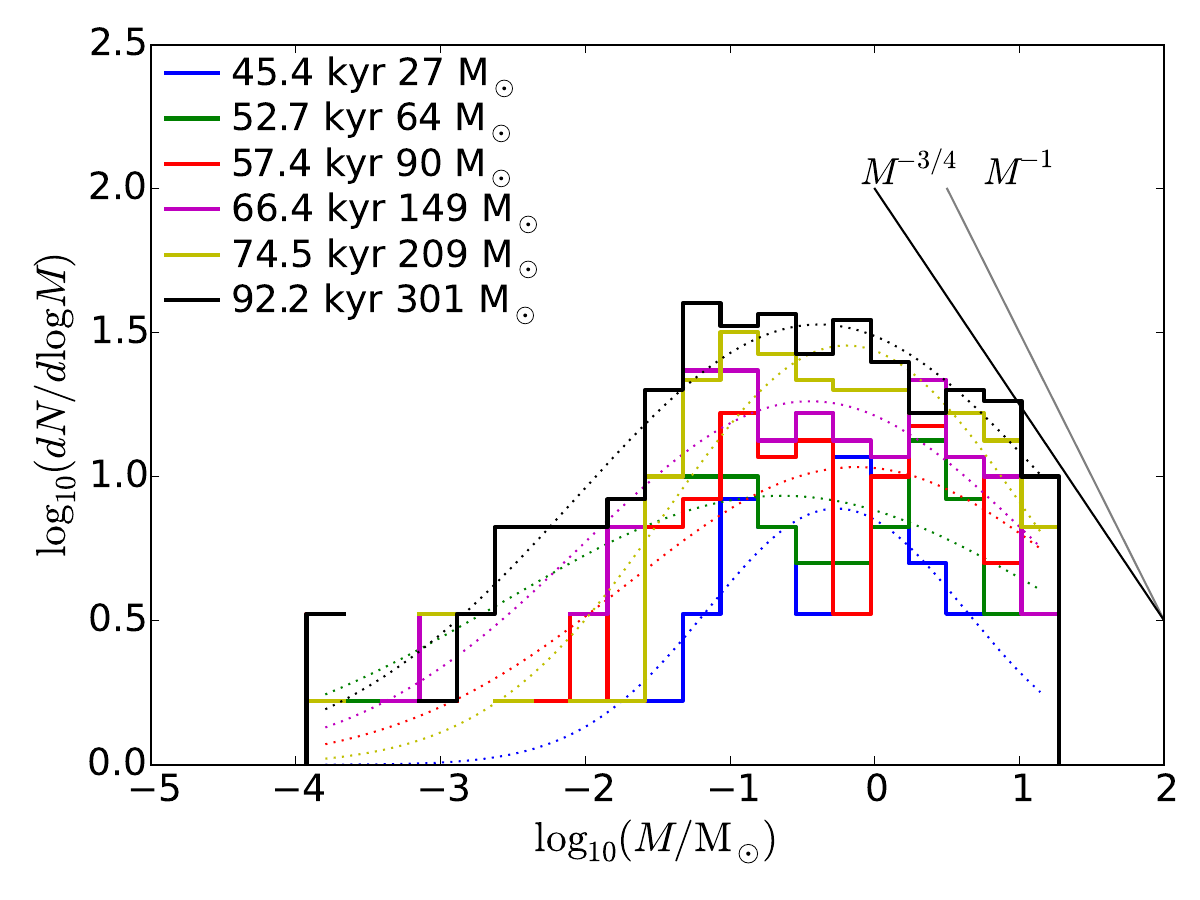}}
\put(1.2,8){\includegraphics[width=5cm]{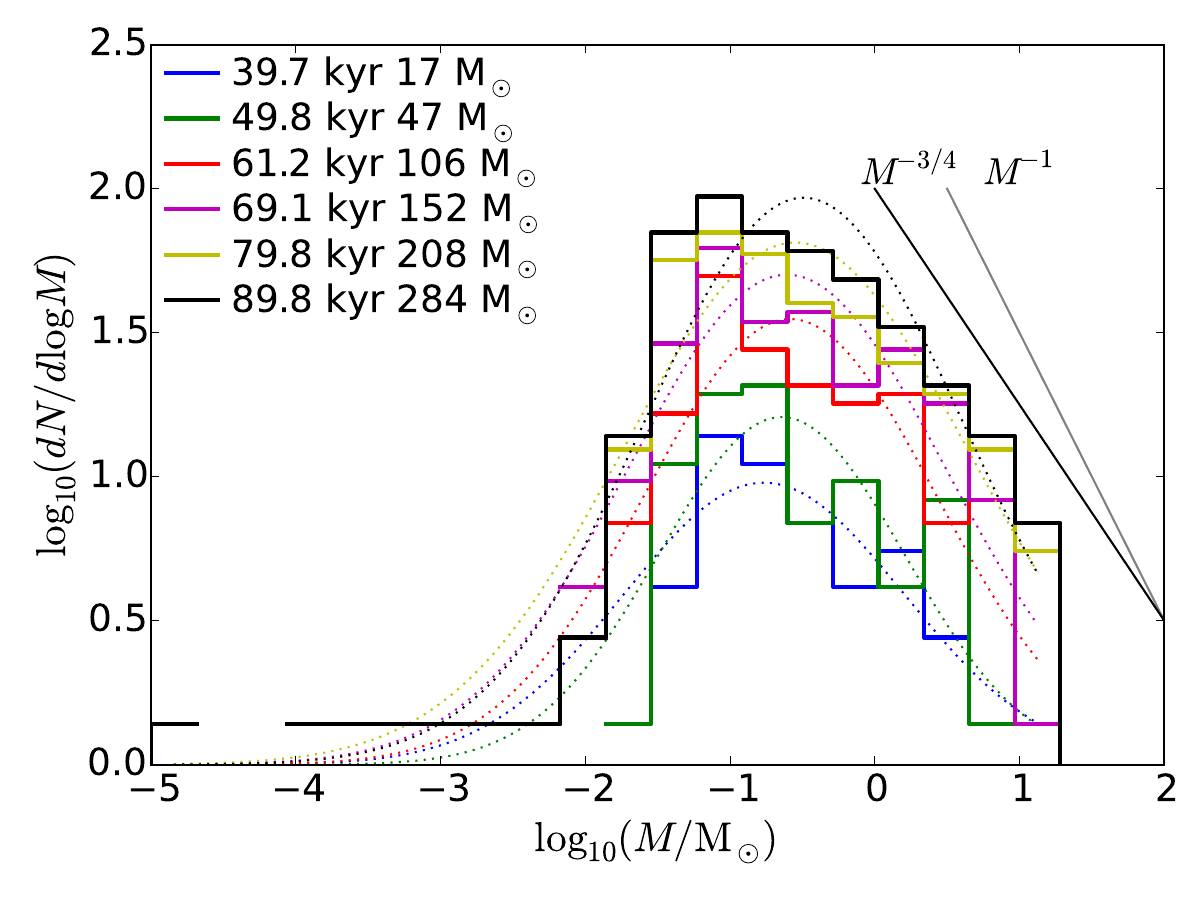}}
\put(-4,8){\includegraphics[width=5cm]{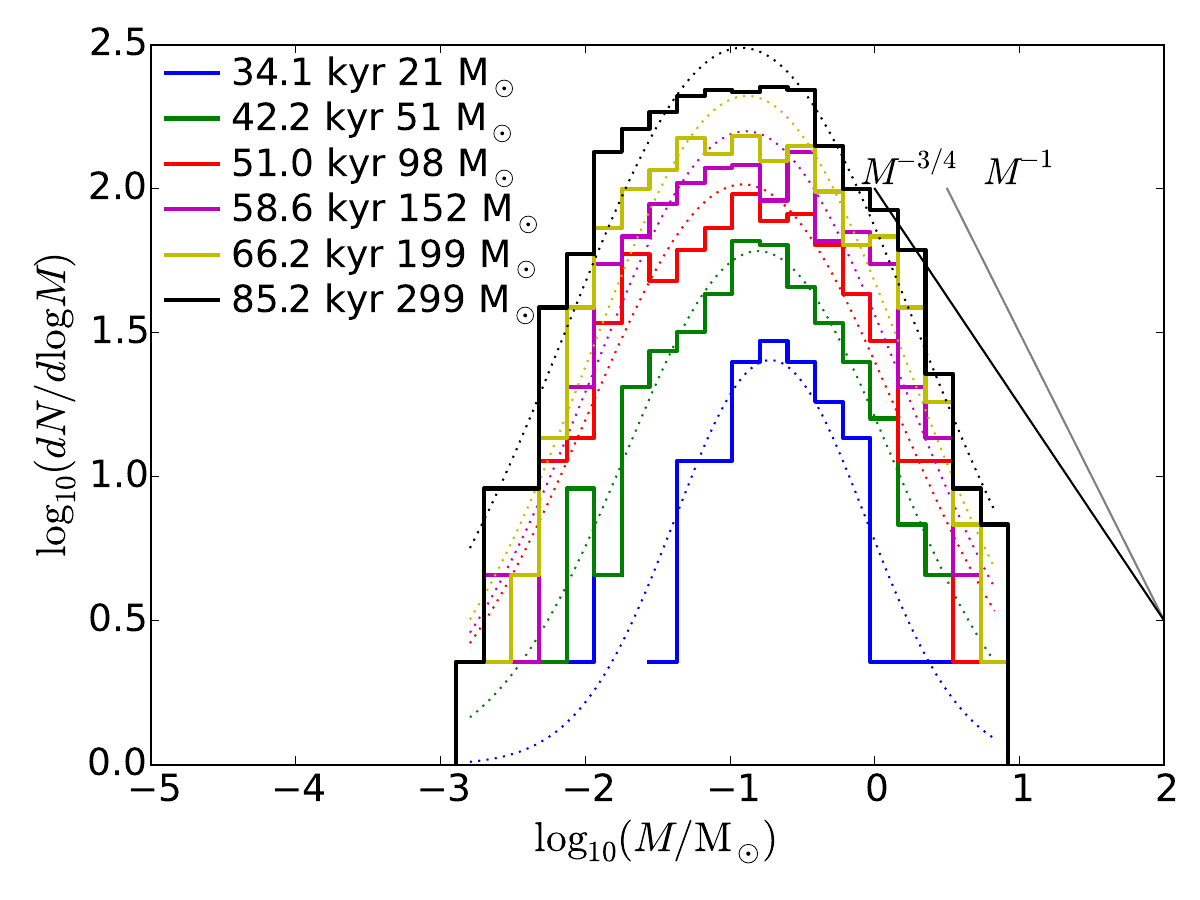}}
\put(3.6,11.7){B1++, 4 AU}
\put(8.6,11.7){B1+, 8 AU}
\put(13.6,11.7){B1, 17 AU}

\put(6.4,4.){\includegraphics[width=5cm]{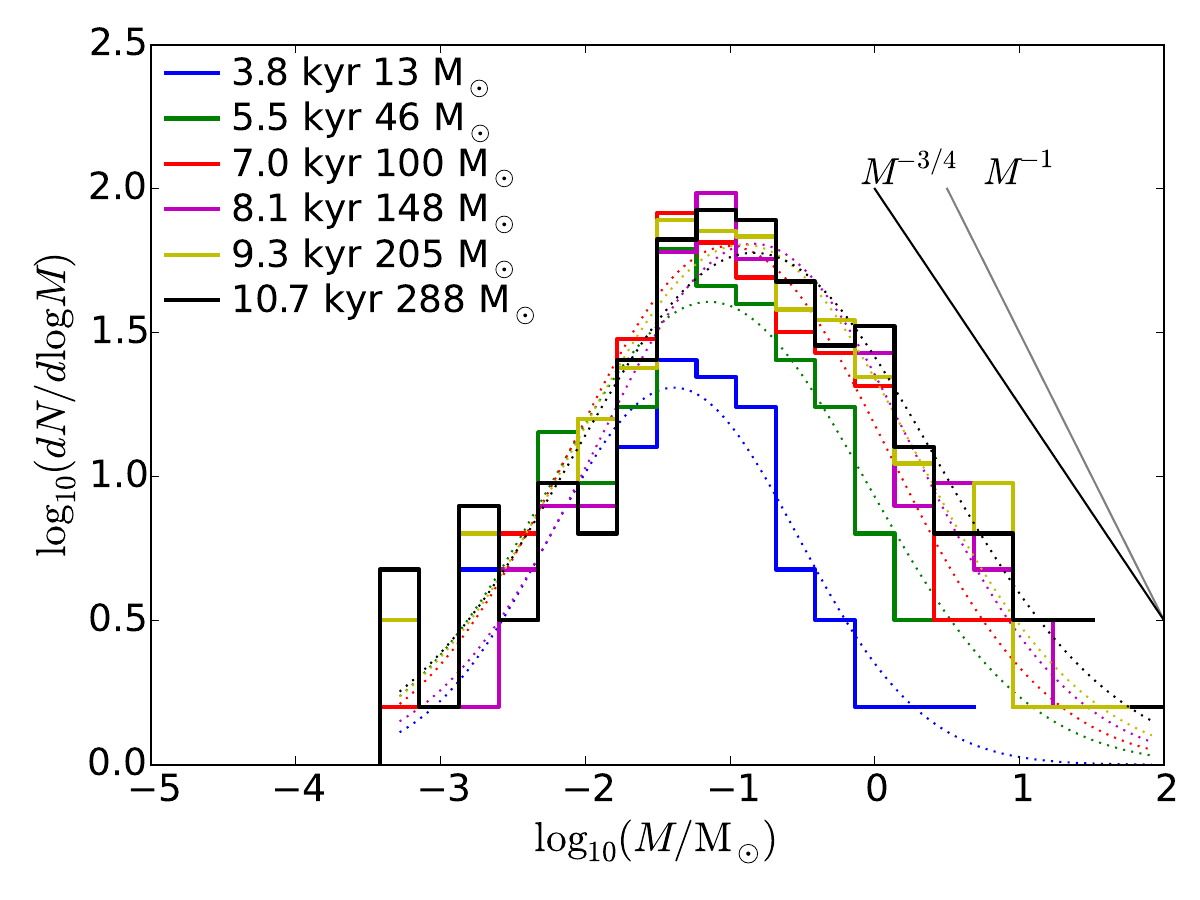}}
\put(1.2,4.){\includegraphics[width=5cm]{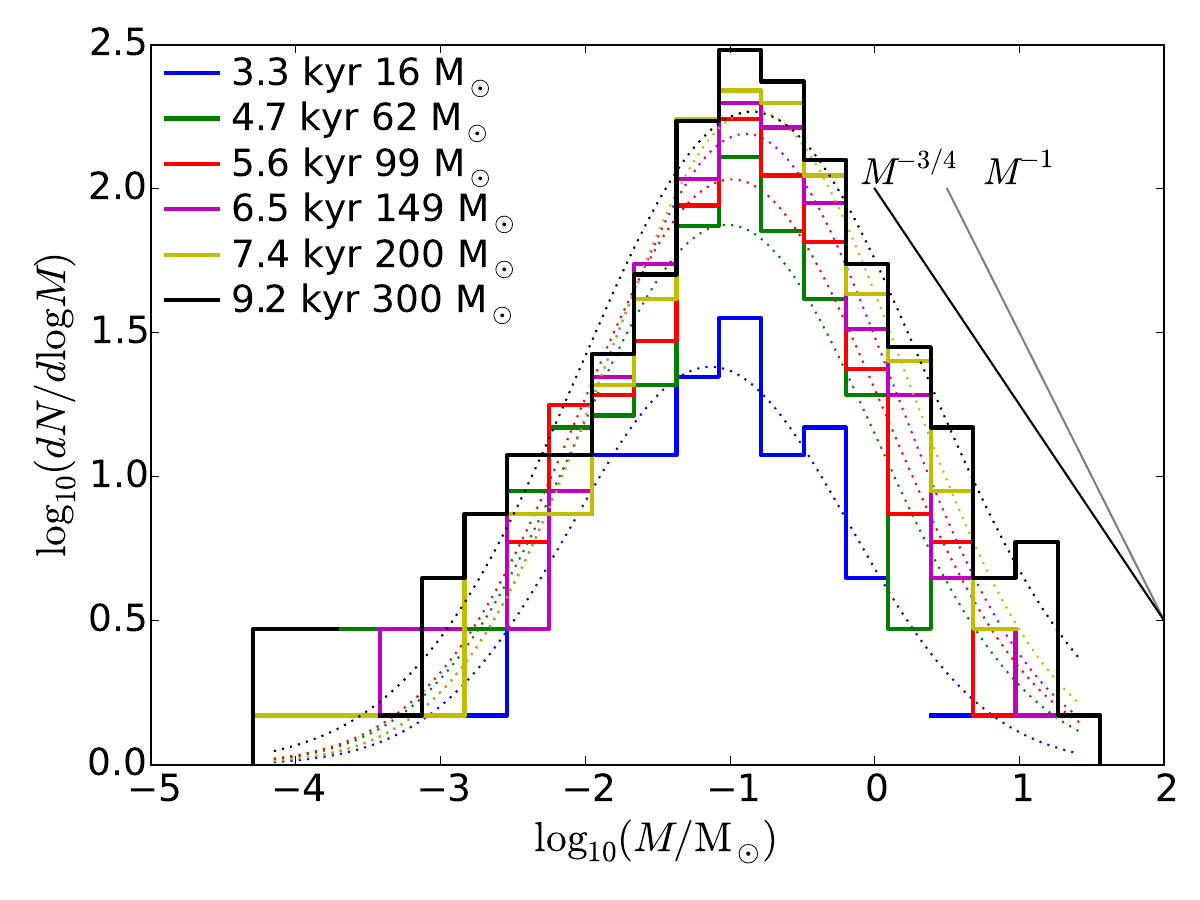}}
\put(-4,4.){\includegraphics[width=5cm]{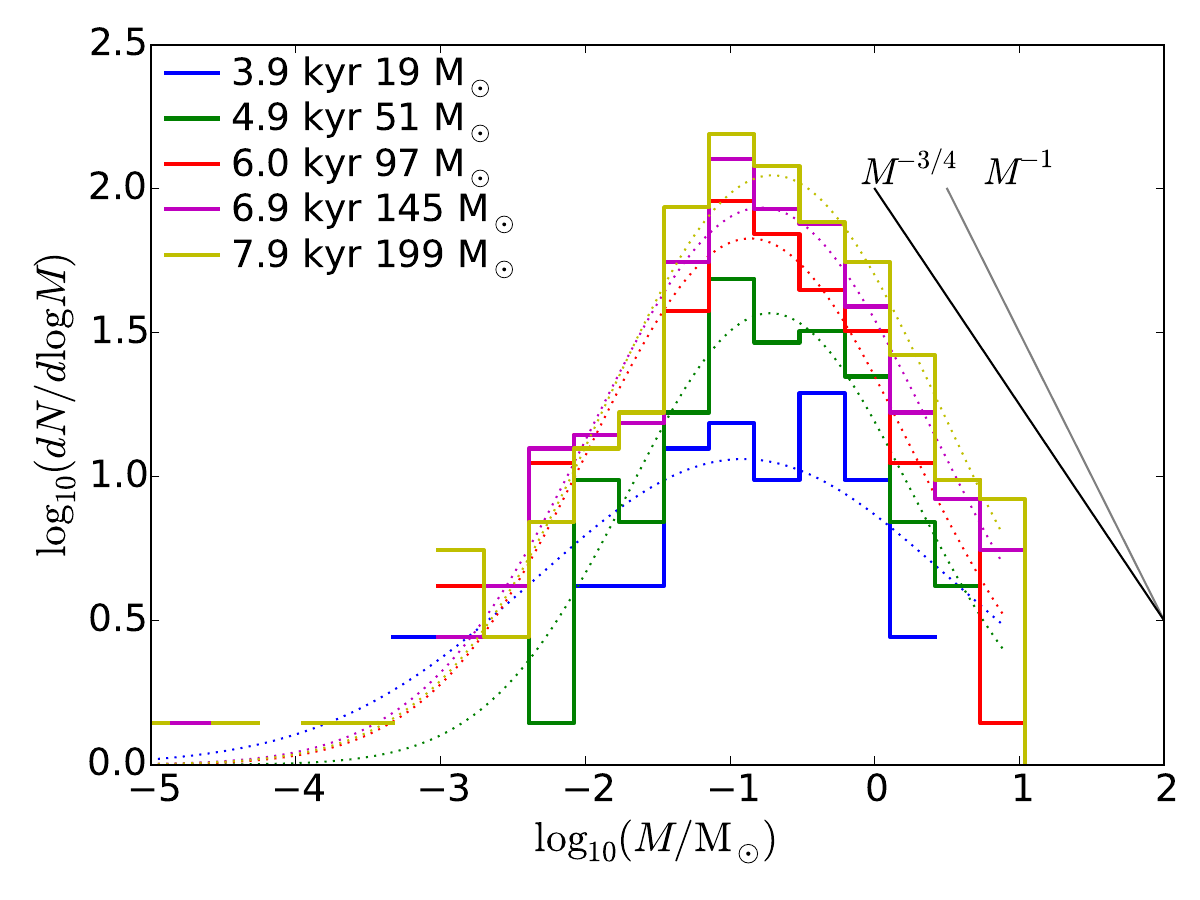}}
\put(3.6,7.7){C1+, 2 AU}
\put(8.6,7.7){C1, 4 AU}
\put(13.6,7.7){C1--, 8 AU}

\put(6.4,0){\includegraphics[width=5cm]{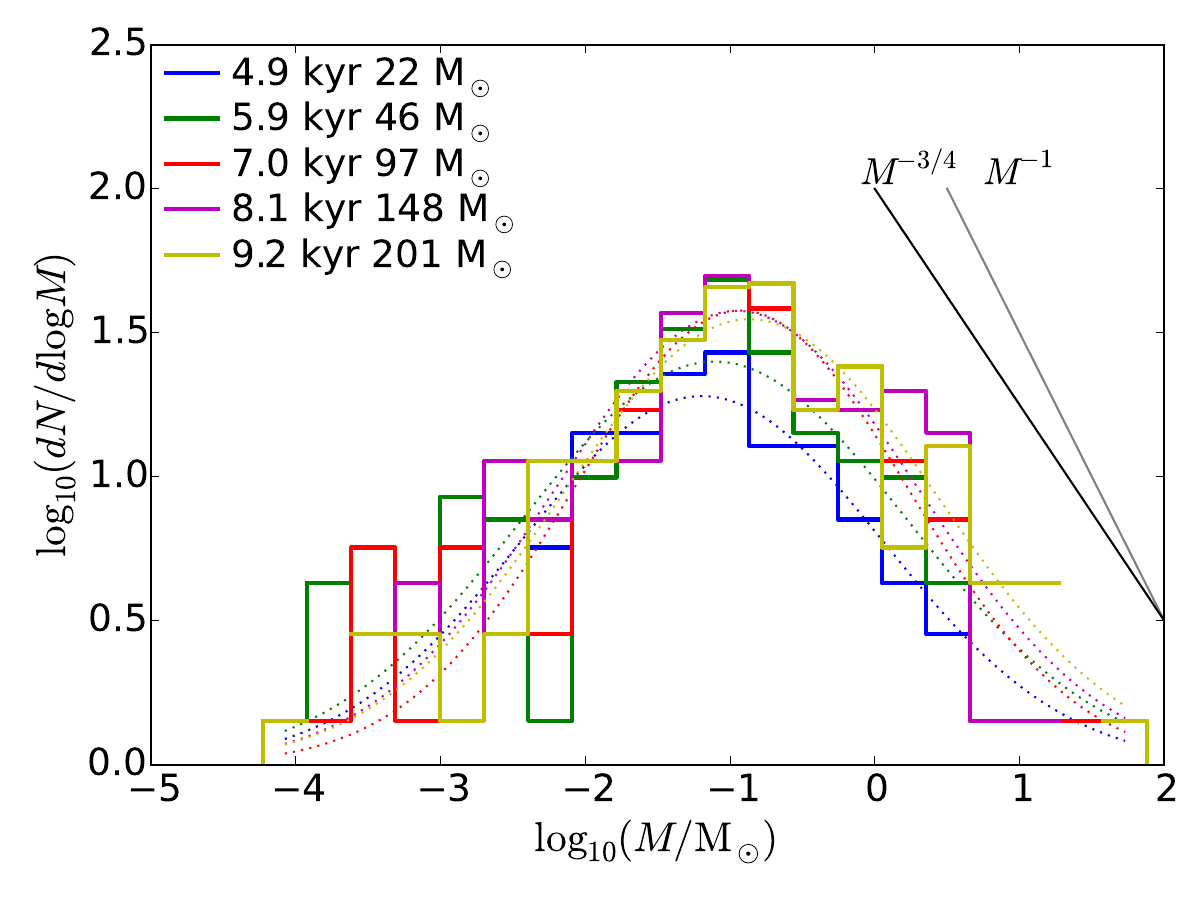}}
\put(13.6,3.7){C1-- --, 16 AU}

\put(-4,0){\includegraphics[width=5cm]{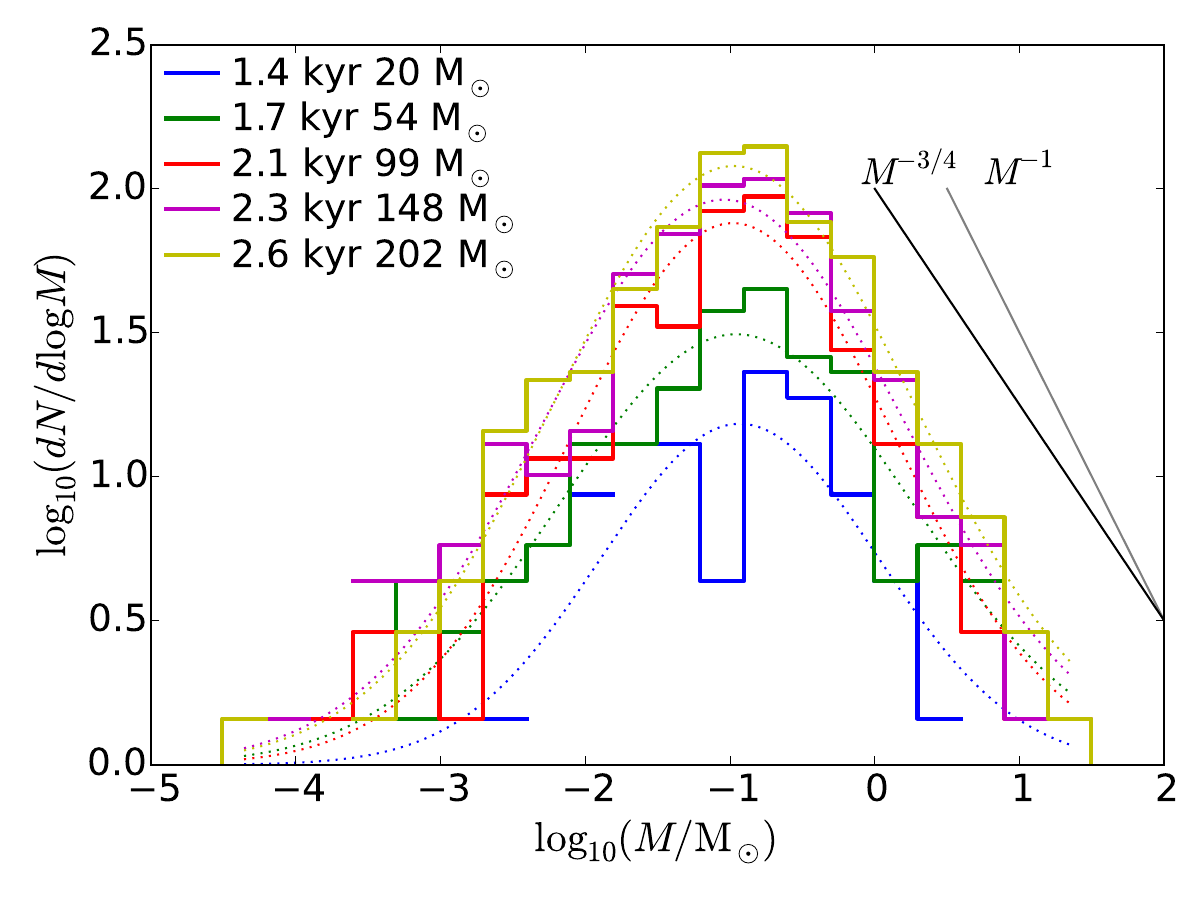}}
\put(3.6,3.7){D1, 2 AU}
\end{picture}
\caption{A series of numerical simulations of a 1000 M$_\odot$ clump from \citet{leeh2018a}.
    Runs A, B, C and D correspond to four initial densities (respectively 8$\times 10^4$, 9$\times 10^5$, 6$\times 10^7$, 5$\times 10^8$ cm$^{-3}$) and several 
    numerical resolutions. Run A presents a plateau with $\Gamma_{IMF} \simeq 0$, 
    Runs C and D show a peak at about 0.1 M$_\odot$ which is independent of numerical resolution and a powerlaw regime with $\Gamma_{IMF} \simeq -3/4-1$. Run B 
    is intermediate between runs A and C and presents the two regimes of $\Gamma_{IMF}$. Reproduced with the permission of $A\&A$. }
  \label{fig:leeh2018a}
\end{figure*}

%\begin{figure}
  %  \centering
   % \includegraphics[width=\textwidth]{Figures/Lee_h_2018.png}
   % \caption{  A series of numerical simulations of a 1000 M$_\odot$ clump from \citet{leeh2018a}.
   % Runs A, B, C and D correspond to four initial densities (respectively 8$\times 10^4$, 9$\times 10^5$, 6$\times 10^7$, 5$\times 10^8$ cm$^{-3}$) and several 
   % numerical resolutions. Run A presents a plateau with $\Gamma_{IMF} \simeq 0$, 
   % Runs C and D show a peak at about 0.1 M$_\odot$ which is independent of numerical resolution and a powerlaw regime with $\Gamma_{IMF} \simeq -3/4-1$. Run B 
   % is intermediate between runs A and C and presents the two regimes of $\Gamma_{IMF}$. Reproduced with the permission of $A\&A$. 
     %   }
    %\label{fig:leeh2018a}
%\end{figure}

\citet{leeh2018a} have explored the influence of initial conditions performing 
a series of calculations for which the gas remains isothermal below $\simeq 10^{10}$ cm$^{-3}$. 
In particular the initial density of their 1000 
M$\odot$ runs is varied by 4 orders of magnitude, keeping the ratio of kinetic over gravitational energy constant. This implies that in these runs, the thermal over gravitational energy is varied by a factor
25 and the Mach number by about a factor of 10, between 7 and 50. 
Figure~\ref{fig:leeh2018a} portrays the results. 
An important change of behaviour
arises for a Mach number of about 10 (which corresponds to run B in figure~\ref{fig:leeh2018a}). Above this value, all runs display a mass spectrum 
with $\Gamma _{IMF}$ of about  -3/4 to -1. However for a Mach number of $\simeq$7, the mass spectrum 
has been found to be extremely flat, $\Gamma _{IMF} \simeq 0$. Such {\it plateau} IMF, which have also been 
observed in other calculations \citep[e.g.][]{bonnell2006,jones2018}
\footnote{The value of $\Gamma _{IMF} \simeq 0$ is also commonly reported 
in simulations attempting to model primordial stars,  \citep[e.g.][]{klessen.glover:2023.pop3.sf} where 
a change with high Mach number is also reported, see for instance Figure 11 of \citet{chon:2022.imf.metallicity.cmb}.} are naturally explained by the 
gravo-turbulent theory applied to a collapsing cloud PDF
as presented in Section~\ref{density_pdf} and a quantitative comparison is presented
in figure~8 of \citet{leeh2018a}. The change of $\Gamma _{IMF}$ from  0 to $\simeq$-1 
comes from the fact that at low Mach number, the thermal term dominates the support of the mass reservoirs, whereas for high Mach  numbers the dominant support is due to turbulence. This latter, unlike the former, is scale dependent. 
Another important quantity which can be inferred from the simulations is the accretion rate onto the 
stars/sink particles. In Figure~5 of  \citet{leeh2018a} a comparison is made between 
the mass of the sinks and the accretion time measured in the simulations and predicted 
by the gravo-turbulent theory and a good agreement is found. 

Interestingly, \citet{BallesterosParedes15} have been interpreting their own results, that is to say 
their inferrence of $\Gamma _{IMF} \simeq -1$ as a consequence of  competitive accretion  
as described in section~\ref{compet_accret}. They stress that the basic assumption
of competitive accretion, namely $\dot{M} \propto M^\alpha$, $\alpha = 2$ is not observed in their simulations where $\alpha$ ranges from 0 to 0.8-1 (see their Figure 2). They however argue that 
competitive accretion occurs locally rather than globally and correcting for the local environment 
they then infer $\alpha \simeq 2$.  This result may suggest that both gravo-turbulent fragmentation 
and competitive accretion may well be operating at the same time. 
%For instance, gravo-turbulence 
%may be responsible of organising the cloud in accretion reservoirs.
%, that is to say gravitationally 
%bound regions out of which stars build their mass. 
%However, there must be a hierarchy of reservoir, 
%large ones contain smaller ones. Therefore 
Within a reservoir in which several stars have formed, 
there may possibly be, at least some, {\it competitive accretion}, between the stars. Such a 
scenario is compatible with the results reported by \citet{smith2008} where the usage of the SPH
Lagrangian scheme allows a full tracking of the mass. Identifying the accretion reservoir where the sink particles formed, they studied the correlation between the sink  and the reservoir masses. 
They concluded that a very good correlation between the two is found up to about 3-5 local freefall times
(as revealed by their Figure 10).
At later time, the correlation persists but becomes progressively less tight (see their Figure 9).

Generally speaking, determining the correspondance between a 
specific gas reservoir and a star is not an easy task. 
It has often been  considered that the prestellar cores (see Section~\ref{core})
should correspond to the reservoir out of which stars built their mass, but detailed studies have found that the correspondance between simply defined reservoirs, 
as considered in gravo-turbulent theories, 
and the mass effectively accreted by stars may be imperfect \citep{pelkonen2021} as 
 only a fraction of the material ends up in the sink.
One possibility is  that this is due to the complex velocity field not taken into the reservoir definition and to the fact that most of the core mass is located in their outskirt. 
Indeed  studies which have investigated the importance of geometry, the tidal field, and surface terms 
of the virial theorem, which are usually not accounted for, have found 
 that their contributions are significant \citep{2006MNRAS.372..443B,dib2007,hennebelle_2019}. 
It is however likely the case, that the statistics inferred for the idealised reservoirs, or say the cores,
may actually be close to the statistics that would be inferred for 
the actual reservoirs because both populations should indeed be closely related. This 
clearly requires future investigations. 
%\\

To summarize, the mass spectra in isothermal calculations of massive 
collapsing clumps tends to predict various asymptotic values for $\Gamma _{IMF}$, which depending on the initial conditions
can be equal to 0 or $\simeq$-0.8 to -1. 
Note that in observations, the range of mass for which $\Gamma_{IMF} \simeq 0$ is reported
appears to be quite narrow (see Figure~\ref{fig:alphaplot}). Therefore it is surprising that prominent {\it plateau} IMF, i.e.
$\Gamma_{IMF} \simeq 0$ ranging up to masses of several $M_\odot$, 
are found 
for initial conditions that do not appear to be unusual in the ISM (i.e. run A of Figure~\ref{fig:leeh2018a}). As discussed in Section~\ref{jets}, it is likely  that protostellar jets and outflows may be solving this apparent contradiction. 

Finally, we reiterate that the peak of the mass spectrum, i.e. the existence of a mass range where $\Gamma _{IMF} > 0$, obtained in isothermal calculations is likely numerical.
%Two main questions arise. What is the origin 
%of the peak of the IMF and what produces a value $\Gamma _{IMF} \simeq -1.3$? 

\subsubsection{Influence of a hard eos, $\Gamma_{ad} > 4/3$}
\label{gamma_sup43}
As discussed in Section~\ref{opacitylim}, at high density (i.e. 
above $10^{10}$ cm$^{-3}$) transport of dust emission becomes inefficient 
and the eos becomes adiabatic. Therefore several works have employed 
an eos that is isothermal at low densitities and adiabatic at high densities
\citep[e.g.][]{Bate03,bonnell2011,bate2012}. For instance, \citet{Bate03} 
employed $\Gamma _{ad}=1$ for $\rho < \rho_{ad} = 10^{-13}$ g cm$^{-3}$ and $\Gamma_{ad}=7/5$
for larger $\rho$. 

\citet{leeh2018b} have performed a series of calculations 
to investigate the role of the eos on the peak of the IMF. 
Both the values of $\Gamma_{ad}$ and $\rho_{ad}$ have been varied. 
Several important conclusions have been inferred. First of all, 
numerical convergence of the stellar mass spectrum can be established 
when such an eos is used as revealed by Figure~\ref{fig:leeh2018a} \footnote{Note that the convergence regarding the smallest objects that form may not have been reached for run A for which the best resolution is only of 9 AU}. 
This is a major difference with the 
case of an isothermal eos for instance. Second of all, the position of the peak of the 
IMF depends on both $\Gamma_{ad}$ and $\rho_{ad}$.
Since these values can be  associated to a specific value of $M_{FHSC}$ (see Equation~\ref{Mfhsc}), it is 
possible to investigate the dependence on the IMF peak onto 
$M_{FHSC}$ by varying the values of $\Gamma_{ad}$ and $\rho_{ad}$. \citet{leeh2018b} found that 
$M_{peak} \simeq 10 \times M_{FHSC}$ as revealed by their Figure 8. 
This finding has led to the model exposed in Section~\ref{FHSC}. In essence due to the hard eos, 
the gas piles up until the central mass equals   $M_{\rm FHSC}$,  providing a robust 
minimum mass for the stars that form. The factor of $\simeq 10$ between $M_{peak}$ and 
$M_{FHSC}$ is less straighforward to understand and according to the model 
of \citet{hennebelle_2019} is due to the tidal forces, which in the neighbourhood of
an existing self-gravitating object prevent further fragmentation and favor 
further accretion into the object. \citet{colman2019} conducted a systematic investigation 
of the influence of tidal forces in the neighbourhood of their sink 
particles at their birth, by computing the mass of gas within the regions where tidal 
forces prevent the formation of new fragments (see their figures~15 and 16). The mass distribution of 
these tidally protected gas reservoirs has been found to be very close to  
the  sink particle distribution seemingly suggesting that the two are very closely related. 
%\begin{figure}
%    \centering
%%%    \includegraphics[width=\textwidth]{Figures/Colman_2020.png}
%    \includegraphics[width=7cm]{Figures/Colman_2020.png}
%    \caption{  IMF formed within a collapse calculation (blue curve) of a massive clumps
%    from \citet{colman2019}.  The orange curve shows the mass function of the reservoirs  that are tidally protected, around new borned sink particles. The agreement between the orange and blue curves  suggests 
%    the importance of tidal forces in regulating the peak of the IMF. Reproduced with the permission of the authors.
%            }
%    \label{fig:colman2020}
%\end{figure}
%\\

The proposed origin for the peak of the IMF, which is due to $i)$ the 
mass of the first hydrostatic core and $ii)$ the stabilizing effect of the 
tidal forces that prevent close fragmentation, leading to further accretion 
onto the existing object, is an appealing scenario because it does not depend 
very sensitively on large scale initial conditions, as indeed verified by \citet{leeh2018a} and 
therefore it predicts a relatively universal peak for the IMF \footnote{In simulations for which a plateau IMF develops, 
the FHSC leads to a lower mass limit rather than a peak. }.   
Let us stress that since these calculations employed an eos, it is necessary for full self-consistency, 
to perform calculations that properly treat radiation as 
discussed in Section~\ref{radiation}.

\subsubsection{Influence of a soft eos, $\Gamma_{ad} < 4/3$}
\label{gamma_inf43}
As recalled in Section~\ref{jeans_mass_peak}, 
\citet{larson2005} proposed that the peak of the IMF 
could be determined by the Jeans mass of the density ($\simeq 10^5$ cm$^{-3}$)
at which the effective eos transitions from $\Gamma _{ad}\simeq 0.7$ to $\Gamma _{ad} \simeq 1.1$ due to the transition from molecular cooling to dust cooling. This picture has been investigated by \citet{Jappsen2005}
 \citep[see also][]{bonnell2006}, who have presented a comprehensive set of  numerical simulations with 
such an eos.  In particular, they varied the density $n_{ad}$ at which the transition between 
the two values of $\Gamma_{ad}$ occurs. They found that, indeed the median stellar mass
(expected to be close to the peak mass), is proportional to $\simeq n_{ad}^{-0.5}$ (see their Figure 6). 
However these results are somewhat inconclusive in light of the possible convergence issues demonstrated in the works discussed in \S~\ref{isotherm}; \citet{Jappsen2005} did survey a variety of mass resolutions, but did not explicitly demonstrate robustness to resolution in a controlled comparison. In fact, their Table 1 indicates that higher-resolution simulations do tend to produce more objects at lower final SFE, consistent with results in Section~\ref{isotherm}. This scenario should thus be revisited  in light of the findings of subsequent studies, particularly regarding the issue of numerical convergence.
%the primary effect of decreasing $n_{ad}$ is to reduce the smaller Jeans mass in the simulations. Higher
%numerical resolution would however likely leads to further fragmentation and to a peak that would occur
%at smaller masses. 

%Recently, \citet{lee2019} vary the effective eos 
%chosing an exponent equal to $\Gamma_{ad}=0.7$ or $\Gamma_{ad}=1.2$ 
%for $n < 10^{10}$ cm$^{-3}$ ($\Gamma_{ad} > 4/3$ was employed at  densities higher than 10$^{11}$ cm$^{-3}$) 
%and found that this has no influence on the sink mass distribution. This seemingly reinforces the 
%conclusion that what matters is 
% the transition from $\Gamma_{ad} < 4/3$ to $\Gamma_{ad} > 4/3$.

%\mike{I think we could omit the above paragraph, because YN's experiments did not really explore the Larson scenario with the transition in $\Gamma_{ad}$ to 1.1 at moderate density - by design $n_{\rm ad} =10^{10}$ was the only characteristic density/Jeans mass possible in her setup.}

\subsection{  The role of the magnetic field}
We now discuss the role magnetic field may have in shaping the IMF. We restrict the 
discussion to isothermal or barotropic runs and postpone the discussion of magnetic 
field in the presence of radiative transfer to the Section~\ref{radiation} or in the 
presence of jets to the Section~\ref{jets}.

\begin{figure}
    \centering
    \includegraphics[width=6cm]{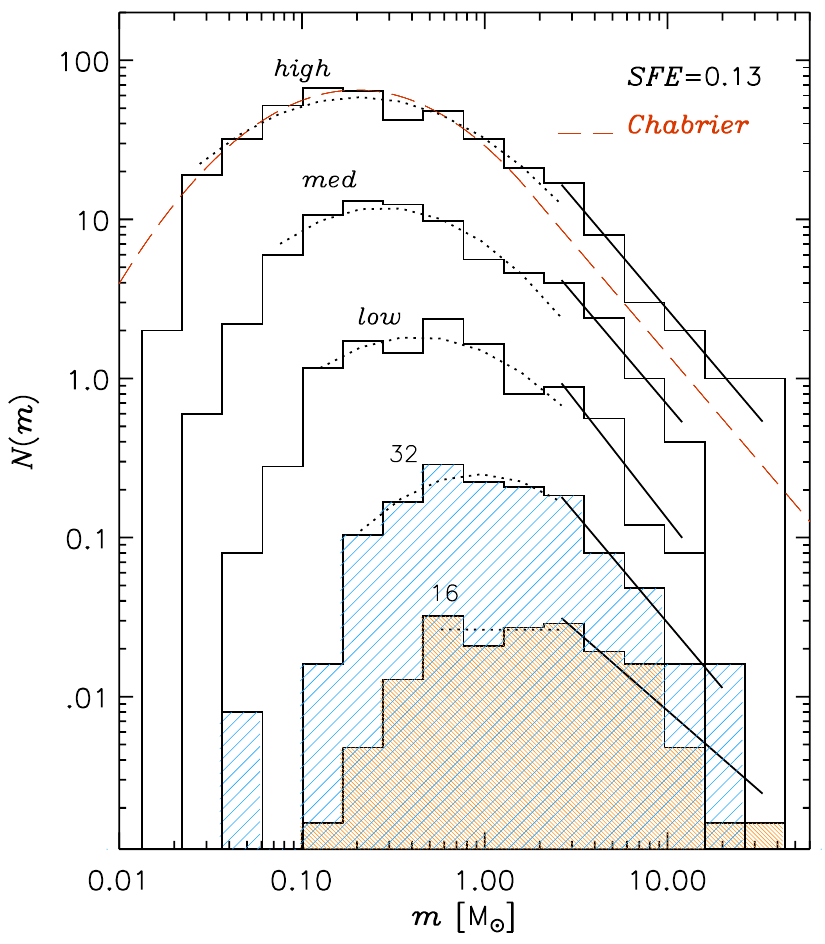}
    \caption{  IMF formed within magnetized isothermal collapse calculations of a turbulent box
    with different numerical resolution
    from \citet{haugbolle2018}. Note that the relative bin heights are shifted for visualization purposes. Reproduced with the permission of the authors.
            }
    \label{fig:haugbolle2018}
\end{figure}
\citet{haugbolle2018} have performed  isothermal MHD runs in a driven periodic 
box setup with side length $4\rm pc$ containing 3000 M$_\odot$ (if the inherently scale-free system is placed on scaling relations corresponding to $C_{\rm s} \sim 0.19 \rm km s^{-1}$). The Mach number is about 10 and the Alfv\'enic Mach number is equal to 4. 
In terms of magnetic field strength this would correspond to a field of about 
7$\mu G$ at a density of about 700 cm$^{-3}$.
The numerical resolution is systematically varied from 800 to 50 AU. 
At the highest resolution, a relatively good agreement with the 
Chabrier's IMF (see figure~\ref{fig:haugbolle2018}) is obtained except around 1-3 M$_\odot$. As the resolution 
is increased from 800 AU to 50 AU, the peak of the IMF is observed 
to shift to lower mass 
\citep[as it is also the case for instance in][]{guszejnov_isothermal_collapse}.
The question as to whether numerical convergence has been reached in 
these simulations is therefore critical. The authors show that 
as resolution increases the position of the peak keeps increasing 
but seems to decrease less rapidly with resolution, suggesting some degree of convergenge or robustness
to numerical resolution. These authors paid particular attention to the particulars of
the sink-particle implementation, finding that a fine-tuned accretion algorithm was essential to
avoid the formation of spurious sink particles.
If this result proves to be robust, it will be essential to understand why this is happening
whereas hydrodynamical simulations do not seem to show sign of convergence.
After all, ideal MHD does not introduce any particular scale in the problem that 
would explain why recursive fragmentation stops to occur at some scales or 
some density. On the other hand, the presence of magnetic fields could quite conceivably
alter the dynamics of filamentary collapse \citep[e.g.][]{lee_H_C_2017}.
\citet{haugbolle2018} also perfomed runs with 4 different
total masses, corresponding to 4 virial parameters. They find that the 
shape of the IMF shifts to lower mass as the virial parameter decreases.
This is qualitatively in good agreement with the gravo-turbulent theories 
since the mean Jeans mass is lower in the simulations that contain more mass. 

\citet{lee2019} have carried out MHD barotropic simulations 
of a collapsing 1000 M$_\odot$ virialised cloud with a maximum spatial resolution 
of about 4 AU. Two 
different initial densities have been explored, respectivelly 
8 10$^4$ (diffuse) and 6 10$^7$ cm$^{-3}$ (dense). The initial Alfv\'enic Mach number 
has been varied from 11 to $\simeq$0.9 for the dense clump whereas for the 
diffuse one it has ben chosen to 2.75. For the dense clump initial condition, the 
IMF presents a marked peak at about $\simeq$0.1 M$_\odot$ whereas at higher mass 
the IMF exhibits a powerlaw with an exponent $\Gamma_{IMF} \simeq -1$. This is the case 
for the Alfv\'enic Mach numbers, which have been explored. Some possible variations 
for the lower mass objects are possibly observed but altogether 
for the dense initial conditions, the magnetic field 
appears to have a modest influence on the IMF. The
physical reason may be that the magnetic field strength is observed to converge at 
densities larger than about 10$^9$ cm$^{-3}$ toward values which 
are nearly independent of the initial magnetisation. For the diffuse clump, the 
IMF, even in the hydrodynamical case, is such that $\Gamma _{IMF} \simeq 0$, which 
as explained in Section~\ref{density_pdf} arises when the thermal support 
dominates at the scale of mass reservoirs. In this context, magnetic field does 
not significantly alter the mass spectrum since the Alfv\'en speed, as the sound speed, 
is roughly scale independent
(see Section~\ref{mag}). Interestingly however, there are less
low mass objects in the MHD than in the hydrodynamical runs indicating that 
in this configuration, magnetic field reduces fragmentation.

\citet{guszejnov2020} have performed a series of scale-free, magnetized
isothermal GMC simulations with the Meshless Finite Mass method, surveying a wide range of Mach numbers ($\mathcal{M} \sim 5-50$), corresponding to masses ranging from $\sim 10^3-10^6$ M$_\odot$ for clouds on Larson's relations. 
They also surveyed a range of $\mu$ values from more than 
10 to about 1 with 4 being the standard value, and a range of virial parameters $\alpha \sim 0.5-4$.  Several trends
are inferred. For large masses they get $\Gamma_{IMF} \simeq -1$ or steeper whereas  at 
low masses, $\Gamma_{IMF} \simeq 0$. They surveyed numerical resolution systematically and did not claim strong convergence of the overall mass spectrum, as finer mass resolutions did always result in more low-mass sinks. But they did find that {\it mass-weighted} quantiles of the IMF could be remarkably well-converged over many decades in mass resolution, in contrast to their previous isothermal hydrodynamics results which did not converge in any sense (c.f. Section~\ref{isotherm}). While also finding an IMF from isothermal MHD with some degree of robustness to resolution like \citet{haugbolle2018}, they did {\it not} concur that this formula is sufficient to explain the IMF: their models, when scaled to real clouds, predicted stellar mass quantiles an order-of-magnitude too large. They therefore concluded that only additional physics capable of {\it reducing} stellar masses -- such as protostellar outflows -- are also necessary in these conditions (see Section~\ref{jets}).
%A reasonable transition between the two regimes, 
%which do not simultaneously exist in all  simulations, is given by the 
%mass weighted median mass. This latter is not universal and varies from runs to 
%runs but for the chosen initial conditions, aiming at being realistic, the transition 
%is found to appear at masses larger than 10 M$_\odot$. This does not appear to be 
%in good agreement with observations for which a value of about 1 M$_\odot$ is 
%observed. 
%\citet{guszejnov2020} therefore conclude 
%that more physics is requested to explain the IMF and in \citet{guszejnov2021} it is proposed that the protostellar jets are playing an important role in this 
%respect (see Section~\ref{jets}). 

\subsection{How turbulence and numerical setup influence the IMF}
As mentionned above, two main numerical setups have been traditionally used to compute the IMF, namely 
isolated turbulent collapsing clumps \citep[e.g.][]{Bate03,BallesterosParedes15,leeh2018a} and 
periodic boxes in which turbulence is driven \citep[e.g.][]{haugbolle2018,mathew2023}.
In the former configuration, turbulence is initially inprinted and is not further driven. 
\citet{leeh2018a} have varied the initial virial  parameter (their figure 7) between 1.5 and 0.1 and found 
that unless its value is smaller than $\simeq$0.3,  the influence of its variation remains limited. In particular, 
$\Gamma_{IMF}$ is not significantly affected and remain close to -0.8 to -1. 
The values of $\Gamma_{IMF}$ reported in the periodic boxes tend to be steaper and closer to -1.3.
For instance, figure~\ref{fig:haugbolle2018} \citep{haugbolle2018} reveals that the high mass part of the stellar distribution, 
between $\simeq 2$ and 10 M$_\odot$ appears to be compatible with $\Gamma_{IMF} \simeq -1.3$. 
Similar trends may be found in the work of \citet{mathew2023}, where turbulence driving is also performed and values of $\Gamma_{IMF} \simeq -1.3$ are also seemingly obtained (their figure 7). 
The difference between the two configurations clearly appears in figure 16 of \citet{guszejnov2022} 
where periodic boxes with and without turbulent driving, as well as collapsing clouds have been 
investigated. All cases have $\Gamma_{IMF} \simeq -1$ except the turbulent driven simulation
for which $\Gamma_{IMF} \simeq -1.3$.

The reason of these differences is presently not clear. One possibility is  
that they are a consequence of the density PDF as stressed in Figure~\ref{fig:recap}. 
Likely enough strong driving limits the formation of massive collapsing clumps in which a 
significant fraction of the mass follows a collapse PDF (Equation~\ref{pdf_grav}) rather
than a lognormal PDF (Equation~\ref{Pr0}).

\subsection{Radiative transfer and protostellar heating}
\label{radiation}
As stressed in Section~\ref{protostellar}, the radiation that emanates from the protostars leads to substantial heating 
of the collapsing envelope and it must be accounted for to properly handle 
the formation of new fragments. 
Moreover, the FHSC itself forms because the dust becomes opaque to its own radiation
(see Section~\ref{opacitylim}) and a self-consistent treatment at scale below few AU requires to 
perform radiative transfer. 
Several calculations have been performed to study the IMF with self-consistent radiative transfer.
\citet{urban2010} have performed SPH calculations of a 670 $M_\odot$ clumps. They introduce sink particles at a density of about
10$^8$ cm$^{-3}$, and add radiative feedback onto the sink particles. They found that calculations which treat radiative feedback 
are markedly different from
the isothermal ones. The number of stars is 
reduced by more than an order of magnitude and
the stars are consistently more massive when radiation is considered.

\citet{bate2009} and \citet{bate2012} carried out high resolution  SPH calculations introducing the sink particles at very high density, i.e.
$n > 10^{19}$ cm$^{-3}$. %Therefore the simulations self-consistently include the optically thick regime at densities  $n > 10^{10}$ cm$^{-3}$. 
In the simulations no stellar feedback is further added
onto the sink particles.
 The resulting IMF  presents a peak at about 0.3 $M_\odot$ and a powerlaw mass spectrum
at high mass.

\citet{krumholz2012} carried out AMR calculations with a numerical 
resolution of 20-40 AU. The sink particles are
created when the Jeans length is not resolved any more  and only objects more massive than 0.05 $M_\odot$
are being considered. Smaller objects are allowed to merge. Both intrinsic and accretion luminosity are  added to the sinks as well as fast winds, which are also introduced in some of the calculations.  
The resulting  mass spectra are nearly flat, that is to say $\Gamma_{IMF} \simeq 0$, when winds are not considered whereas
when winds are added, the radiation can escape along the cavities opened by the winds.
 In these circumstances, the mass spectra present a peak around 0.3 $M _\odot$
and a powerlaw, $dN/d \log M \propto M^{\Gamma _{IMF}}$, with $\Gamma_{IMF} \simeq -0.5$ to -1. 

\citet{hennebelle2020} conducted hydrodynamical AMR simulations of 1000 $M_\odot$ clumps
with a spatial resolution of 4 to 1 AU.
 Stellar and accretion luminosities are both included  and various efficiencies, $f_{acc}$
 (see Equation~\ref{Lacc}),
  ranging from 0 to 50$\%$ have been considered.
Two sets of initial conditions corresponding to an initial radius of about 0.1 and 0.4 pc
have been explored. For the most compact clumps and when $f_{acc}$ is high, a flat mass
 spectrum  is formed. 
 Otherwise in all runs mass spectra that present a peak around 0.3-0.5 M$_\odot$ and a power law
 at higher masses appear.
 This is the case even when no radiative feedback applies, i.e. when $f_{acc}=0$. 
 This is also the case when radiation is entirely neglected and that a barotropic equation
 of state is used. 
 High efficiency radiative feedback runs however tend to present a broader distribution, both at the low mass and high mass end.
 The most massive stars are up to two to three times more massive than in the barotropic and low-feedback efficiency runs.
\citet{hennebelle2022} have performed MHD runs with a spatial resolution down 
to 1 AU and which also take into account ambipolar diffusion. As previous authors 
\citep{commercon2011b,Myers_2013_ORION_radiation_IMF}, it has been found that magnetic field and 
radiation reduce fragmentation, particularly when they are both included. As a result, 
strong magnetic field and high radiation efficiency models, present top-heavier IMF. A comparison 
with the analytical model discussed in Section~\ref{density_pdf} is presented and it has been concluded that 
the various IMF inferred can be satisfactorily reproduced (down to objects of masses
0.1-0.2 M$_\odot$) once the mean sound speed 
and the mean Alfv\'en speed are used in the analytical model. \\

As a general conclusion, there is a consensus amongst the various studies that radiative feedback 
is playing a significant role regarding the stellar mass spectra that form in relatively dense
and massive clouds, likely cluster progenitors. Radiative feedback tends to reduce the number of objects
promoting the formation of massive stars and in some circumstances (compact and/or highly magnetized clouds)
to even top-heavy IMF. In the context of the gravo-turbulent model discussed in Section~\ref{density_pdf} and the thermally dominated regime where $\Gamma_{IMF}=0$ (see figure~\ref{fig:recap}), clearly 
the presence of intense radiative stellar feedback would favor this mode and 
may explain why a flat IMF is often found in simulations that include radiative stellar feedback.

However the origin of the peak of the IMF itself remains debated.
In the works where the FHSC is resolved, the peak seems to be a consequence of the dust opacity and the 
existence of the FHSC rather than to the heating of the protostars. 
This is the case in the simulations presented in \citet{leeh2018b}, \citet{colman2019}, \citet{hennebelle2020} and \citet{hennebelle2022}.
This is also likely the case in the studies presented by \citet{bate2009}, \citet{bate2012} and 
\citet{bate2019} where no stellar feedback is being explicitly added and where the FHSC is very 
nicely described by introducing the sink particles only when the second collapse
has started.
It is nevertheless important to caution here that whereas numerical convergence has been 
convincingly established when an hard eos is being used, 
establishing convergence when radiative transfer is properly accounted for 
appears to be far more difficult \citep{hennebelle2020} and remains a challenge. 

In the work where the FHSC is not resolved, it is claimed that the origin
of the peak has to be attributed to stellar radiative feedback (see Section~\ref{protoheat}). Detailed analysis 
performed by \citet{krumholz2016,Cunningham_2018_feedback} 
have found  agreements with the picture proposed in \citet{krumholz2011} where radiation sets the peak. We stress here that the issues regarding numerical convergence 
and sink algorithms may be particularly critical and should be thoroughly investigated.
Finally we note that the simulations presented in \citet{hennebelle2020}, do not support the validity 
of the radiative feedback setting the IMF's peak 
since in these simulations, the peak is found to remain at the same mass even when the accretion luminosity 
vanishes.

\subsection{The influence of metallicity}
As discussed in Section~\ref{cooling}, the abundance of heavy elements and 
of dust (and their {\it relative} ratios, \citealt{sharda:2023.co.ratio}) are key quantities regarding the thermal balance of the gas. Since the gas 
metallicity can vary significantly,  typically from $Z_d=0.01$ to $Z_d=3$ in observed galaxies,
investigating the dependence of the IMF on this parameter is important.

\begin{figure}
    \centering
    \includegraphics[width=9cm]{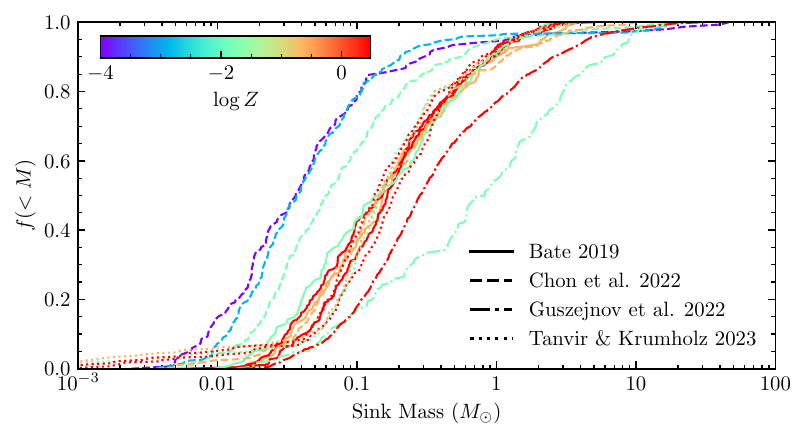}
    \caption{ % IMF formed within radiative collapse calculations of a massive clump
%    with various metallicities
%    from \citet{bate2019}. 
    Cumulative sink-particle mass distributions reported by multiple numerical parameter studies surveying metallicity: \citet{bate2019}, \citet{chon:2022.imf.metallicity.cmb}, \citet{guszejnov2022}, and \citet{tanvir2023}. Line styles plot the different studies, and color encodes the Solar-scaled metallicity as indicated in the colorbar. Note the wide range of results at $Z _d=0.01$ (green).
%\mike{got the data from multiple studies and made a plot, thought maybe we could show a more complete picture than just Bate 19.}
            }
    \label{fig:metallicity}
\end{figure}

In Figure~\ref{fig:metallicity} we plot the results from a selection of different recent numerical IMF studies that surveyed metallicity. In $z\sim 0$ conditions, some initial metallicity studies were performed by 
\citet{myers11a} and \citet{bate2014}, studying the collapse and fragmentation of 
several $100 M_\odot$ for a wide range 
of metallicity. Both works found that the IMF was remarkably insensitive to metallicity, 
a result later reproduced by \citep{bate2019} and \citet{tanvir2023} with the same respective codes but with 
various improvements and additional physics. This agreement is despite large differences in the respective methodologies:
with \citet{bate2014} neglecting radiation from the protostellar surface but resolving $\lesssim 1 \rm AU$ scales and the FHSC, and \citet{myers11a} accounting for protostellar radiation fully but reaching a maximum resolution of $7 \rm AU$.

This insensitivity to metallicity is not always found: \citet{guszejnov2022} found that reducing metallicity increased the mean stellar mass, due to the resulting increased temperature (\S\ref{cooling}). Here the initial conditions and cloud density could be important: their fiducial cloud had a mean density of $\sim 10^{3}\,\rm cm^{-3}$, where the temperature structure was determined largely by processes {\it other} than dust cooling, unlike the denser ($\gtrsim 10^5 \rm\,cm^{-3}$) clouds simulated by the above authors. \citet{chon:2022.imf.metallicity.cmb} found an opposite trend: in their simulations reducing metallicity {\it decreased} the typical stellar mass, while also enabling the formation of a few very-massive stars. Unlike the other works mentioned in this section, they did not account for any radiation other than the CMB, so by neglecting protostellar radiation it is possible that they overestimated disk fragmentation (c.f. \citealt{offner2009}).

There is some interplay between the effect of metallicity variations and the background radiation field, e.g. the CMB. Surving $z=0-20$, \citet{chon:2022.imf.metallicity.cmb} reported generally increasing stellar mass with increasing $z$. \citep{bate:2023.hiz.imf} found that in a $z=5$ CMB their results were essentially the same as those at $z=0$ in \citet{bate2019}, except for their $Z_{\rm d}=1$ $z=5$ model, which was shifted upward by a factor of $\sim 3$ in mass. This result is in reasonable agreement with the scaling with $Z_{\rm d}$ and the radiation field $u_{\rm rad} \propto (1+z)^4$ predicted by Equation~\ref{Mfhsc}, assuming that the FHSC mass is setting the overall IMF mass scale. 
%Both works concluded that the resulting IMF formed 
%in their calculations, are almost identical
%when they modify the metallicity , despite significant differences 
%between the two types of simulations. Indeed 
%\citet{myers11a}  have a resolution of about 7 AU 
%and explicitly treat the radiative feedback. On the other hand 
%\citet{bate2014} resolved much smaller scales
%(the sink particles have a radius of about 1 AU) but do not add radiative energy 
%to account for feedback explicitly.

%Moreover \citet{bate2014} also employed 
%an equation of state to account for the H$_2$ excitation and dissociation is, explicitly describing the isothermal to adiabatic transition and the FHSC, so 

%We therefore believe that the most likely explanation of why 
% little variations of the IMF with metallicity is obtained in 
% \citet{bate2014} is due to the weak dependence of
% $M_{\rm FHSC}$ (see Equation \ref{Mfhsc}). On the other hand,  
% \citet{myers11a} present an argument to show that the 
% temperature around the vicinity of a radiating star, has 
% a weak dependence on the metallity (similar to Section~\ref{protoheat}).

%\citep{batebonell2005} % varying the opacity limit with metallicity

 % perfectly-controlled experiment varying only dust opacity and not temperature
%\mike{
%The density of the cloud may be an important factor for the question of metallicity effects on the IMF, because different cooling and heating processes dominate
%at different densities (e.g. Section~\ref{cooling}). \citet{bate2019} and \citet{tanvir2023} found weak metallicity effects in relatively dense ($\gtrsim 10^5\,\mathrm{cm}^{-3}$) clouds, where the balance of radiative dust emission and absorption determines the temperature structure in a way that is only weakly sensitive to dust abundance. In more-diffuse clouds the effect of metallicity can be more pronounced because the temperature structure will be determined by the balance of fine-structure cooling, cosmic rays, and FUV radiation, which will generally vary with metallicity. \citet{guszejnov2022} simulated a $0.01Z_\odot$ cloud with mean density $\sim 10^3 \,\rm cm^{-3}$, and found that it did produce a measurably more top-heavy IMF than the fiducial case, with the mean stellar mass scaling in a manner roughly consistent with the temperature scaling of a Jeans-type fragment mass, $\propto T^{3/2}$.
%}

The effect of metallicity on the formation of massive stars and the upper tail or cutoff of the IMF is nuanced, due to several competing effects. As mentioned above, the higher temperatures expected at low metallicity should suppress fragmentation in at least some cases. Lower-$Z$ stars have weaker stellar winds, and dust-poor environments are less subject to the effects of radiation pressure which could regulate massive stellar growth and disrupt the natal clumps. All of these factors would facilitate the formation of very-massive stars. However, the absence of significant metal line cooling in metal-poor HII regions can make them a factor of $\sim 2$ warmer ($\sim 20,000 \rm K$) than at Solar metallicity. This greatly reduces the ionizing flux, and hence stellar mass, required to create an expanding HII region that can disrupt a star-forming GMC, with $\mathcal{Q} \propto T^{-2.85}$ \citep{grudic:does.god.play.dice}. This is why the $0.01 Z_\odot$ cloud in \citet{guszejnov2022} had a significantly reduced star formation efficiency and maximum stellar mass compared to the fiducial case, $\sim 25 M_\odot$ versus $\sim 45 M_\odot$. Which of the competing feedback scalings sets the upper IMF cutoff is likely to depend on initial conditions, because photoionizing feedback becomes irrelevant compared to radiation pressure at high densities \citep{krumholz.matzner:2009.hii.regions}.

\subsection{The influence of protostellar jets}
\label{jets}

\begin{figure}
    \centering
    \includegraphics[width=\textwidth]{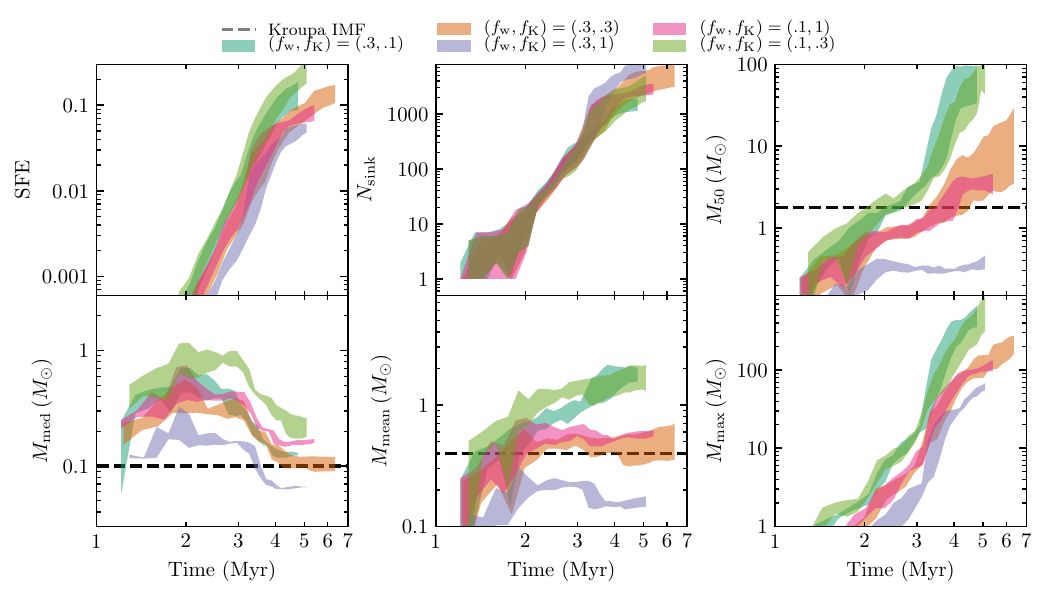}
    \caption{Result of a parameter study of $2\times 10^4 M_\odot$ GMC simulations with protostellar jet feedback, varying only the feedback parameters $f_{\rm w}$ and $f_{\rm K}$ \citep{guszejnov2021}. Plotted as a function of time are the fraction of gas mass converted to stars (SFE), the number of sink particles ($N_{\rm sink}$), the mass-weighted median stellar mass $M_{\rm 50}$, median and mean sink mass $M_{\rm med}$ and $M_{\rm mean}$, and the maximum sink mass $M_{\rm max}$. Note the large overlap of $N_{\rm sink}$ versus time for all models; the strength of jet feedback does not fundamentally affect fragmentation physics. Also note the correspondance between models with similar values of the jet momentum loading $f_{\rm w} f_{\rm K}$, indicating that jet feedback operates in a momentum-driven regime.}
    \label{fig:guszejnov.jets.imf}
\end{figure}

The influence of protostellar jets on the evolution of clusters and more 
specifically on the IMF has been studied by various groups, and several effects have been 
claimed. First, jets certainly contribute to inject kinetic energy into the collapsing
clumps and therefore constitute a significant source of turbulence although the exact 
amount remains a bit uncertain due to the inaccuracies with which jets properties are
hampered \citep[e.g.][]{cunningham2009,guszejnov2021} as illustrated in figure~\ref{fig:guszejnov.jets.imf}. 

Second, jets limit the 
star formation rate within star forming regions and lead to values that are in much better 
agreement with observations than simulations without jets \citep[e.g.][]{Wang10,federrath_2015_inefficient_sf,verliat2022}.

Third, jets appear to shift the peak of the IMF and the mass of the smallest objects which form 
compared to when they are not included by a factor of several. This effect has been 
consistently reported in several studies \citep{li2010,guszejnov2021,mathew2021}. It is generally 
admitted that the reason is due to the generation of smaller mass gravitationally unstable density
fluctuations induced by the jets. This could also be due to less efficient accretion 
from the individual mass reservoir, say the cores, due to the jet feedback. The two 
explanations are not exclusive from each other and may operate simultaneously. Indeed the effect 
observed on figure 6 of \citet{guszejnov2021} indicate a substancial shift in the IMF's peak by almost a factor 
of 10. Note that this may solve a possibly important problem. As discussed previously, several 
authors \citep{leeh2018a,jones2018,guszejnov2020} found that if the thermal support of the star forming 
clump is too high, a plateau IMF, i.e. $\Gamma_{IMF} \simeq 0$, tends to develop. However the clump initial
density values at which this happens are not particularly low compared to observations and therefore, jets 
likely constitute a solution by regulating kinetic energy in collapsing clumps. 
Let us stress a possible caveat. In most of works which so far have been investigating the impact of jets, the FHSC is actually not resolved and therefore fragmentation
is limited by numerical resolution. Very recently \citet{lebreuilly2023} have performed the first simulations with jets and about 1 AU resolution. They found 
that whereas jets do not affect the number of objects with mass below or comparable to a few times 
the mass of the FHSC, they indeed reduce the mean object mass in the simulation. 

Finally, jets may also be playing an important role regarding stellar radiation feedback since they open up cavities along 
 which photons may escape but also because they reduce the accretion rate onto the stars, both effects concour to diminish the effective radiation feedback that the collapsing clump is 
experiencing \citep[e.g.][]{hansen_lowmass_sf_feedback}. These effects, which add up to the ones discussed above,  are clearly present in the simulations presented for instance in figure 8 of \citet{krumholz2012}
where the simulation without wind exhibits a clear plateau ($\Gamma_{IMF}=0$) whereas the ones with wind 
present a clear powerlaw with $\Gamma_{IMF} \gtrsim -1$.

\subsection{Massive stellar feedback and cloud disruption}

It is generally agreed that winds and radiation from massive stars are the key processes regulating star formation on the scale of giant molecular clouds and ultimately dispersing them (Section~\ref{sec:winds} and \ref{sec:radiation}). This directly affects the gas supply for accretion and fragmentation, and the dynamics of star cluster assembly and dispersal, all of which affect the IMF. The mass spectra predicted by calculations neglecting these processes are literally inconclusive, because the stellar masses must be extracted at some arbitrary time or final SFE, and further evolution of the mass spectrum cannot be ruled out. The usual choice of a constant $\sim 5-20\%$ SFE is probably accurate in detail, because cloud-scale calculations with feedback unanimously find that SFE varies with cloud/clump properties, most notably surface density \citep{chevance:2023.pp7.gmcs.review}.

Massive stellar winds/radiation and cloud disruption have only recently begun to be accounted for in star cluster formation simulations intended to model the IMF. \citet{Gavagnin_2017_SF_feedback} and \citet{chong2019} both performed RHD star cluster formation simulations with sink particles emitting ionizing radiation and no other feedback, and both found a flat and/or top-heavy mass spectrum incompatible with the observed IMF, suggesting that photoionization alone is not sufficient to regulate the entire IMF, although it can regulate the high-mass regime. But in the past year, the first studies accounting for winds and/or radiation in concert with protostellar outflows have emerged \citep{verliat2022,grudic2022}. The STARFORGE studies have demonstrated the complementary roles of different feedback processes in setting the IMF: protostellar jets have a dominant effect on the $\sim 0.1-1M_\odot$ range, but cannot disrupt $\gtrsim 10^4 M_\odot$ clouds or regulate runaway accretion of the most massive stars \citep{guszejnov2021}. Accretion onto the most-massive stars continues via filamentary, $\sim 1-10 \rm pc$ scale flows until halted by feedback; this tends to coincide with the disruption of the cloud as a whole \citep{guszejnov2022}. Until the cloud is disrupted and star formation is quenched, the stellar mass spectrum evolves continuously. This mode of self-regulated massive star formation inevitably results in a maximum stellar mass that depends on the cloud bulk properties and composition \citep{grudic:does.god.play.dice}. This sensitivity to initial conditions highlights the importance of characterizing the initial conditions for star formation further, and exploring simulation setups that are more realistic than the usual idealized periodic box or isolated clump setups.

\begin{summary}[SUMMARY POINTS]
\begin{enumerate}
\item Tremendous progresses have been accomplished during the last two decades 
in our understanding of the IMF. It is now possible to self-consistently treat 
most of the physical processes thought to play a role in setting up the IMF and 
to better cover the necessary large range of spatial scales, from pc to AU, though further lines
of necessary improvements remain. 
\item Self-gravity, turbulence, and feedback are responsible 
for setting the stellar mass spectrum for masses above few M$_\odot$.
Gravo-turbulent theories as well as competitive accretion make
specific predictions for the powerlaw exponent of the IMF, $\Gamma_{IMF}$,
which depending on the physical conditions is predicted to be equal to 0, -3/4, -1 or 
$\simeq -1.3$. Whereas these different regimes may be relatively 
universal, the stellar masses for which they apply depend on 
large scale environments such as magnetic field, Mach number and temperature.
$\Gamma_{IMF}=0$ is predicted to develop in collapsing clouds in which thermal and magnetic supports 
dominate over turbulence.
\item Protostellar jets, by injecting kinetic energy in star forming clumps, may 
limit powerlaw exponant $\Gamma_{IMF} =0$ and favor $\Gamma_{IMF} = -3/4$ to -1. 
\item By significantly increasing the temperature of the 
star forming clumps favor the formation of massive stars,  radiative stellar feedback, depending of the circumstances, 
may favor the development of 
a powerlaw exponent $\Gamma_{IMF} =0$ for stellar masses up to a few M$_\odot$.
Strong magnetic fields  are expected to produce similar effects. 
\item The characteristic mass of stars, that is to say, the peak of the 
IMF, is likely, at least in some circumstances, a consequence of the first hydrostatic core
around which tidal forces reduce fragmentation allowing further accretion.
\item The influence of numerical resolution and of the sink particle
algorithms may have drastic consequences on the computed IMF and 
their influences should be carefully verified. This is particularly 
critical for the characteristic mass or peak of the IMF.
\end{enumerate}
\end{summary}

% Future Issues
\begin{issues}[FUTURE ISSUES]
\begin{enumerate}
%\item JWST observations \mike{elaborate on this point?}
\item The links between CMF and IMF should be further investigated. This requests a thorough 
definition and distinction between an observed {\it core} and an effective {\it reservoir}, that is to say the gas mass that is eventually accreted by a star. The statistics of the {\it core} and the {\it reservoir}
should then be compared.
\item  An important issue for future simulations will be to improve spatial 
resolution and systematically check for numerical convergence. In particular, describing well
the first hydrostatic core and making sure that the results are independent of 
sink particle algorithms is crucial.
\item It is fundamental to  improve statistics by performing simulations of more 
massive clumps and by running series of independent realisations of similar initial conditions 
but also to systematically explore the dependence  of the IMF on a large variety of initial 
conditions. The spatial resolution of these simulations must be sufficient to resolve AU scales and 
produce low mass objects whereas the largest spatial scale should be sufficient to form the 
most massive stars.
\item It is necessary to keep developing analytical modeling and to perform comparisons with simulation results.
This is the only way to obtain generic results and to understand the physical processes. 
\item  The physics that is being treated should be futher 
improved. This is particularly the case for  dust, radiative transfer and  non-ideal MHD.
Presently, the opacities and the resistivities still suffer large uncertainties.
Let us stress also that whereas carrying out 
more and more realistic  simulations is obviously an important goal, intentionally-simplified  models are also necessary to understand the physical processes. 
\item A key achievement will be detailed and complete modeling of well observationally constrained  specific regions both for the gas and the stellar populations. 
\end{enumerate}
\end{issues}

\section*{DISCLOSURE STATEMENT}
 The authors are not aware of any affiliations, memberships, funding, or financial holdings that
might be perceived as affecting the objectivity of this review. 

% Acknowledgements
\section*{ACKNOWLEDGMENTS}
We are grateful to Lynne Hillenbrand for making available her compilation of IMF measurements, Sunmyon Chon and Tab Tanvir for sharing their simulation IMF data, and to Matthew Bate, Mark Krumholz, Philippe Andr\'e, Gilles Chabrier, Shu-ichiro Inutsuka, Yueh-Ning Lee and Fr\'ed\'erique Motte for helpful discussions. We thank Estelle Moraux and Ugo Lebreuilly for a critical reading of the manuscript and  David Guszejnov for extended discussions about this review. We are very grateful to Eve Ostriker for a detailed reading of the manuscript and numerous constructive comments which led to significant improvements. 
This research has received funding from the European Research Council synergy grant ECOGAL (Grant : 855130). Support for MYG was provided by NASA through the NASA Hubble Fellowship grant \#HST-HF2-51479 awarded  by  the  Space  Telescope  Science  Institute,  which  is  operated  by  the   Association  of  Universities  for  Research  in  Astronomy,  Inc.,  for  NASA,  under  contract NAS5-26555.

% References
%
% Margin notes within bibliography
%\section*{LITERATURE\ CITED}

%To download the appropriate bibliography style file, please see \url{https://www.annualreviews.org/page/authors/general-information}. \\

%\noindent
%Please see the Style Guide document for instructions on preparing your Literature Cited.

%The citations should be listed in alphabetical order, with no titles. For example:

\bibliography{merged_bibliography}{}
\bibliographystyle{ar-style2}

\end{document}